\documentclass[aip,cha,reprint,
secnumarabic,
nofootinbib, tightenlines,
nobibnotes, showkeys, showpacs,
groupedaddress
]{revtex4-1}

\usepackage{multirow} 
\usepackage[pdftex]{graphicx}
\usepackage{mathtools}
\usepackage{bbold}
\usepackage[caption=false]{subfig}
\usepackage{tabularx}
\usepackage[normalem]{ulem}
\usepackage{comment}

\newcommand{\arXiv}[1]{{\tt \href{http://arXiv.org/abs/#1}{arXiv:#1}}}
\usepackage[usenames,dvipsnames,svgnames,table]{xcolor}
\definecolor{mygreen}{rgb}{0,0.5,0}
\definecolor{myred}{rgb}{0.5,0,0}
\definecolor{myblue}{rgb}{0,0,0.5}

\newcommand{\rgedit}[1]{\textcolor{black}{#1}}

\begin{document}

\title{Exact Coherent Structures and Chaotic Dynamics \rgedit{in a Model} of Cardiac Tissue}

\author{Greg Byrne}
\affiliation{School of Physics, Georgia Institute of Technology, Atlanta, Georgia 30332-0430, USA}
\email{roman.grigoriev@physics.gatech.edu}
\author{Christopher D. Marcotte}
\affiliation{School of Physics, Georgia Institute of Technology, Atlanta, Georgia 30332-0430, USA}\author{Roman O. Grigoriev}
\affiliation{School of Physics, Georgia Institute of Technology, Atlanta, Georgia 30332-0430, USA}

\begin{abstract}
Unstable nonchaotic solutions embedded in the chaotic attractor can provide significant new insight into chaotic dynamics of both low- and high-dimensional systems. In particular, \rgedit{in} turbulent fluid flows, \rgedit{such} unstable solutions \rgedit{are referred to as} exact coherent structures (ECS) \rgedit{and} play an important role in both initiating and sustaining turbulence. The nature of ECS and their role in organizing spatiotemporally chaotic dynamics, however, is reasonably well understood only for systems on relatively small spatial domains lacking continuous Euclidean symmetries. Construction of ECS on large domains and in the presence of continuous translational and/or rotational symmetries remains a challenge. This is especially true for models of excitable media which display spiral turbulence and for which the standard approach to computing ECS completely breaks down. This paper uses the Karma model of cardiac tissue to illustrate a potential approach that could allow computing a new class of ECS on \rgedit{large domains of arbitrary shape} by decomposing them into a patchwork of solutions on smaller domains, or tiles, which retain Euclidean symmetries locally.
\end{abstract}



\maketitle

\begin{quotation}
Spiral turbulence in excitable systems, which features multiple interacting spiral waves that repeatedly break up and merge, is of substantial practical interest because of its relation to life-threatening cardiac arrhythmias such as ventricular fibrillation. While mathematical models of cardiac tissue vary in complexity, even the simplest, such as the Karma system, are too high-dimensional to yield much insight, limiting our understanding of the dynamical mechanisms that sustain spiral turbulence. A growing body of recent work in fluid turbulence supports the conjecture that high-dimensional, spatiotemporal chaos can be understood as a series of transitions between unstable recurrent patterns, also known as exact coherent structures, embedded in the chaotic set on which these dynamics take place. If successful, a similar approach could allow construction of a dynamical description of spiral turbulence in cardiac tissue in terms of models with dramatically reduced dimensionality. Such a reduced-order description may significantly advance our understanding of spiral turbulence and pave the way towards more effective defibrillation techniques. However, the local Euclidean symmetries, inherited by spiral waves from the respective global symmetries of the underlying evolution equations, significantly complicate the job of finding any exact coherent structures. The objective of this paper is to develop a formalism for local symmetry reduction that would enable efficient computation of these structures.
\end{quotation}

\section{Introduction}

\rgedit{Cardiac arrhythmias, such as atrial and ventricular fibrillation,} are characterized by spatially complex, high-dimensional dynamics that are generated as \rgedit{multiple} excitation waves propagate through cardiac tissue, \rgedit{merging and breaking up.}  Despite substantial advances in computing power and the development of detailed ionic models of cardiac cells, quantitatively accurate direct numerical simulation (DNS) of cardiac tissue remains computationally expensive and provides limited dynamical insight into mechanisms that initiate and sustain the spatiotemporal chaos that underpins these arrhythmias.

\rgedit{Substantial progress in understanding some types of spatiotemporal chaos has been made over the past two decades using an idea that is now over a century old. In developing celestial mechanics, Poincar\'e~\cite{poincare1899} realized that unstable equilibria and periodic orbits provide a skeletal structure which organizes chaotic dynamics. His idea was later developed in the context of quantum chaos by Gutzwiller~\cite{gutzwiller71} and subsequently applied to high-dimensional chaos generated by nonlinear partial differential equations (PDEs) such as the Kuramoto-Sivashin\-ski equation~\cite{Christiansen97, lanCvit07} and Ginzburg-Landau equation~\cite{lop05rel}.}

\rgedit{Although in one spatial dimension unstable periodic solutions could be computed using brute-force Newton iteration, this numerical approach becomes intractable for two- and three-dimensional PDEs whose discretizations routinely involve millions of degrees of freedom. In this case both nonchaotic solutions and their spectra can be computed efficiently~\cite{Viswanath:2007re} using a combination of Newton descent, Krylov subspace/GMRES solution of the Newton equations, and ``trust-region'' heuristic for the magnitude of the Newton steps. Newton-Krylov methods facilitated recent studies of various fluid flows at intermediate Reynolds numbers~\cite{GHCW07, Meseguer2009, deLozar2012, Chandler2013}, which produced valuable new insight into the mechanisms that generate and sustain fluid turbulence -- arguably the most challenging unsolved problems of classical physics. Although periodic orbit theory has never been used to analyze spatiotemporally chaotic dynamics in excitable systems, its success in uncovering the mysteries of fluid turbulence gives us hope that it can also generate new insights into the problem of fibrillation and thereby help develop new and improved methods of defibrillation~\cite{IdZhKn95,WaKilIde03,Luther2011}.}

\rgedit{Despite the progress that has already been made in using unstable
nonchaotic solutions to understand chaotic dynamics, many open problems remain. In particular, it is not always clear what types of unstable nonchaotic solutions play a dominant role in spatiotemporal chaos. For spatially extended systems, nonchaotic solutions are characterized not only by their temporal properties (e.g. equilibria, periodic orbits), but also by their spatial structure. In fact, the spatial structure received far more attention in the studies of fluid flows, which provided the motivation for the recent development of periodic orbit theory. As a result, nonchaotic solutions of nonlinear PDEs embedded in the chaotic set on which the dynamics take place have become known as exact coherent structures (ECS), reflecting their connection with the empirically observed coherent structures in fluid flows.}

\rgedit{Spatially extended systems often respect continuous Euclidean symmetries, which complicate the dynamical description based on periodic orbit theory. In particular, continuous symmetries give rise to several other dynamically relevant classes of nonchaotic unstable solutions, such as relative equilibria and relative periodic orbits, which reduce to equilibria and time-periodic orbits in a co-moving reference frame. Notably, the numerical methods for computing such solutions in co-moving frames were developed in the context of excitable/oscillatory systems such as the Barkley model~\cite{barkley1992,Henry2002}, FitzHugh-Nagumo, $\lambda-\omega$, and complex Ginzburg-Landau model \cite{BeTh04}. In two (or three) dimensions, however, rotational symmetry requires that the computation be performed on a circular (or cylindrical) domain, using a polar grid, which severely limits the usefulness of this approach. We recently showed~\cite{Marcotte2014} that, for systems with local Euclidean symmetries, relative equilibria and relative periodic orbits could be computed on domains of arbitrary shape, using arbitrary grids, with the help of weighted Newton-Krylov method.} 

In this work, we use a prototypical model of cardiac tissue to determine which types of nonchaotic solutions form a skeleton for the chaotic set on which spatiotemporally chaotic dynamics that underlie fibrillation take place. \rgedit{Although numerous nonchaotic solutions involving few unstable spirals with well-separated cores were computed using weighted Newton-Krylov method, these solutions were found to be located away from the chaotic set.} Attempts to find unstable solutions embedded in the chaotic set, which feature multiple spirals with closely-spaced cores, were unsuccessful. The analysis of converged few-spiral solutions suggests that this failure is due to the local Euclidean symmetries which cause slow relative drift and/or rotation of the spiral cores. If this is indeed the case more generally, it may be possible to decompose multi-spiral solutions into tiles~\cite{BohHubOtt96,BohHubOtt97}, each of which contains a single spiral described by either a periodic or a relative periodic solution. 
The ability to reduce local symmetries via such a decomposition will provide an important first step towards constructing a reduced-order dynamical description of spiral turbulence.

This paper is organized as follows.  In Section~\ref{sec:two} we describe the modified Karma model of cardiac tissue used in this study.  In Section~\ref{sec:three} we discuss the relationship between Euclidean symmetries and different classes of nonchaotic solutions.  Section~\ref{sec:four} describes our failed attempts to find dynamically relevant \rgedit{nonchaotic solutions} embedded in the chaotic set.  In Section~\ref{sec:five} we explore the effects of local symmetries using artificially constructed unstable nonchaotic solutions with multiple spirals. The construction of tiles and their use for symmetry reduction are discussed in Section~\ref{sec:six}. Our conclusions and directions of further research are presented in Section~\ref{sec:eight}.

\section{The Karma Model}
\label{sec:two}

\rgedit{Mathematical models of cardiac tissue vary greatly in complexity. The most detailed models of cellular kinetics include tens of ordinary differential equations describing the evolution of intracellular concentrations of various ions and the state of ionic channels as well as the transmembrane potential (i.e., the difference between the intra- and the extracellular electric potential). Mono\-domain tissue models also include the electrical coupling between the cells. Bidomain models go one step further, accounting for the spatial and temporal variation of both the intracellular and the extracellular potential.}

The simplest monodomain models can have as few as two local variables that describe the transmembrane potential and the internal state of a cardiac cell (cardiomyocyte).  Despite this simplification, they can produce dynamics that are as rich and complicated as those of the bidomain models \rgedit{including a detailed description of the cellular kinetics.} Most monodomain models are in the class of reaction-diffusion systems described by coupled nonlinear PDEs of the form
\begin{equation}\label{eq:rde}
\partial_t{{\bf w}}=D \nabla^2 {\bf w}+\tau_u^{-1}f({\bf w}),
\end{equation}
where the vector field ${\bf w}=(u,{\bf v})$ incorporates the trans-membrane voltage $u$ and gating variable(s) ${\bf v}$, $D$ is a diagonal matrix of diffusion coefficients, $\tau_u$ is the characteristic time scale, and the nonlinear function $f({\bf w})$ describes the \rgedit{cellular kinetics}.

Although detailed multi-variable models may provide an accurate description of single cell dynamics, \rgedit{they often fail to accurately describe the dynamics of cardiac tissue. In addition, they require substantial computational resources to obtain numerical solutions with appropriate spatial and temporal resolution.} 
Furthermore, when compared to some algebraically simpler, lower-dimensional models, the added complexity provides little additional insight into the {\it dynamical} mechanisms responsible for generating and maintaining complex arrhythmic behaviors \rgedit{such as fibrillation}.

\begin{figure}
    \begin{tabular}{ccc}
      \subfloat[]{\includegraphics[width=0.4\columnwidth]{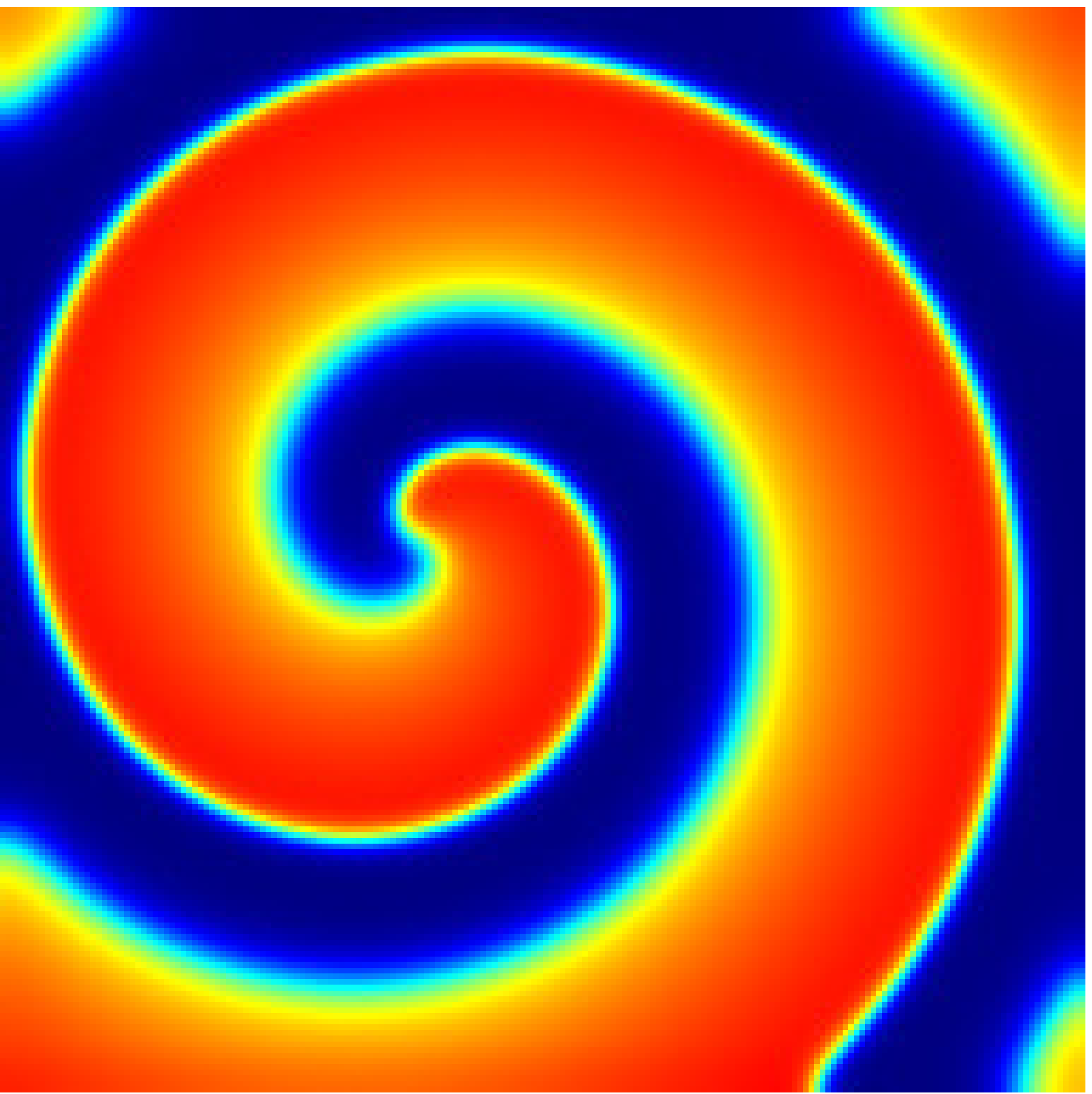}}
      & \subfloat[]{\hspace{2mm}\includegraphics[width=0.4\columnwidth]{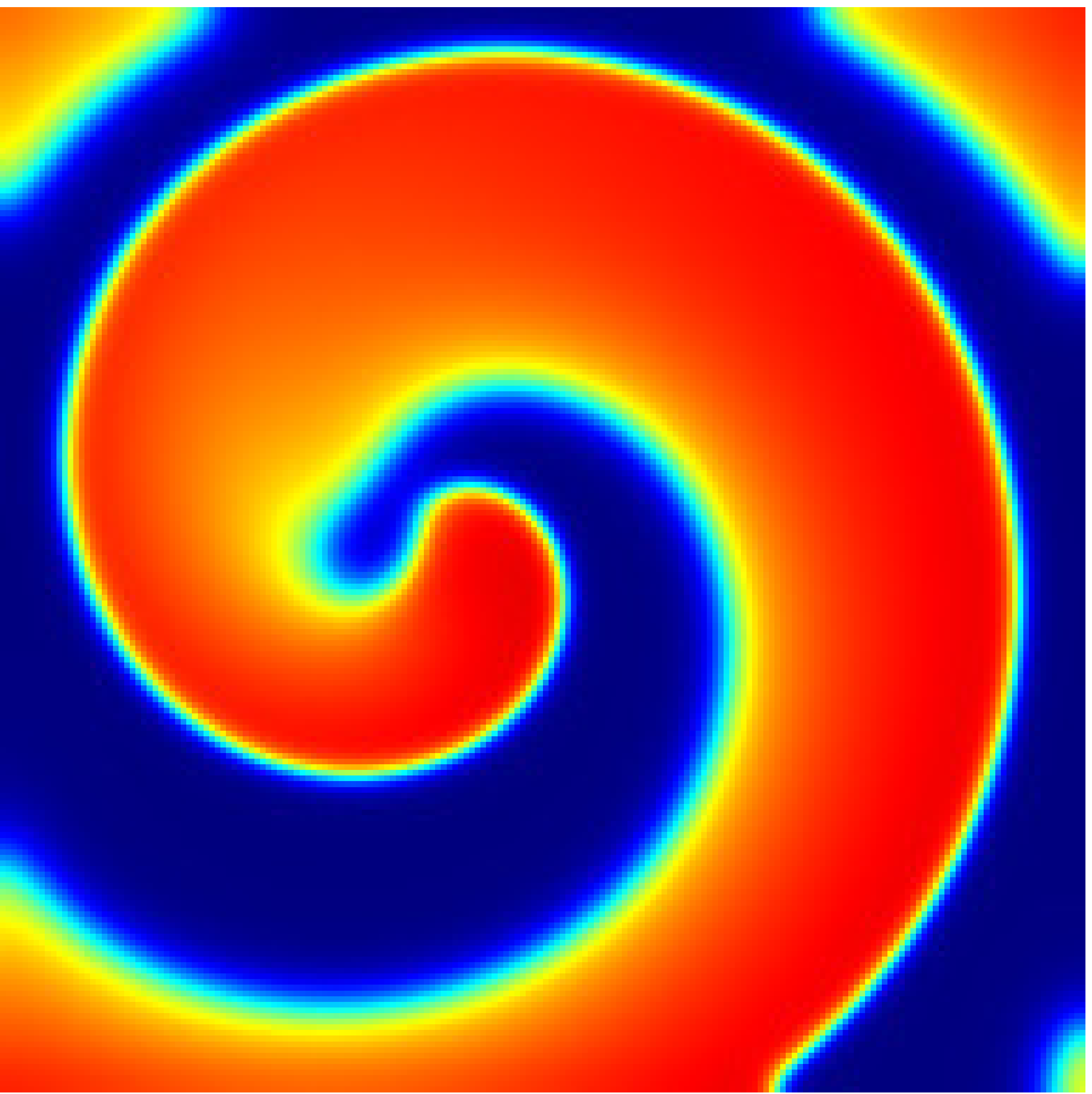}}
      & \includegraphics[height=3.45cm]{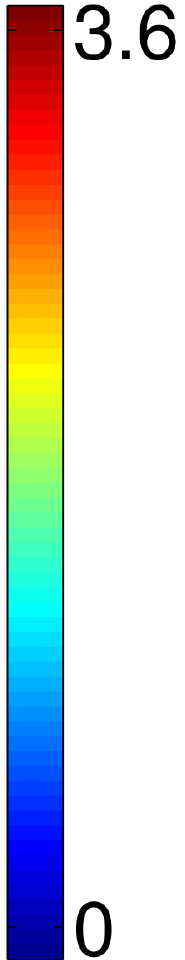}\\
      \subfloat[]{\includegraphics[width=0.4\columnwidth]{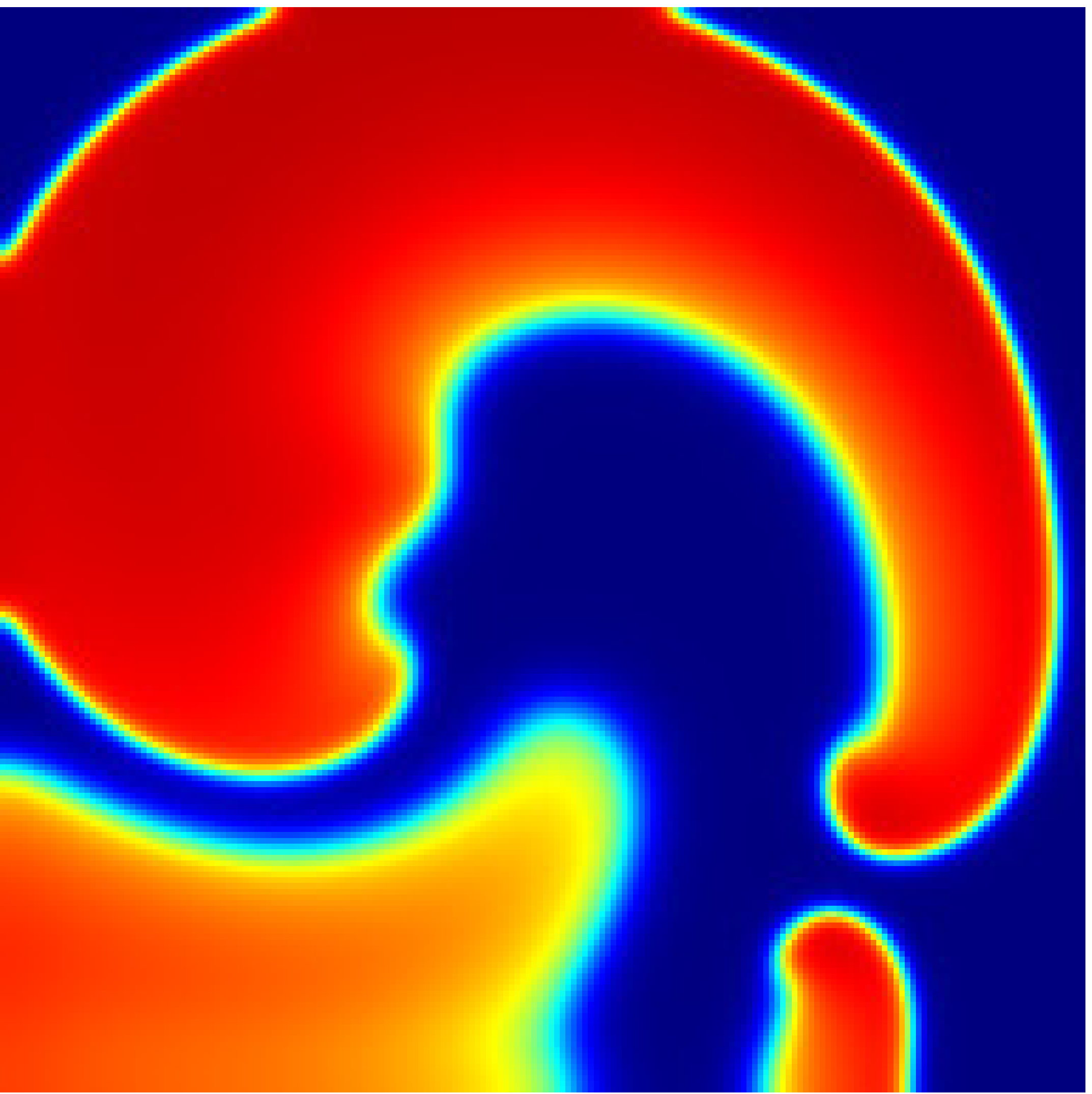}}
      & \subfloat[]{\hspace{2mm}\includegraphics[width=0.4\columnwidth]{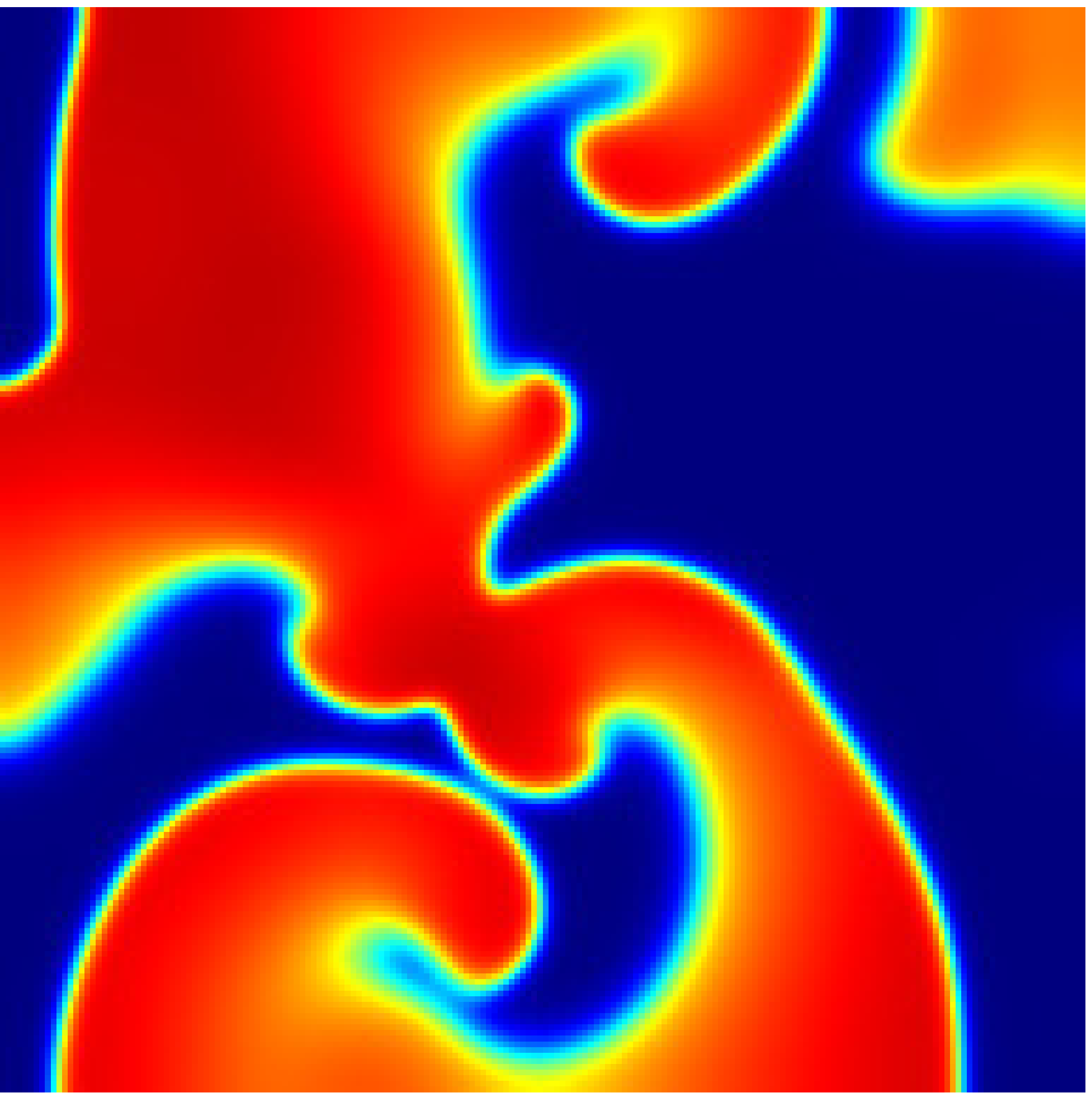}} & \\
    \end{tabular}
  \caption{An unstable spiral wave shown in (a) \rgedit{undergoes an alternans instability (b) and breaks up into multiple spiral segments (c), eventually evolving into spiral turbulence (d),} in a caricature of the transition from atrial tachycardia to atrial fibrillation on a square domain ($L=50.3$ mm) with no-flux boundary conditions. The same color bar is used for the voltage $u$ in this and all subsequent Figures.}
  \label{fig1}
\end{figure}

In contrast, the two-variable Karma model~\cite{Karma1994} provides a greatly simplified description of cardiomyocyte excitation dynamics while still reproducing an essential alternans instability.  This instability is \rgedit{believed to be} responsible for the breakup of a single-spiral solution associated with the arrhythmic state of tachycardia and, ultimately, the transition to fibrillation in two or three dimensions, \rgedit{as illustrated in Fig.~\ref{fig1}.}

We use a slightly modified form of the Karma model for the \rgedit{cellular kinetics}
\begin{equation}
f({\bf w})=
\left(\begin{array}{c}
(u^* - v^{4})[1 - \tanh(u-3)]u^{2}/2 - u \\
\epsilon[\beta \Theta_{s}(u-1) + \Theta_{s}(v-1)(v-1) - v]
\end{array}\right),
\label{eq:karma}
\end{equation}
where $\Theta_{s}(u) = [1 + \tanh(su)]/2$.  The parameter $\epsilon$ describes the ratio of excitation and relaxation time scales, $u^{*}$ is a phenomenologically chosen voltage scale, and $\beta = 1/(1 - \exp{(-R)})$ controls the restitution and susceptibility of traveling excitation waves to alternans.
\rgedit{Following the original study~\cite{Karma1994}}, we choose the parameters $\tau_u=2.5$ ms, $u^*=1.5415$, $\epsilon=0.01$, $D_u = 0.11$ mm$^2$/ms, and $R=1.273$ such that spiral wave solutions break up as a result of the alternans instability.

The modifications (the addition of weak diffusion in the gating variable, replacing the Heaviside step function $\Theta(u)$ with its smoothed version $\Theta_{s}(u)$, and the additional term $\Theta_{s}(v-1)(v-1)$) 
serve to smooth the \rgedit{unphysical discontinuities and singularities of the original model without noticeably changing its dynamics. The stiffness parameter $s$ (in this study we set $s=32$) controls the switching of the gating variable as well as the shortest length and time scales of the spiral wave solutions~\cite{Marcotte2014}. Existing models of cardiac tissue, including the original Karma model, ignore diffusion in describing the dynamics of the gating variable(s). However, it is well-established, although perhaps not well-known, that all relevant ions and small secondary messenger molecules pass through the gap junctions between cardiac cells~\cite{Bevans98,Garcia04}. Hence, weak diffusion (we set $D_v = 5.5 \times 10^{-3}$ mm$^2$/ms) of the gating variable is physiologically justified.}

\rgedit{Our focus here is on spiral turbulence in two dimensions, which can be thought of as a model of atrial fibrillation. Without loss of generality, we can assume that the computational domain $\Omega$ is rectangular.} The PDEs (\ref{eq:rde}) are \rgedit{therefore} discretized on a rectangular uniform two-dimensional grid, with $N$ grid points in the $x$-direction and $M$ grid points in the $y$-direction.  \rgedit{The use of a rectangular grid is consistent with the structure of cardiac tissue, whose properties are generally anisotropic (i.e. different in the direction along, and transverse to, the muscle fibers).}
We use a grid spacing $\Delta x=\Delta y=262$ $\mu$m which corresponds to the typical size of cardiomyocytes and use the physiologically relevant no-flux boundary conditions ${\bf n}\cdot\nabla{\bf w}=0$, unless otherwise specified.

The equations are solved using a fourth-order Runge-Kutta time integrator and a nine-point finite difference approximation for the Laplacian that minimizes the effect of discretization on the rotational symmetry of the evolution equations \cite{Marcotte2014}.  The time step used in the simulations was $\Delta t=0.01$ ms, except for the states shown in Fig.~\ref{fig4} where $\Delta t=0.1$ ms.

\section{Unstable nonchaotic solutions and Euclidean symmetries}
\label{sec:three}

On an unbounded two-dimensional domain $\Omega=\mathbb{R}^2$, reaction-diffusion equations are equivariant under the set of transformations, or actions, $g$ that form a group $G=\mathcal{G} \otimes E^{+}(1)$.  In particular, {\it spatial} transformations from the Euclidean subgroup $\mathcal{G}=E(2)$ include continuous translations, continuous rotations and discrete reflections.  The subgroup $E^{+}(1)$ reflects the fact that the dynamics are invariant under {\it temporal} \rgedit{translations, but are not} time-reversible.

Spatially uniform solutions ${\bf w}$ such that $f({\bf w})=0$ \rgedit{(e.g. the stable resting state ${\bf w}\approx (0,0)$),} invariant with respect to the entire group $G$, are examples of the simplest type of \rgedit{nonchaotic solutions} -- equilibria.  
However, typical \rgedit{nonchaotic} solutions respect only some of the transformations in $G$, lowering effective symmetry to a subgroup of $G$. In particular, a plane wave solution traveling with velocity ${\bf c}$ is a relative equilibrium generated by a continuous translational symmetry which satisfies $\partial_t{\bf w} - {\bf c}\cdot\nabla{\bf w} = 0$. In this case $\partial_t$ is the generator of translations in time and $\nabla$ is the generator of translations in space. Another relative equilibrium generated by continuous rotational symmetry, which satisfies $\partial_t{\bf w} - \omega\partial_{\theta}{\bf w} = 0$, corresponds to a spiral wave. Here $\theta$ is the polar angle in the $(x,y)$ plane, $\omega$ is the rotational frequency, and $\partial_\theta$ is the generator of rotations. These relative equilibria reduce to (nonuniform) equilibria in a reference frame translating with velocity ${\bf c}$ or rotating with angular velocity $\omega$, respectively. For these special solutions, evolution (i.e., time translation) is equivalent to a spatial translation or rotation.

The introduction of physical boundaries has a markedly different effect on these solutions. \rgedit{For no-flux boundary conditions,} the effect of the boundaries is determined \rgedit{primarily} by the shape (and size) of the domain $\Omega$ on which the evolution equations are defined. Some \rgedit{nonchaotic} solutions, such as the uniform equilibria or spiral waves, survive unchanged (or almost unchanged) when computed on domains of very different shapes and sizes, while others, such as plane waves, cease to exist except as a transient.
The effect of the boundaries on spiral waves is of the most interest to us, since spiral waves play an important role in sustaining spatiotemporally chaotic dynamics underlying complex cardiac arrhythmias. 

\rgedit{Although it is easier to compute single-spiral waves described by relative equilibria on circular domains, resulting solutions are not representative of any cardiac rhythms. For instance, a spiral wave describing tachycardia would not correspond to a relative equilibrium because atria are not circular. Similarly, the atria cannot be tiled with circles, so multi-spiral patterns arising during atrial fibrillation cannot be decomposed into relative equilibria, even on arbitrarily short time scales. Hence, it is crucial to understand the properties of spiral wave solutions on domains of arbitrary shape. Rectangular domains are more relevant, since they break all global Euclidean symmetries and, at the same time, can be used to tile a surface. Fig.~\ref{fig1}(a) shows a sample unstable periodic solution computed on a square domain of side $L=50.3$ mm with no-flux boundary conditions using weighted Newton-Krylov solver~\cite{Marcotte2014}. It describes a pinned spiral wave with a period of $T=125.98$ ms and wavelength $\lambda = 19.4$ mm (for reference, the left human atrium has a typical diameter of 30-40 mm).} 

\rgedit{Since this is not a relative equilibrium, the actions of time translation and rotation become distinct, $\partial_t{\bf w}\ne\omega\partial_\theta{\bf w}$, where $\omega=2\pi/T$.  However, the effect of the boundaries on the spatial structure of the spiral wave as well as its temporal properties (e.g. its period $T$) becomes exponentially weak for sufficiently large $L$. The solutions computed on domains of different size become indistinguishable to numerical precision in the overlap region, with the difference becoming significant only near the boundary~\cite{Marcotte2014}. In particular, the periodic solution shown in Fig.~\ref{fig1}(a) is almost identical to a relative equilibrium in the interior of the domain.} Hence, although the \rgedit{boundary conditions break} the global rotational symmetry of the spiral wave, local rotational symmetry is preserved away from the boundaries, where \rgedit{$\partial_t{\bf w}\approx\omega\partial_\theta{\bf w}$ to numerical accuracy.}

For the parameters chosen in this study, single-spiral solutions are unstable with respect to the alternans instability. \rgedit{If allowed to evolve for a sufficiently long time, an initial condition that is arbitrarily close to the spiral wave shown in Fig.~\ref{fig1}(a) will deviate from that unstable solution exponentially fast, developing modulation in the width of the excitation wave (alternation of the action potential duration), as illustrated in Fig.~\ref{fig1}(b). When the amplitude of the modulation becomes sufficiently large, the tissue fails to recover between two successive wave fronts, leading to conduction block and a breakup of the spiral wave (cf. Fig.~\ref{fig1}(c)). After a sequence of breakups, a state of spiral turbulence featuring multiple interacting spiral waves is established (cf. Fig.~\ref{fig1}(d)).}

\begin{figure}[t]
\centering
\includegraphics[width=0.85\columnwidth]{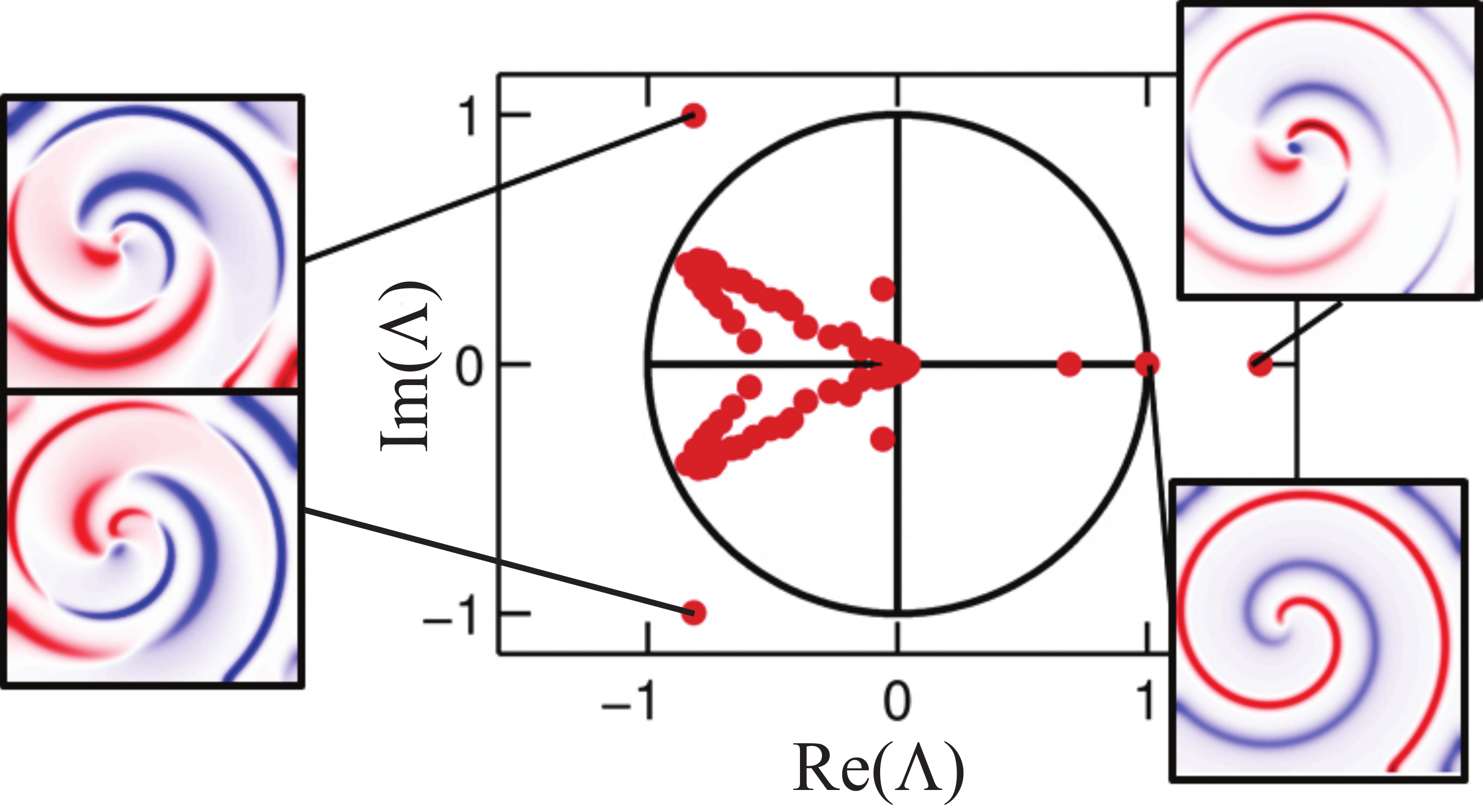}
\caption{The spectrum of the single-spiral solution shown in Fig.~\ref{fig1}(a).  The unstable eigenmodes are associated with the alternans instability.  \rgedit{The Goldstone mode with $\Lambda=1$} is associated with an infinitesimal time translation.}
\label{fig2}
\end{figure}

\rgedit{In the process of computing a solution, the Newton-Krylov solver also generates its stability spectrum. The leading Floquet modes (or simply eigenmodes)} of the single-spiral solution \rgedit{depicted in Fig.~\ref{fig1}(a)} are shown in Fig.~\ref{fig2}. The pair of unstable eigenmodes on the left, corresponding to a complex conjugate pair of Floquet multipliers \rgedit{(or eigenvalues)} outside the unit circle, are associated with the variation in the width of the excitation wave (i.e., alternans instability).  A single unstable eigenmode on the right \rgedit{looks like a frustrated Golstone mode $\partial_y{\bf w}$; it is associated with a weakly broken translational symmetry and corresponds to an almost rigid vertical shift of the entire spiral wave.}  The Goldstone mode $\partial_t{\bf w}$ has a unit \rgedit{eigenvalue} $\Lambda=1$ and is associated with a continuous symmetry (time-translation).

For a spiral wave solution of the PDEs (\ref{eq:rde}) on an unbounded domain we should expect to find three \rgedit{Goldstone} modes corresponding to each of the three continuous symmetries \rgedit{of $E(2)$}: rotation and translations in the $x$ and $y$ directions. The continuous time-translation symmetry does not produce an additional \rgedit{Goldstone} mode because for spiral waves described by relative equilibria $\partial_t{\bf w}=\omega\partial_\theta{\bf w}$.
On finite, but sufficiently large, domains the three continuous spatial symmetries become local, rather than global, but should still generate three \rgedit{unit} eigenvalues, \rgedit{even after (\ref{eq:rde}) is discretized, provided the grid is sufficiently fine. Indeed, this was found to be the case~\cite{Marcotte2014} for small values of the stiffness parameter ($s\lesssim 3$).}
As is typical of solutions in the presence of continuous translational symmetries, in this limit the spiral waves drift, \rgedit{and hence} are described by relative periodic solutions (they become periodic in a reference frame moving with the drift velocity). 

However, for the values of parameters considered here, there is also a local source of continuous symmetry breaking -- the discreteness associated with the computational grid (or the cellular structure of the tissue). The modified Karma model possesses an internal length scale $\ell_v=2\sqrt{D_u\tau_u}/(\beta s)$. For $s=32$, this gives $\ell_v\sim 24$ $\mu$m, i.e., much smaller than the grid spacing $\Delta x=262$ $\mu$m. Hence the computational grid is too coarse to completely resolve the spatial structure of the spiral wave, effectively reducing the continuous translational symmetries to discrete translations, $x\to x+\Delta x$ and $y\to y+\Delta y$, such that \rgedit{$\partial_x{\bf w}$ and $\partial_y{\bf w}$ no longer correspond to eigenmodes}~\cite{Marcotte2014}.

Our numerical results are consistent with this conclusion. On square domains of sufficiently large size, \rgedit{for $s\gtrsim 3$, there is a large, but finite,} number of unstable spiral waves described by periodic solutions \rgedit{with essentially identical periods and stability spectra} and differing by a discrete shift in the $x$ and/or $y$ direction. Fig.~\ref{fig1}(a) shows just one of those solutions. Each \rgedit{one of these spiral waves} possesses a \rgedit{Goldstone} mode $\partial_t{\bf w}$ which \rgedit{coincides} with $\partial_\theta{\bf w}$ (to numerical precision) away from the boundaries, i.e., continuous rotational symmetry is preserved locally.

\section{Exact coherent structures}
\label{sec:four}

Stable and unstable uniform states, plane and spiral excitation waves discussed previously illustrate the types of solutions (equilibria, periodic orbits, relative equilibria and relative periodic orbits) relevant to cardiac tissue dynamics, \rgedit{including fibrillation}. However, none of the solutions are \rgedit{embedded in the chaotic set on which arrhythmic dynamics underlying fibrillation take place, i.e., they do not correspond to ECS. For instance, while locally the excitation waves almost always take the shape of small spirals during fibrillation, they never organize globally into a single large spiral wave.}

In order to find \rgedit{ECS, i.e., global unstable nonchaotic solutions} embedded in the chaotic set, we used the method of close returns~\cite{pchaot} which has been used successfully in the context of fluid turbulence~\cite{CviGib10}. The procedure involves finding near-recurrences in the {\it chaotic} \rgedit{solutions}, which become initial conditions that are subsequently refined into exact {\it nonchaotic} solutions
\rgedit{using the Newton-Krylov solver~\cite{Marcotte2014}.
In the presence of global symmetry,} initial guesses for relative equilibria, periodic orbits, as well as relative periodic orbits can be found as the minima of the recurrence function
\begin{equation}
\label{eq:recf}
E(t,\tau)\equiv\min_{g\in \mathcal{G}}
{\|g{\bf w} (t) - {\bf w} (t -\tau) \|_2},
\end{equation}
where $\|\cdot\|_2$ denotes the 2-norm in $\mathbb{R}^{2NM}$ \rgedit{and the group $\mathcal{G}$ describes the symmetries in the presence of boundaries. However, we found that the minima of (\ref{eq:recf}) are always achieved for $g\approx \mathbb{1}$ and in practice we can set $\mathcal{G}=\{\mathbb{1}\}$.} 

\begin{figure}[t]
\centering
\includegraphics[width=\columnwidth]{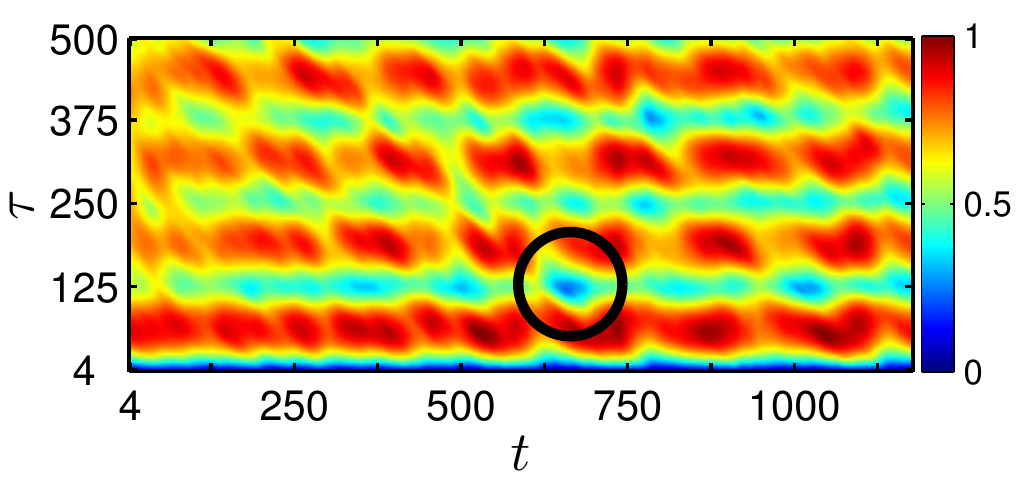}
\caption{A fragment of the normalized recurrence function $E(t,\tau)$, where $t$ and $\tau$ are in units of ms. The black circle identifies a minimum associated with a close return of a chaotic \rgedit{solution}.  Minima such as this one are used to identify initial \rgedit{conditions for refinement into ECS using the Newton-Krylov solver.}}
\label{fig3}
\end{figure}

\rgedit{It should be noted that setting $\mathcal{G}=\{\mathbb{1}\}$ in (\ref{eq:recf}) does not constrain the exact solutions to absolute equilibria and absolute periodic orbits. For instance, a slowly drifting or rotating solution described by a relative periodic orbit or relative equilibrium, such that $g{\bf w}(t)={\bf w}(t-T)$ with $g\approx \mathbb{1}$, will generate a minimum of (\ref{eq:recf}) with $\tau\approx T$ when $\mathcal{G}=\{\mathbb{1}\}$. Solutions that are characterized by fast global rotation or translation (e.g. spiral or plane waves), and for which $g$ is significantly different from $\mathbb{1}$, are characterized by a high degree of spatial coherence. The lack of global spatial coherence is a distinguishing feature of fibrillation, and therefore we should not expect to find any solutions that exhibit fast global rotation or drift. } 

A fragment of the recurrence plot for a chaotic solution computed on a square domain $\Omega$ with side $L=50.3$ mm and periodic boundary conditions is shown in Fig.~\ref{fig3}. 
\rgedit{The periodic boundary conditions lead to chaotic dynamics that are qualitatively identical to those in the presence of no-flux boundary conditions even on relatively long time scales. However, for periodic boundary conditions fibrillation persists indefinitely, while for no-flux boundary conditions it can terminate spontaneously due to the spiral cores colliding with the boundaries and disappearing.} Once a sufficiently low minimum (circled) of the recurrence function $E(t,\tau)$ is identified, the corresponding state ${\bf w}(t-\tau)$ is used as the initial guess for a solution with period close to $\tau$.

\begin{figure}
    \begin{tabular}{ccc}
        \subfloat[]{\includegraphics[width=0.4\columnwidth]{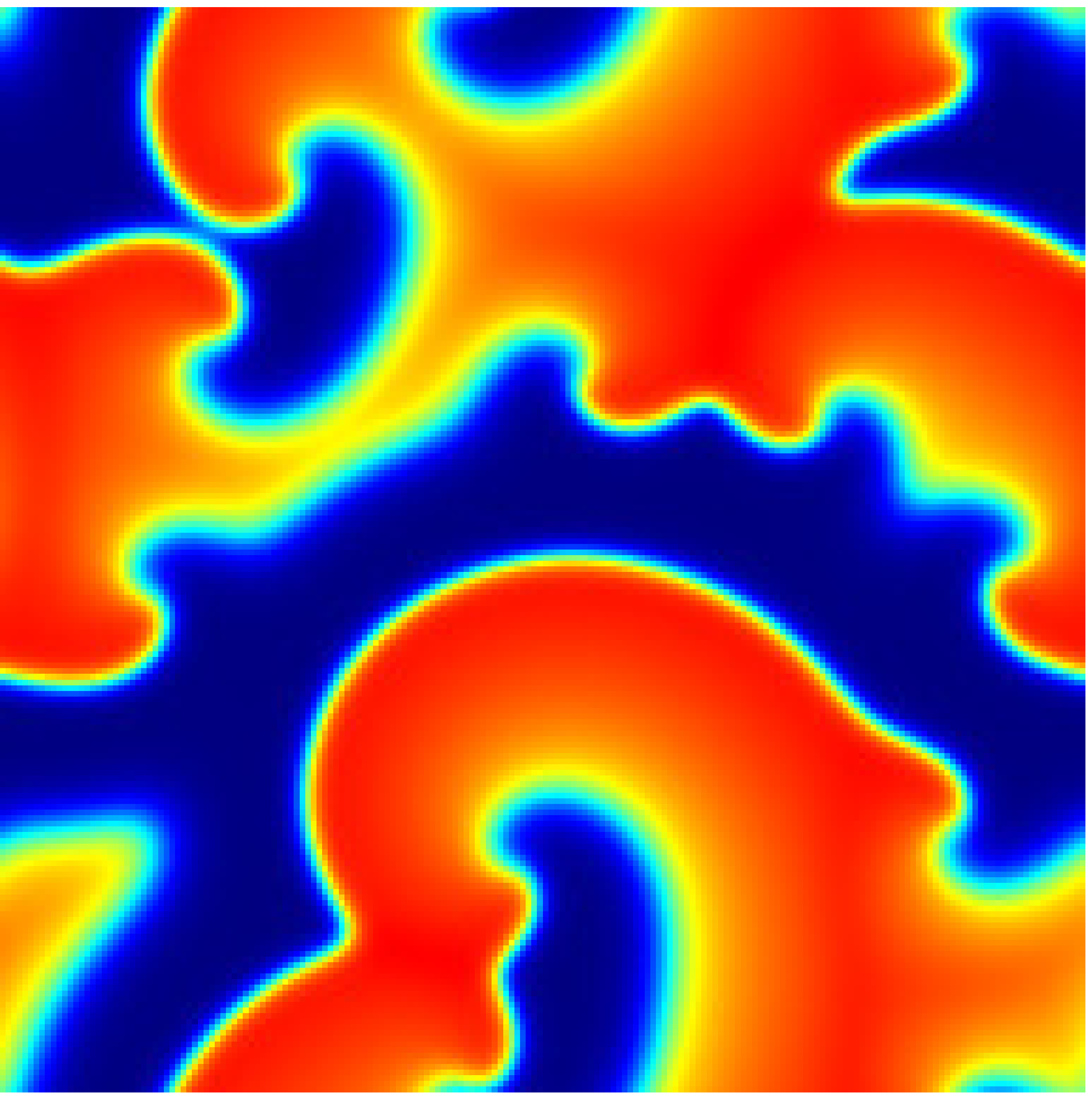}}
      & \subfloat[]{\hspace{2mm}\includegraphics[width=0.4\columnwidth]{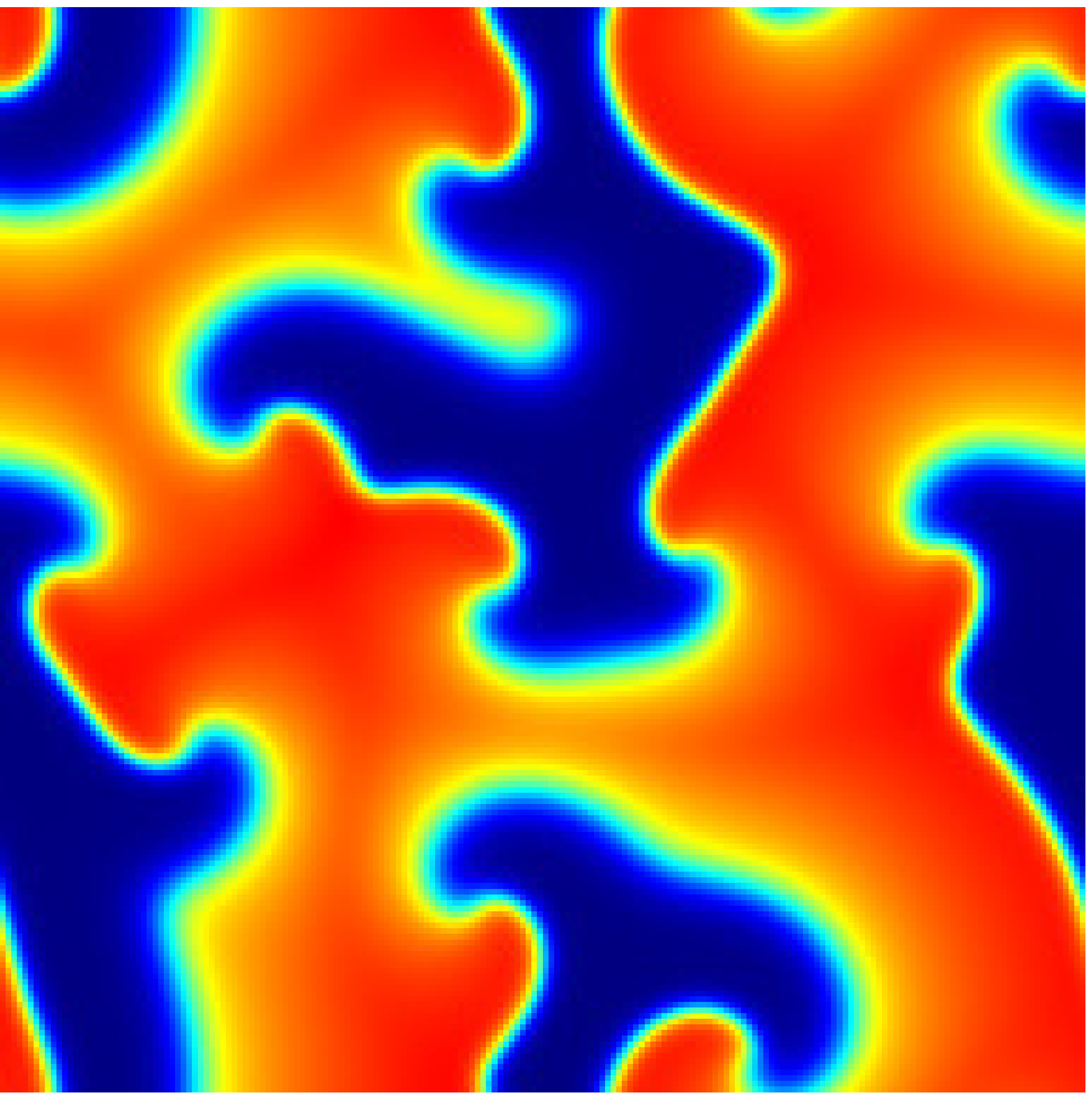}} \\
      \subfloat[]{\includegraphics[width=0.4\columnwidth]{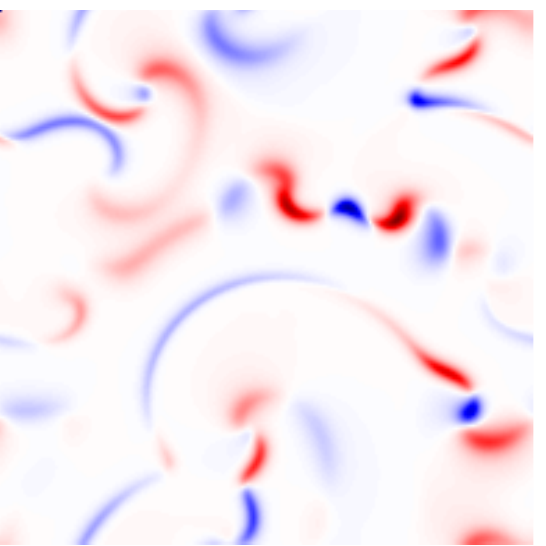}}
      & \subfloat[]{\hspace{2mm}\includegraphics[width=0.4\columnwidth]{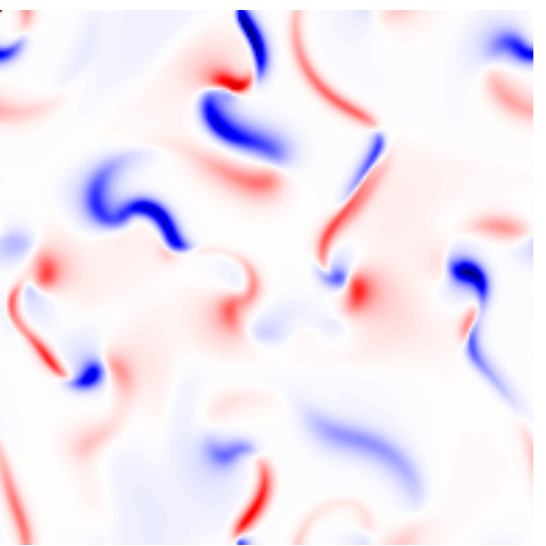}} & \hspace{0.12cm} \includegraphics[height=3.7cm]{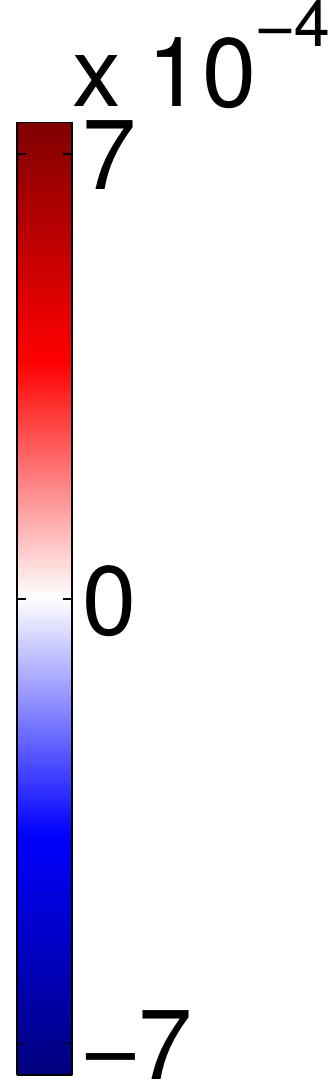}\\
      \subfloat[]{\includegraphics[width=0.4\columnwidth]{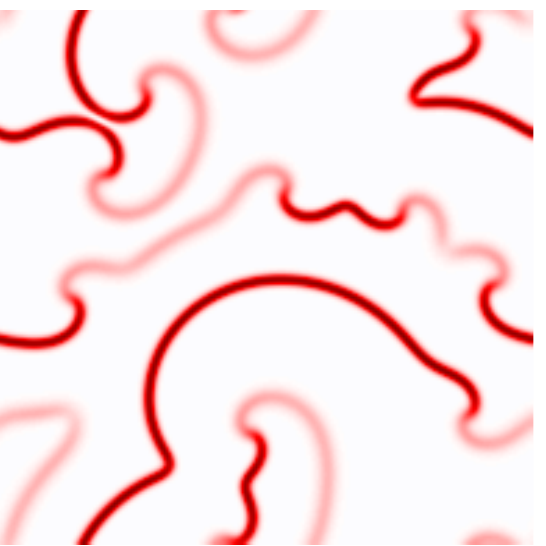}}
      & \subfloat[]{\hspace{2mm}\includegraphics[width=0.4\columnwidth]{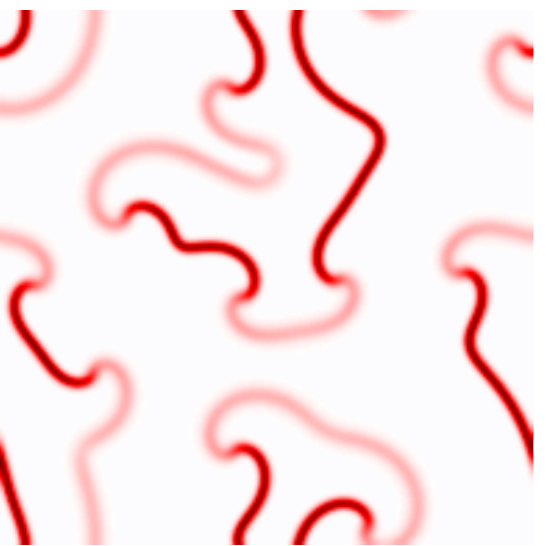}}
      & \hspace{-0.4cm} \includegraphics[height=3.4cm]{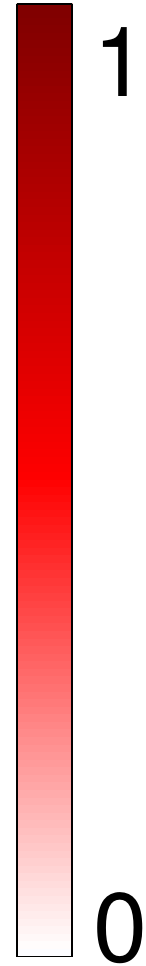}\\
      \subfloat[]{\includegraphics[width=0.4\columnwidth]{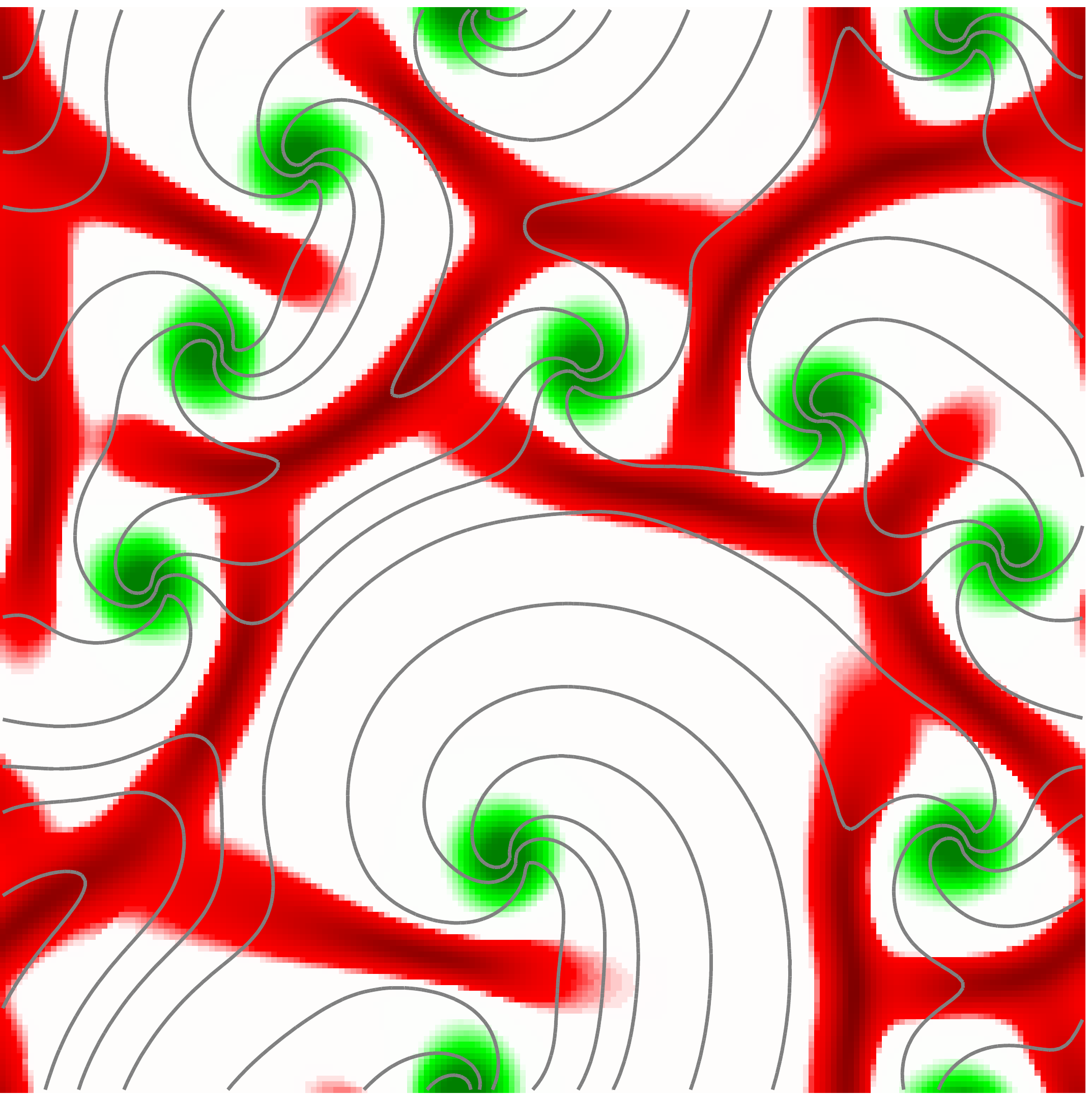}}
      & \subfloat[]{\hspace{2mm}\includegraphics[width=0.4\columnwidth]{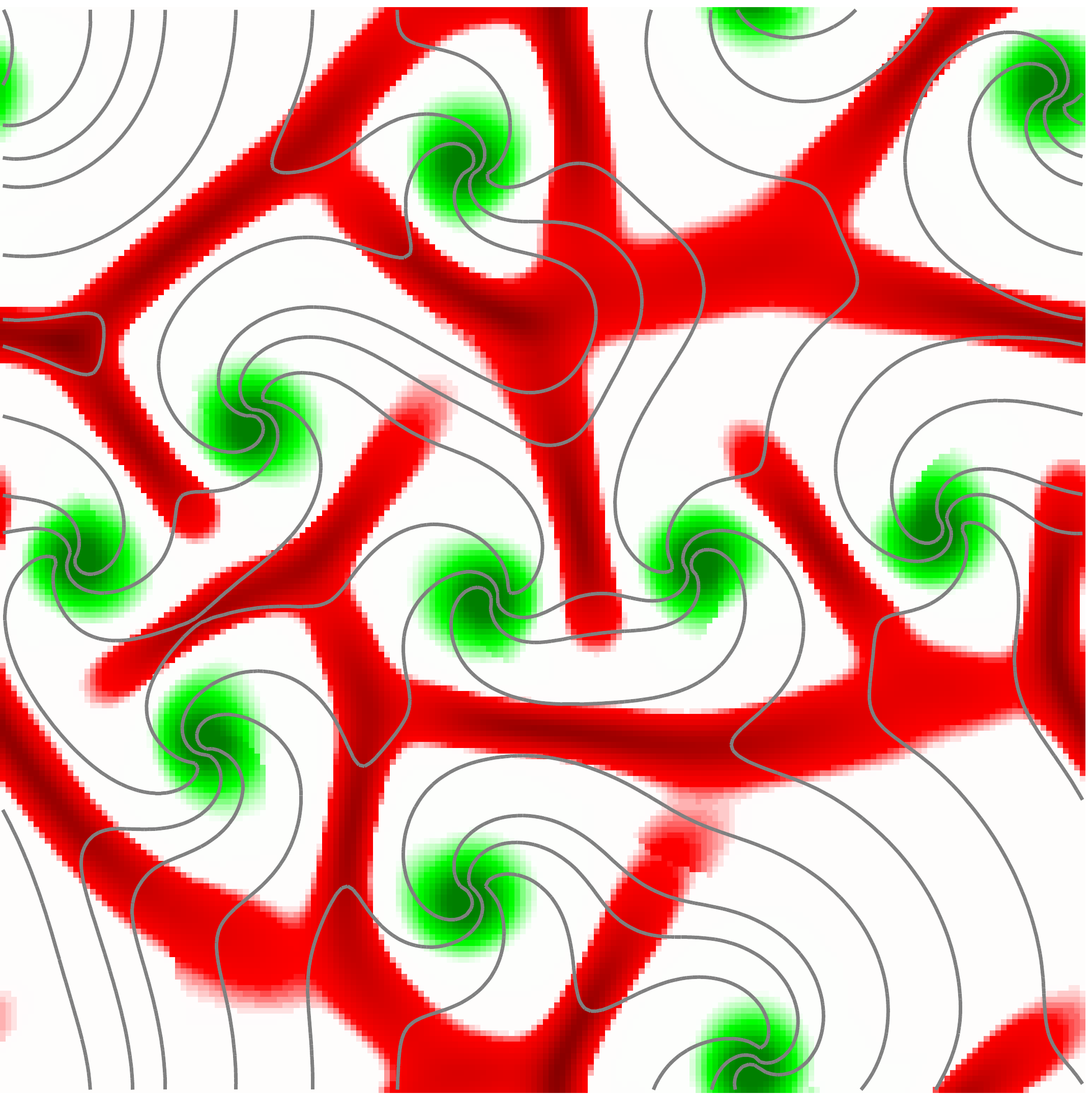}} 
      & \hspace{-0.4cm} \includegraphics[height=3.5cm]{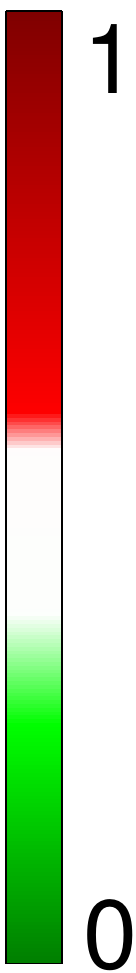}\\
    \end{tabular}
  \caption{(a-b) Snapshots of the voltage $u(t)$ for two initial guesses which correspond to minima of $E(t,\tau)$. (c-d) The relative residual $[u(t)-u(t-\tau)]/\|u(t)\|_\infty$. (e-f) The magnitude of the voltage gradient $|\nabla u(t)|$.  (g-h) The cycle area $I_1$ \rgedit{(to be defined in Section~\ref{sec:six})}. Level sets of $v$ are shown in gray. The domain is a square of side $L=50.3$ mm with periodic boundary conditions.}
  \label{fig4}
\end{figure}

Rather surprisingly, none of the initial guesses we tried converged to either relative equilibria, periodic orbits, or to relative periodic orbits. (We did not search for equilibria, since $\|\partial_t{\bf w}\|$ never becomes small for chaotic solutions.) Figs.~\ref{fig4}(a) and \ref{fig4}(b) show the voltage component $u$ for two typical examples of the numerous multi-spiral states identified using the recurrence analysis.  
They nearly recur after $\tau=251.94$ ms and $\tau=251.83$ ms, respectively. These values of $\tau$ correspond to \rgedit{approximately} double the temporal period of a single spiral shown in Fig.~\ref{fig1}(a), $2T = 251.96$ ms. Newton iterations stagnate at the values of relative residual $\|{\bf w}(\tau) - {\bf w} (0) \|_2/\|{\bf w} (0)\|_2$ equal to $6\times 10^{-3}$ for the state shown in Fig.~\ref{fig4}(a) and $9\times 10^{-3}$ for the state shown in Fig.~\ref{fig4}(b), compared to $O(10^{-13})$ for the converged single-spiral state shown in Fig.~\ref{fig1}(a).  The voltage components $u(t)-u(t-\tau)$ of the corresponding residual are shown in Figs.~\ref{fig4}(c-d).  They are spatially localized in the regions where $|\nabla u(t)|$ has the largest magnitude (cf.  Figs.~\ref{fig4}(e-f)), or near the front and back of the excitation wave. In order to interpret these findings, next we consider multi-spiral solutions that are constructed artificially and do not lie on the chaotic set on which \rgedit{fibrillation takes} place.

\section{Multi-Spiral Solutions with Local Symmetry}
\label{sec:five}

The weak effect of the boundaries on the essential dynamics of the spiral wave suggests that the local translational and rotational symmetries of single-spiral solutions should be inherited by multi-spiral solutions, provided the spacing between the spiral cores is sufficiently large. To verify this and to gain insight into the properties of multi-spiral states on finite domains, we explore the simplest artificially generated solutions, which are constructed by assembling pairs of converged single-spiral solutions.

\subsection{Local Rotational Symmetries}
\label{sec:fiveSS1}

\begin{figure}[t]
 \begin{tabular}{cc}
    \subfloat[]{\includegraphics[width=0.47\columnwidth]{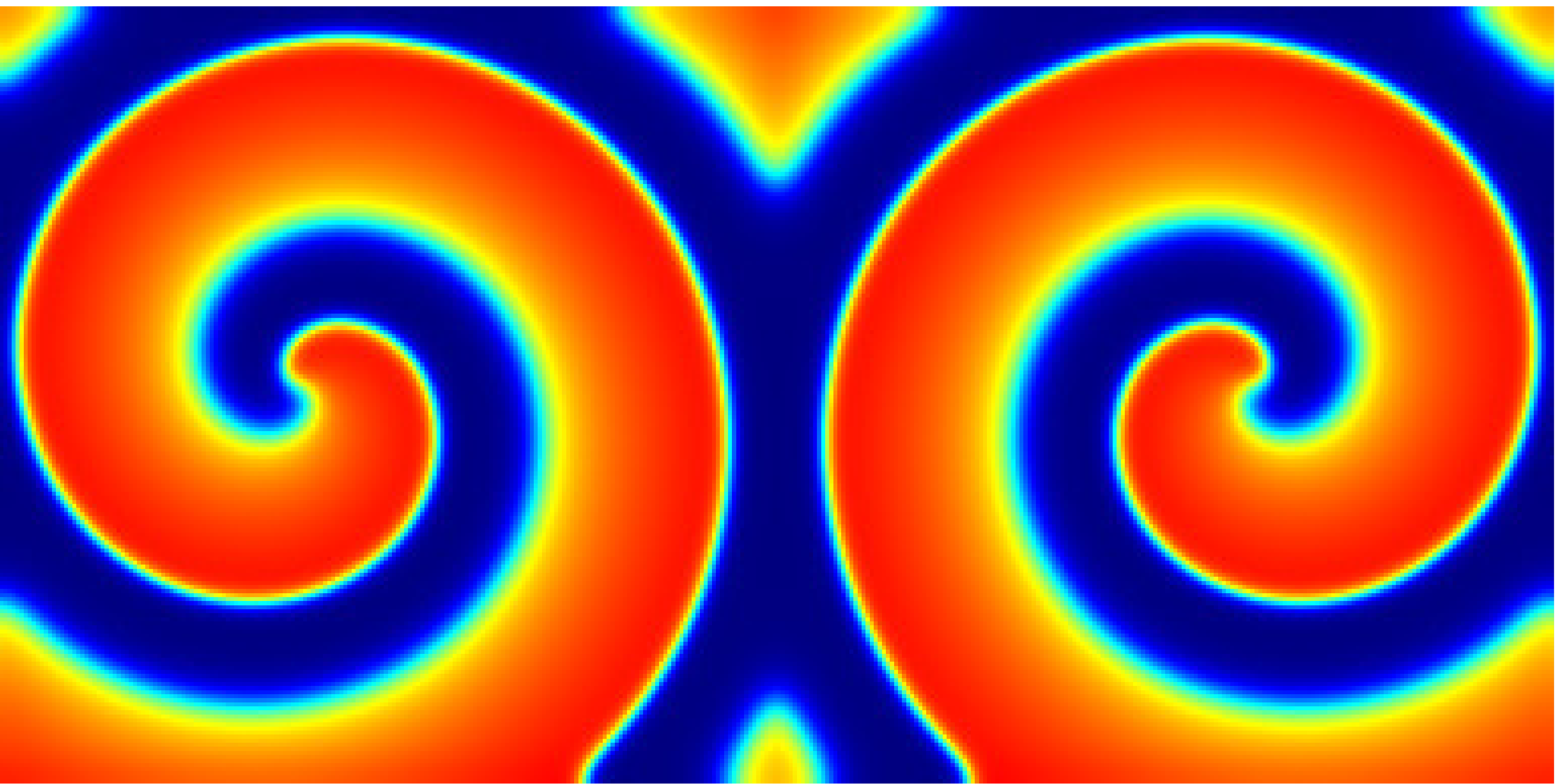}} &
    \subfloat[]{\hspace{2mm}\includegraphics[width=0.47\columnwidth]{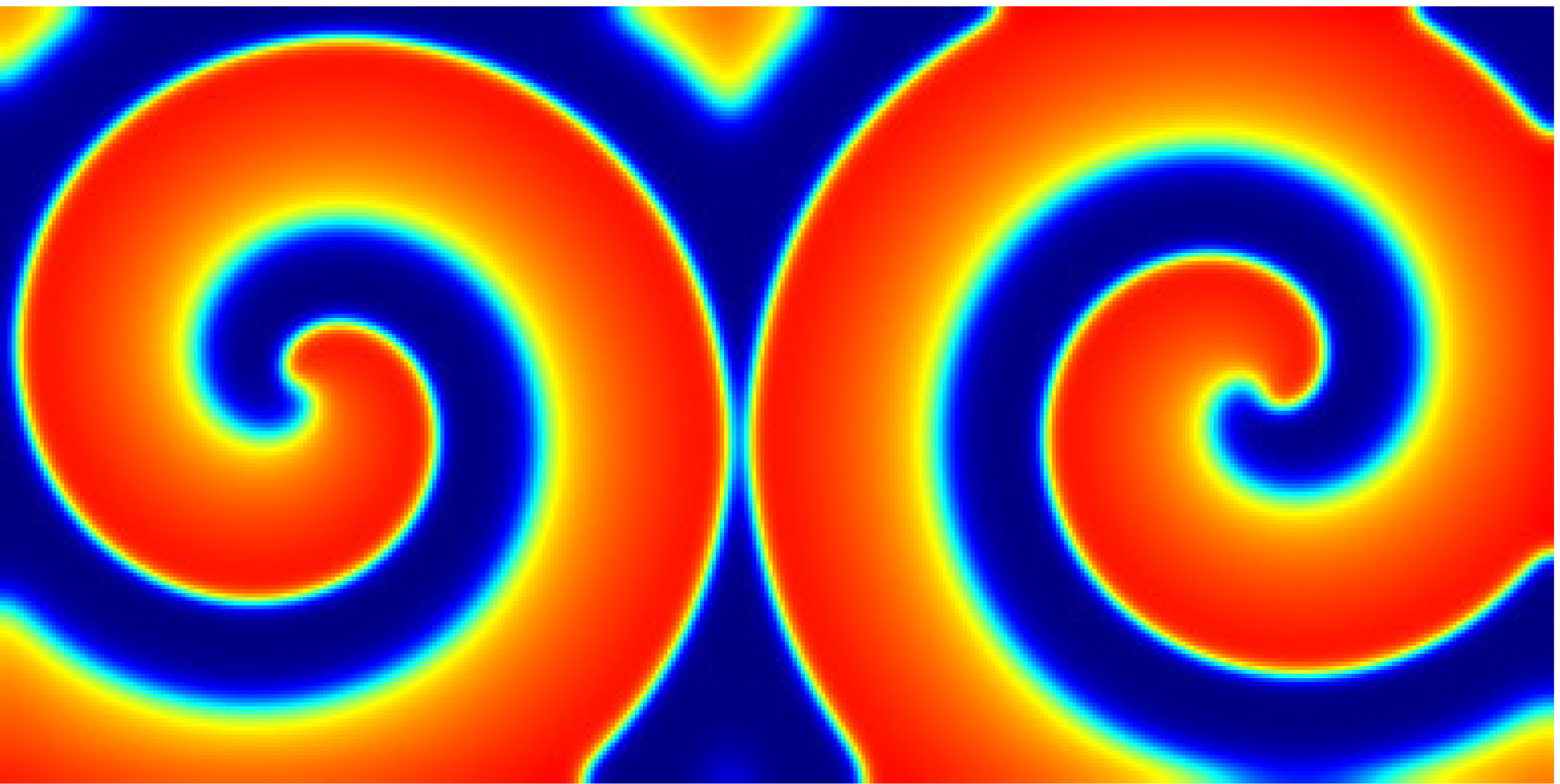}}\\
    \subfloat[]{\includegraphics[width=0.47\columnwidth]{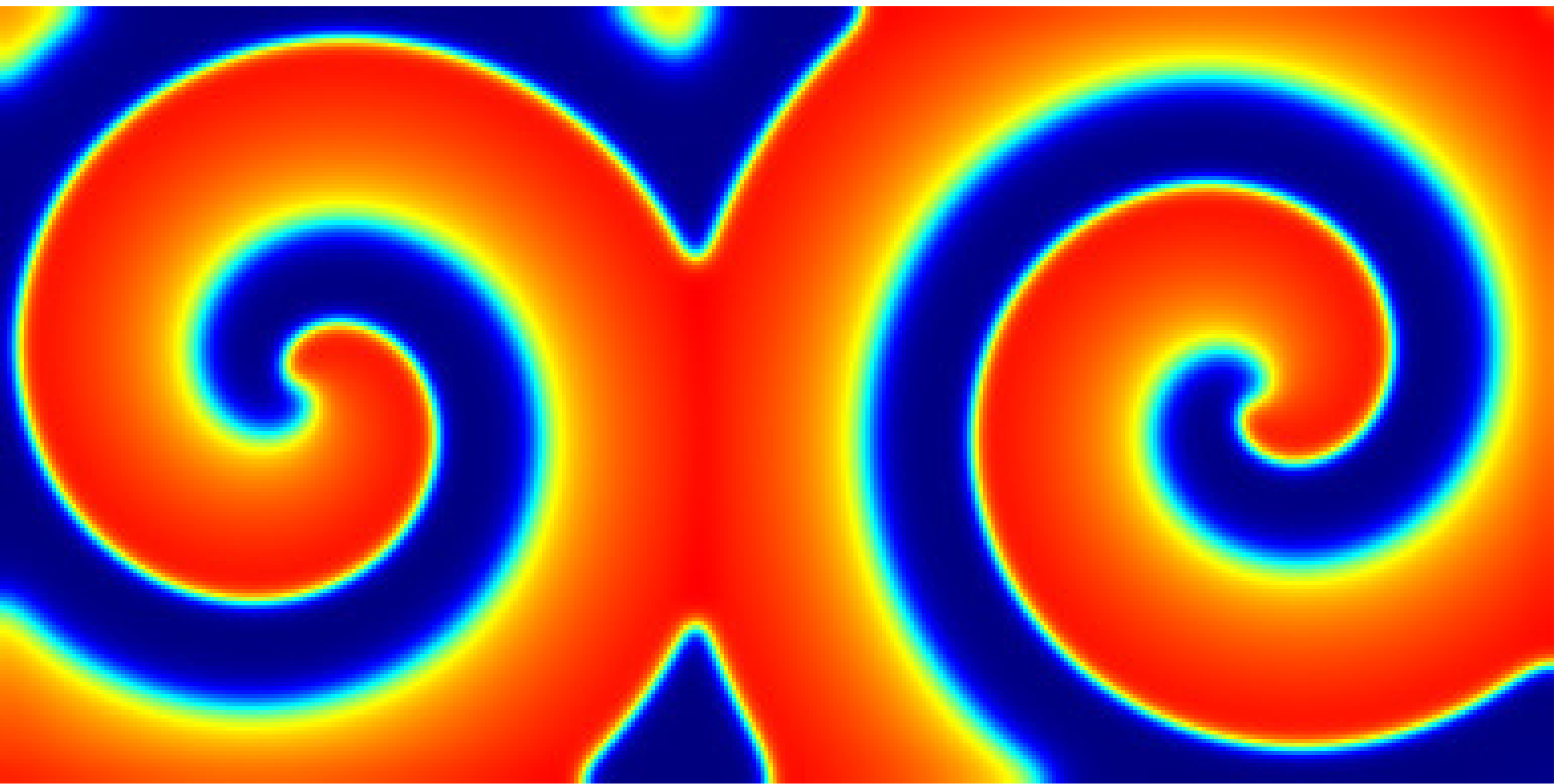}} &
    \subfloat[]{\hspace{2mm}\includegraphics[width=0.47\columnwidth]{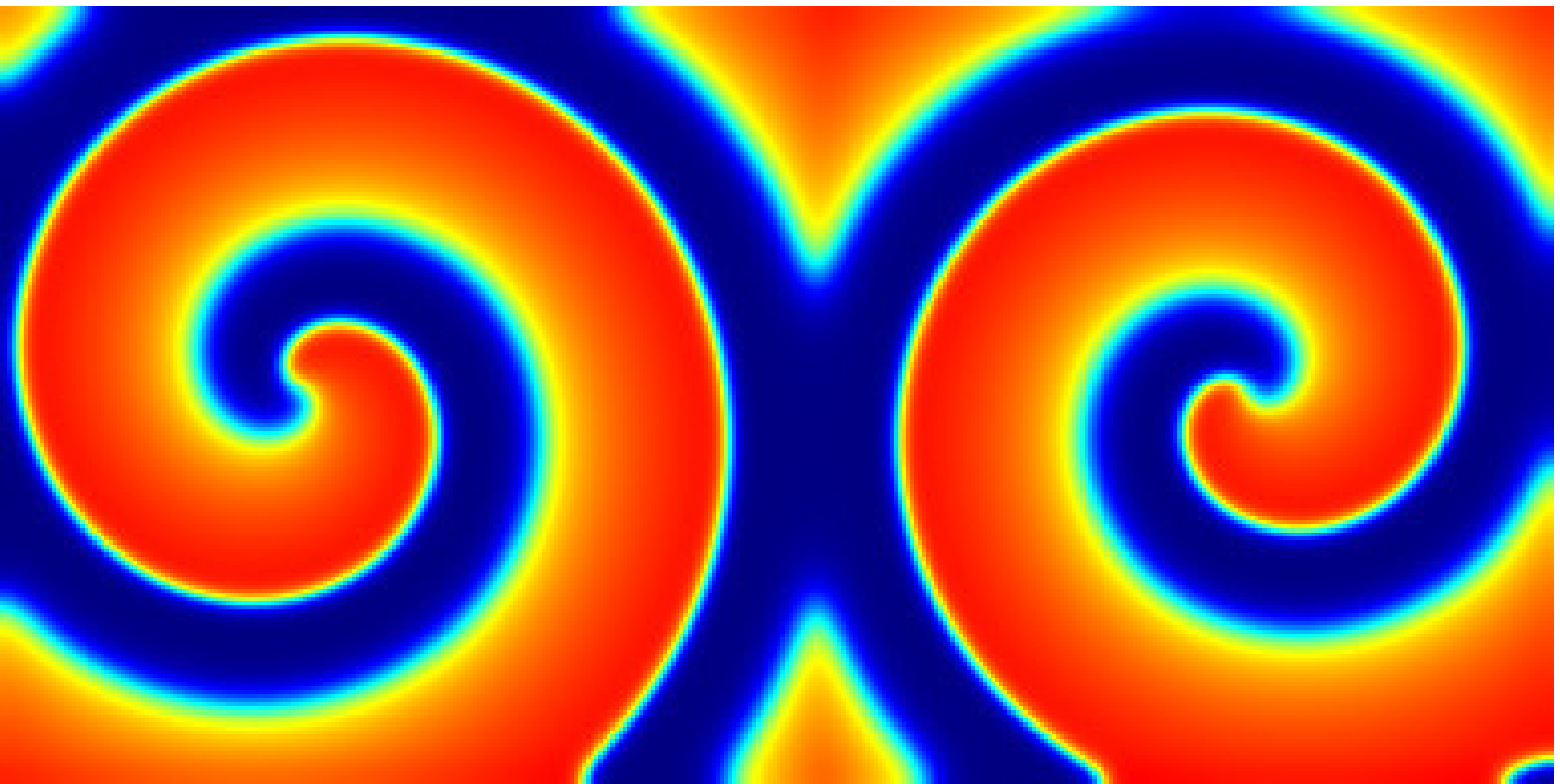}}
  \end{tabular}
  \caption{Snapshots of the voltage $u$ for counter-rotating two-spiral solutions on a rectangular domain of size $50.3$ mm $\times$ $100.6$ mm. The phase shifts $\delta\theta=2\pi(\delta t/T)$ between neighboring spirals are: (a) $0$, (b) $\pi/2$, (c) $\pi$, (d) $3\pi/2$.}
\label{fig5}
\end{figure}

\begin{figure}[t]
  \begin{tabular}{cc}
    \subfloat[]{\includegraphics[width=0.47\columnwidth]{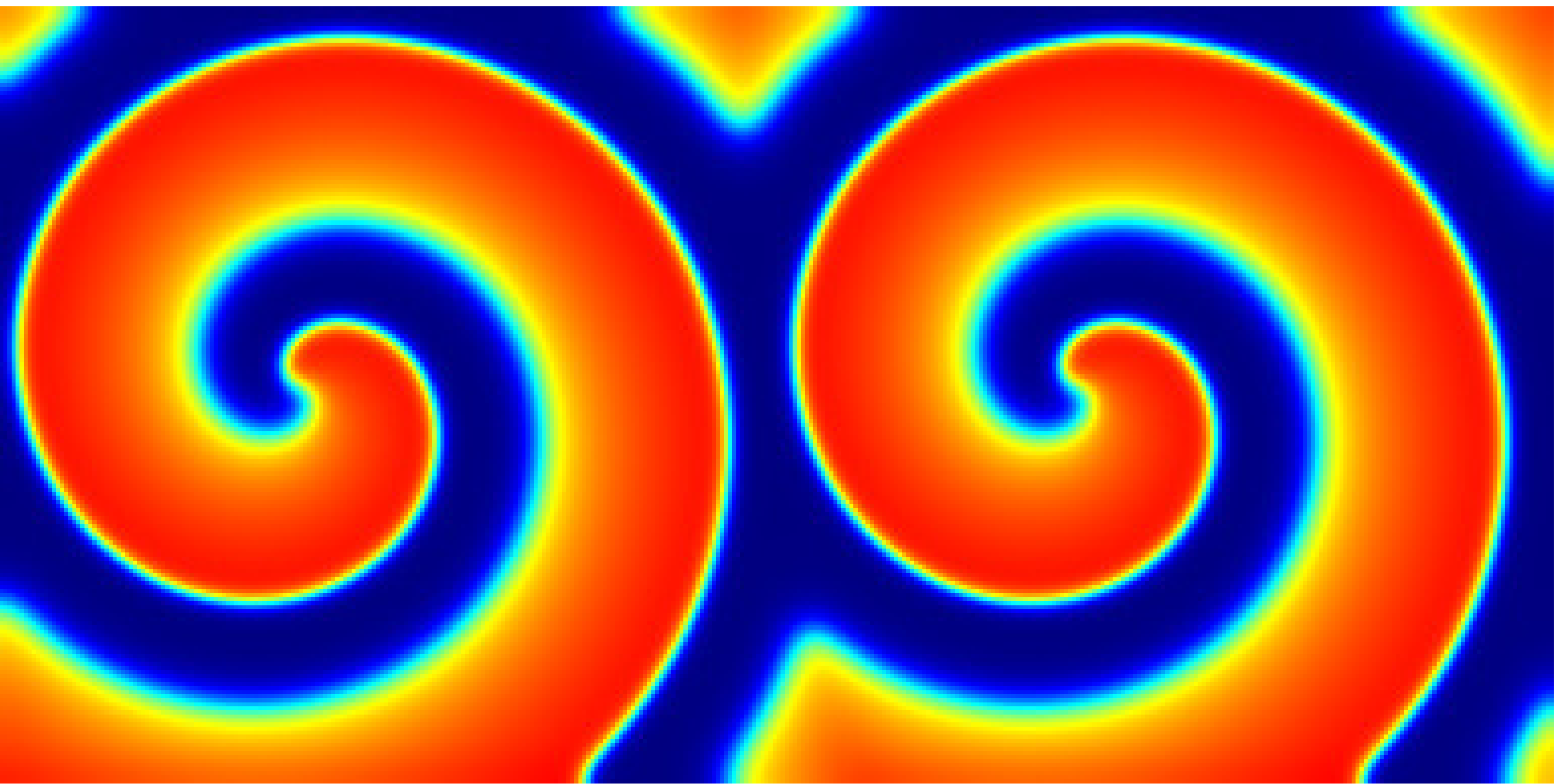}} &
    \subfloat[]{\hspace{2mm}\includegraphics[width=0.47\columnwidth]{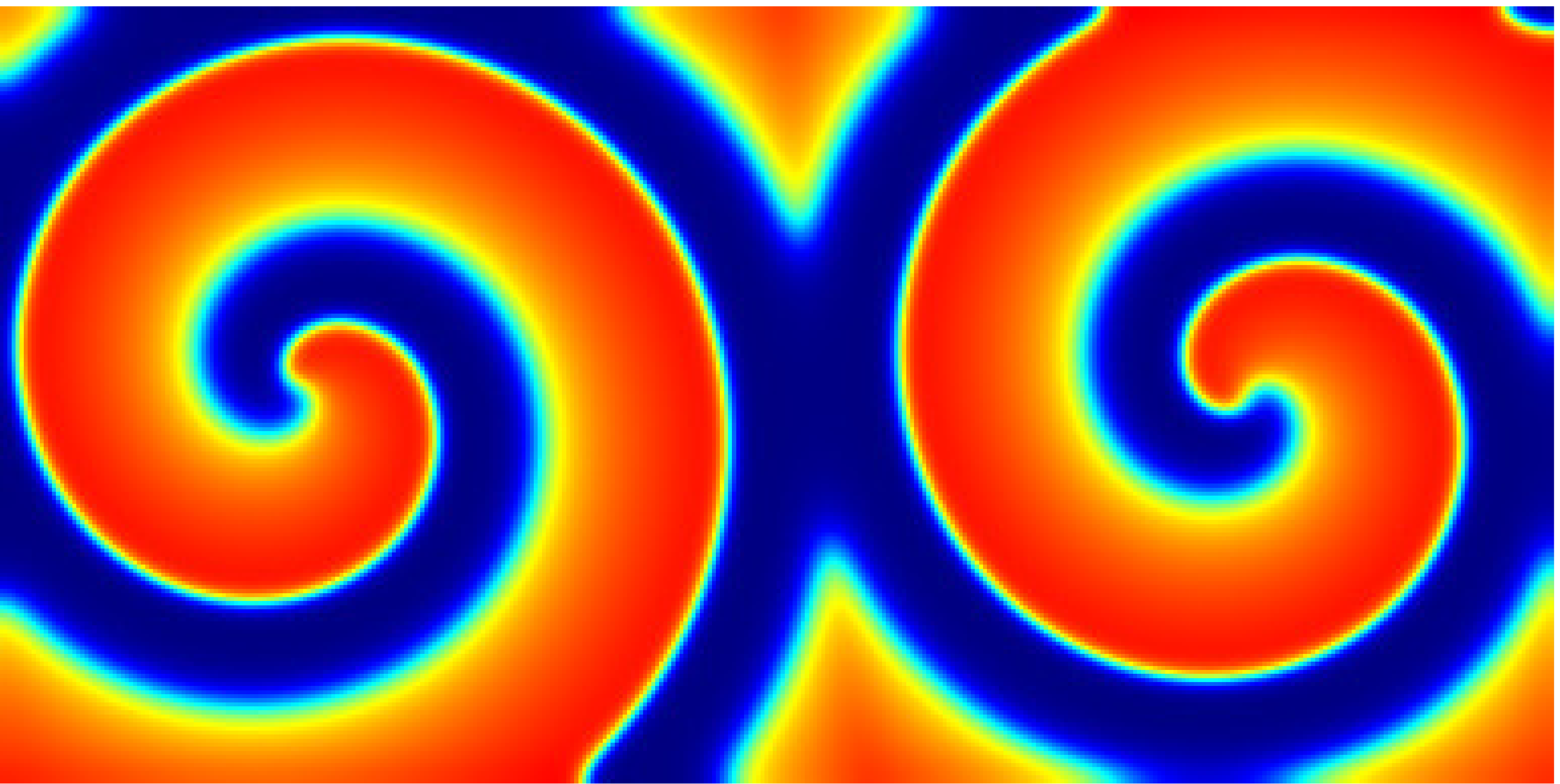}}\\
    \subfloat[]{\includegraphics[width=0.47\columnwidth]{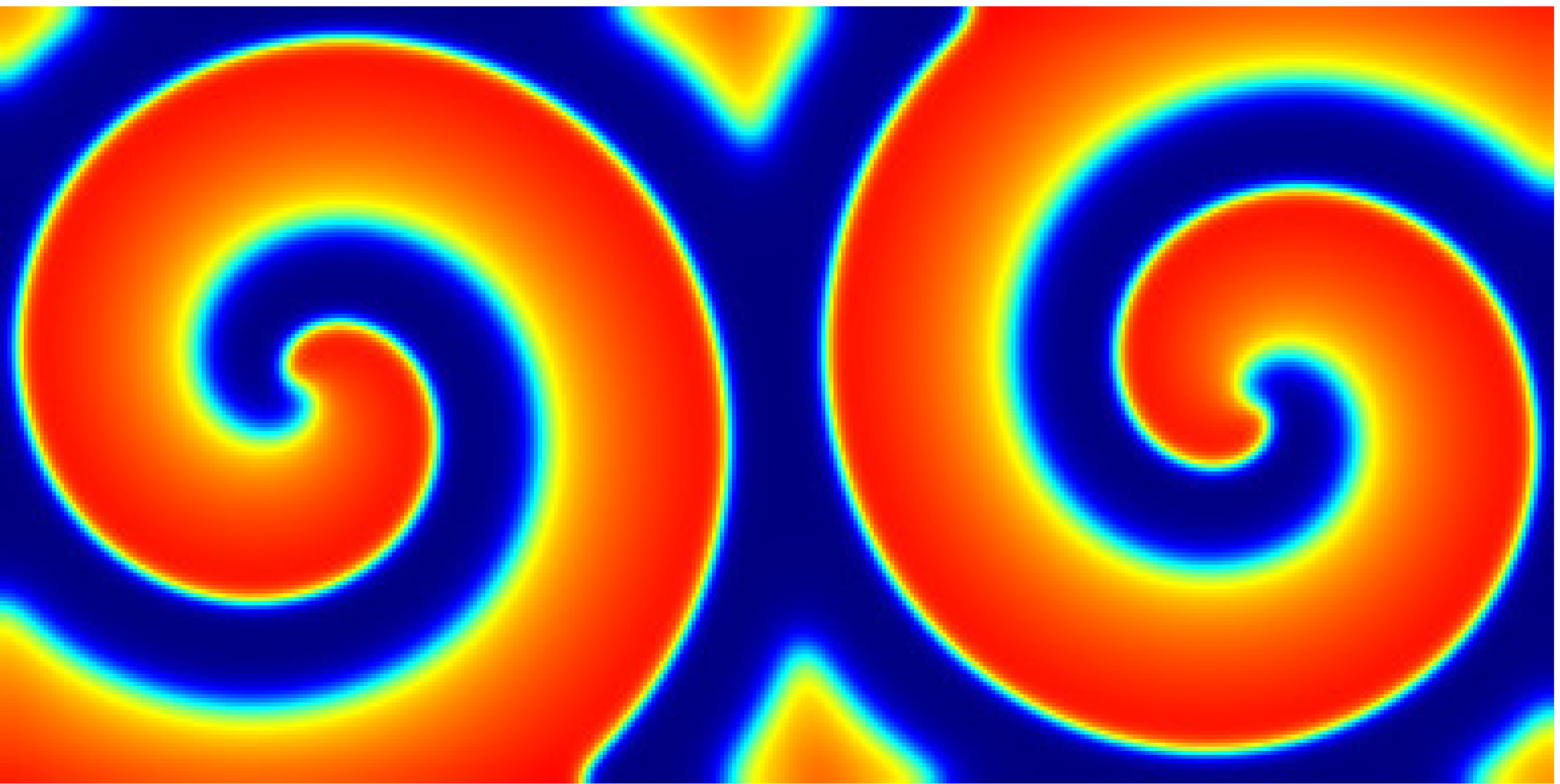}} &
    \subfloat[]{\hspace{2mm}\includegraphics[width=0.47\columnwidth]{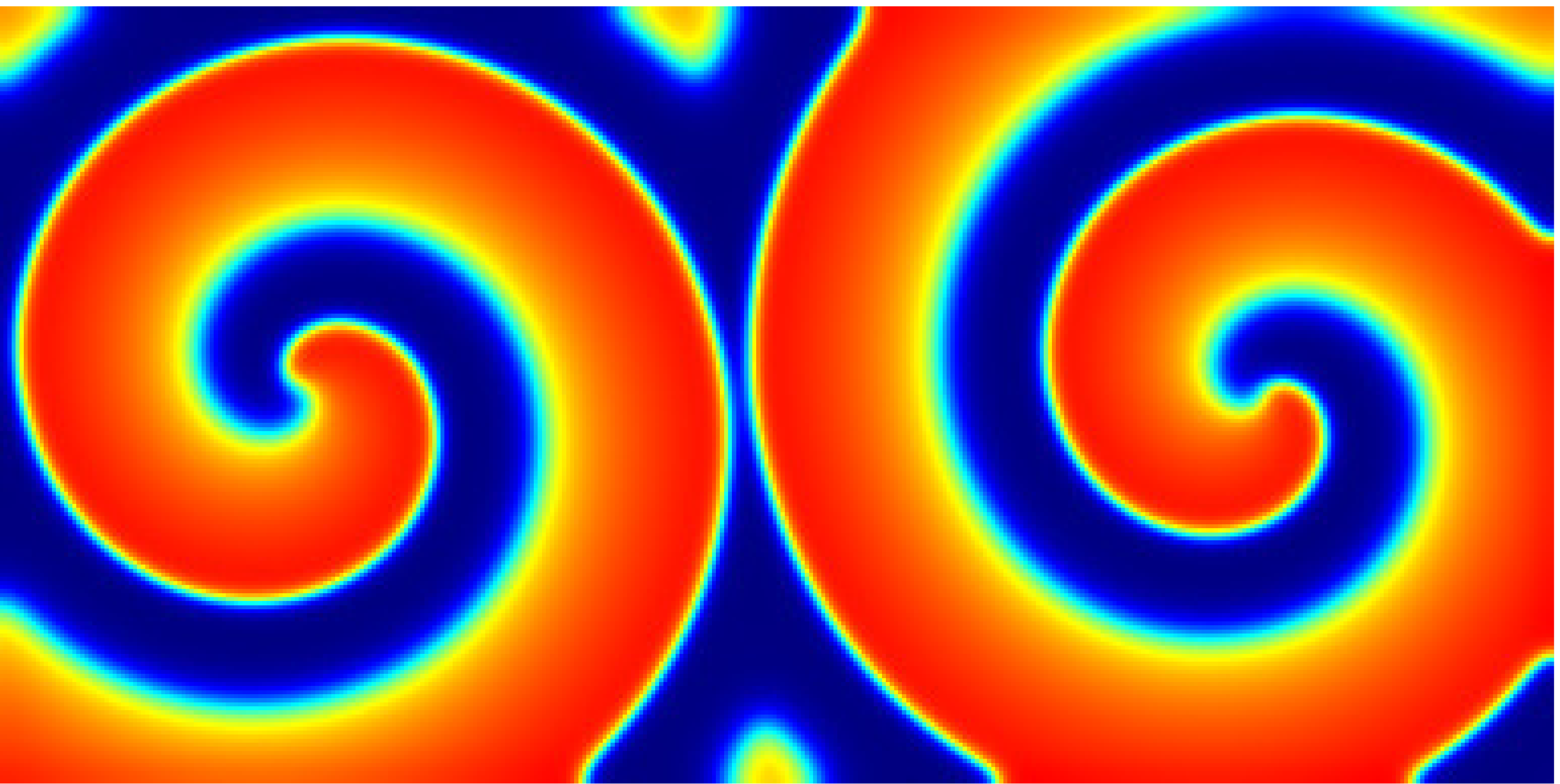}}
  \end{tabular}
  \caption{Snapshots of the voltage $u$ for co-rotating two-spiral solutions on a rectangular domain of size $50.3$ mm $\times$ $100.6$ mm. The phase shifts $\delta\theta=2\pi(\delta t/T)$ between neighboring spirals are: (a) $0$, (b) $\pi/2$, (c) $\pi$, (d) $3\pi/2$.}
  \label{fig6}
\end{figure}

We first explore local rotational symmetry by constructing a set of initial states containing two spirals that are phase shifted with respect to each other.  These initial states are prepared by placing a copy of a
single, isolated spiral solution such as the one shown in Fig.~\ref{fig1}(a)
next to a second identical copy that has been integrated forward in time by
a fraction of its temporal period.  Both co-rotating and counter-rotating spiral states are prepared to account for possible effects of chirality.  \rgedit{These initial states are then refined into exact solutions using the Newton-Krylov solver.}  Converged solutions for four phase shifts $\delta \theta=\{0,\pi/2,\pi,3\pi/2\}$ are shown in Fig.~\ref{fig5} (counter-rotating) and Fig.~\ref{fig6} (co-rotating).

\begin{figure}[t]
\centering
\includegraphics[width=\columnwidth]{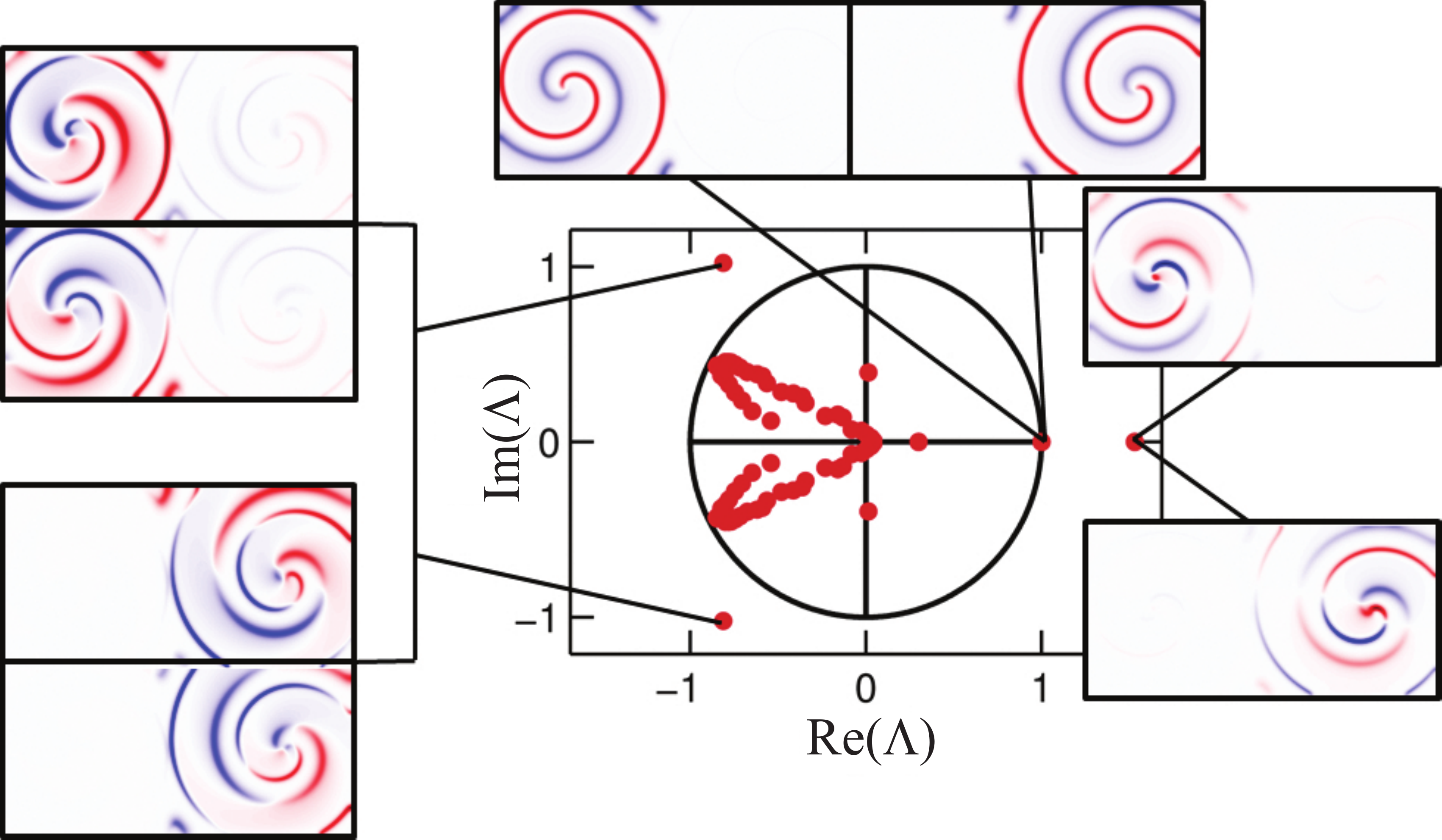}
\caption{The spectrum of the two-spiral solutions shown in Fig.~\ref{fig6}(c). }
\label{fig7}
\end{figure}

The leading \rgedit{eigenvalues} for the co-rotating spirals with $(\delta\theta=\pi)$ shown in Fig.~\ref{fig6}(c) are plotted in Fig.~\ref{fig7}.  This spectrum is representative of the spectra found for the other phase-shifted two-spiral solutions (both co- and counter-rotating) in the sense that all leading eigenmodes \rgedit{could be} spatially localized to the region occupied by one or the other spiral.  \rgedit{Each pair of localized eigenmodes corresponds to one of the eigenmodes shown in Fig.~\ref{fig2} for the single-spiral solution. The corresponding \rgedit{eigenvalues} are degenerate (to numerical precision $O(10^{-6})$), indicating that the dynamics of the two spirals are independent. }
The pair of \rgedit{Goldstone} modes associated with the two unit \rgedit{eigenvalues} ($\Lambda=1$) reflects the freedom of either spiral to undergo small phase shifts relative to the other without destroying the time-periodic nature of the solution globally. 
Just like for single spirals, in the interior of the domain the phase shifts are indistinguishable from rotation of one spiral relative to the other: rotation by angle $\delta \theta$ corresponds to a temporal shift $\delta t=(\delta\theta/2\pi) T$.

\subsection{Local Translational Symmetries}
\label{sec:fiveSS2}

\begin{figure}[t]
  \begin{tabular}{cc}
    \subfloat[]{\includegraphics[width=0.47\columnwidth]{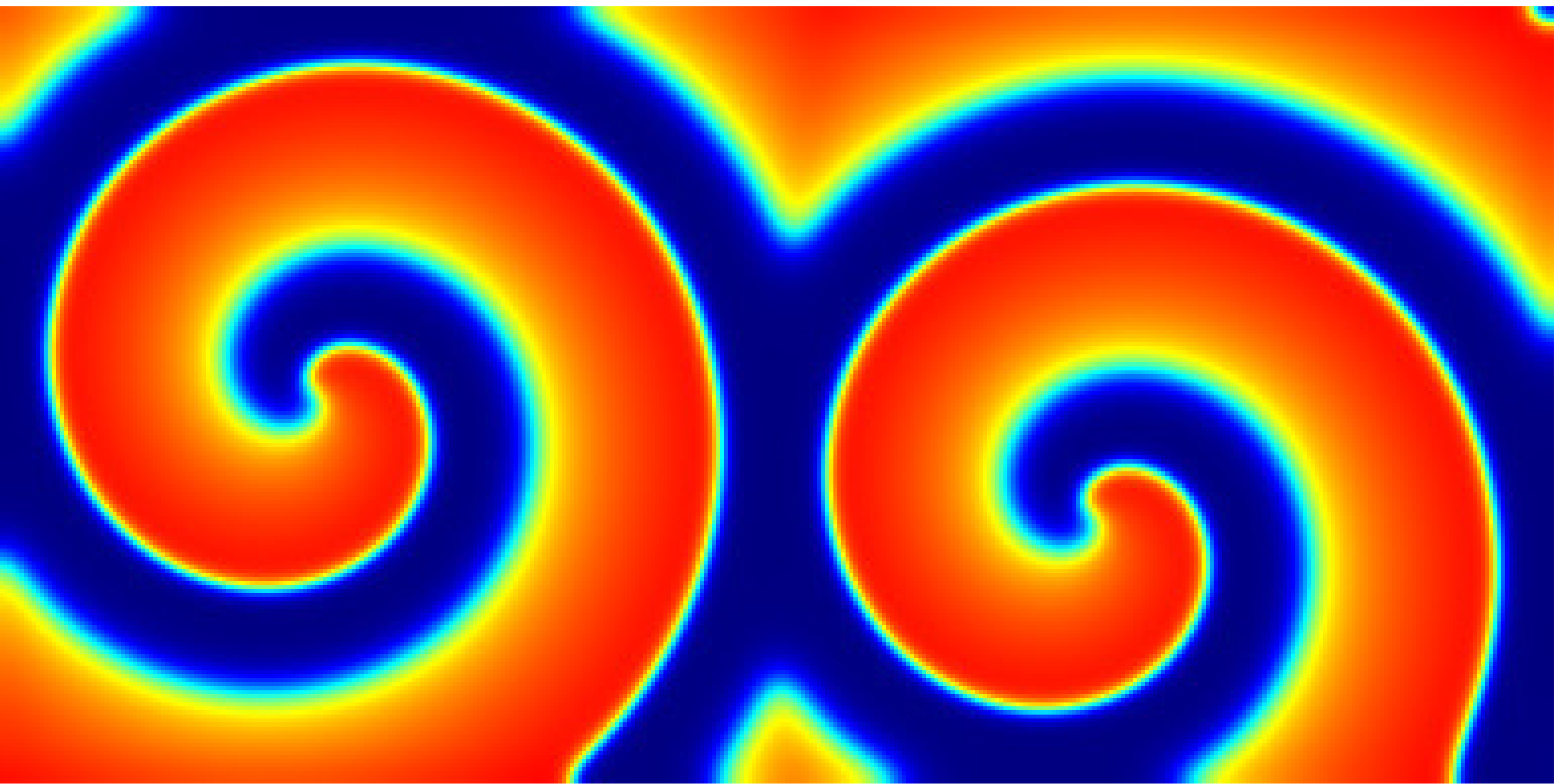}} &
    \subfloat[]{\hspace{2mm}\includegraphics[width=0.47\columnwidth]{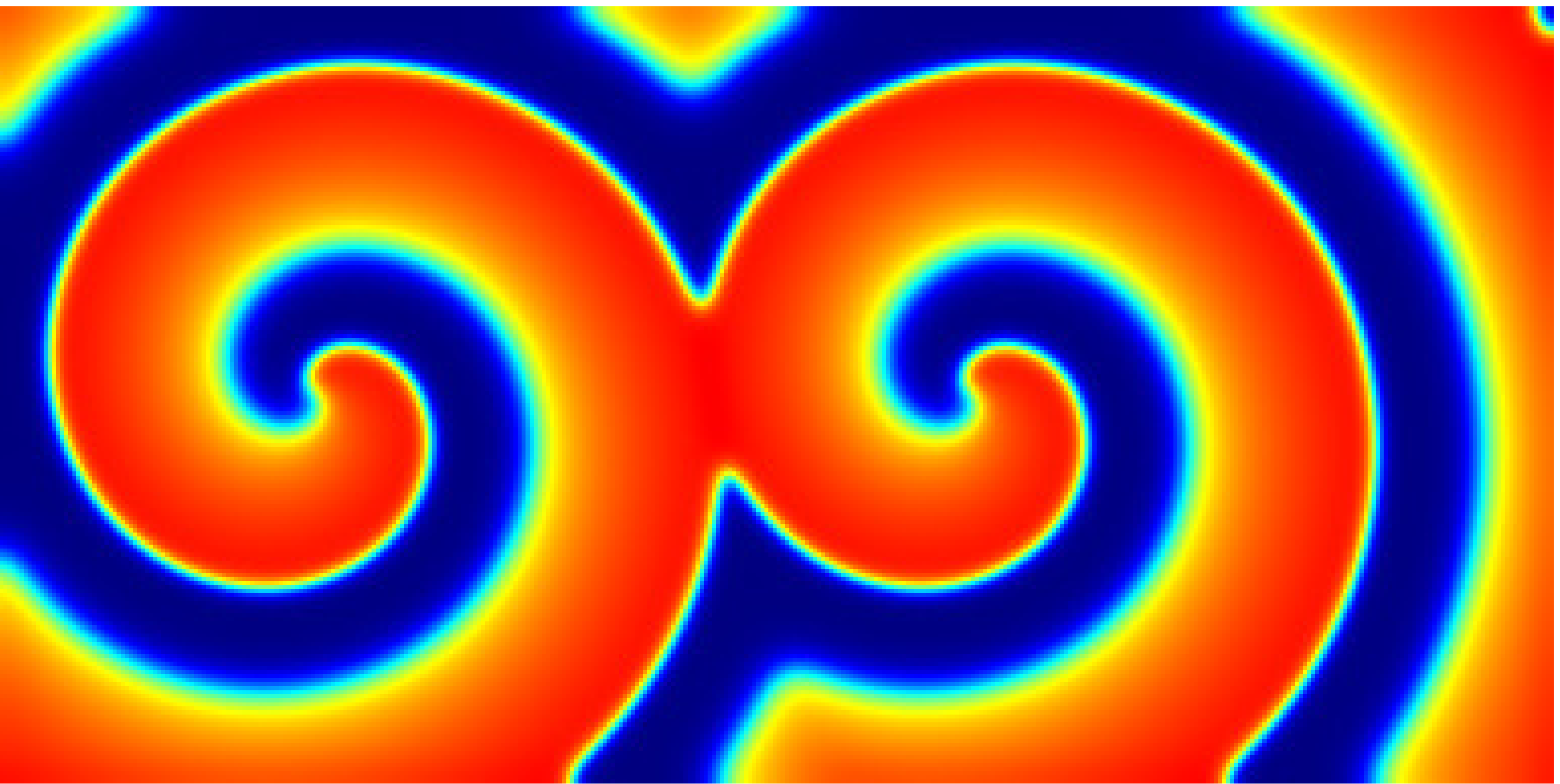}}
  \end{tabular}
  \caption{Snapshots of the voltage $u$ for two-spiral solutions on a rectangular domain with size $50.3$ mm$\times 100.6$ mm.
The spiral on the right side of the domain is shifted (a) vertically by 30 grid points (i.e., $\delta y=-7.86$ mm) or (b) horizontally by 30 grid points (i.e., $\delta x=-7.86$ mm) relative to the spiral on the left side of the domain.}
\label{fig8}
\end{figure}

We explore local translational symmetries, which become discrete on our computational grid, by using two distinct sets of solutions.  The first set is constructed by taking two single, isolated spiral solutions (in phase) and applying discrete shifts in the horizontal or vertical direction to one of the spirals by any integer multiple of the grid spacing before recombining them.  Figure~\ref{fig8} shows a set of converged two-spiral solutions in which the core of the spiral on the right side of the domain has been shifted with respect to the spiral core on the left side of the domain.  The shift for the spiral in Fig.~\ref{fig8}(a) is by 30 grid points along the negative \rgedit{$y$ direction (downward), while the shift for the spiral in Fig.~\ref{fig8}(b) is 30 grid points in the negative $x$ direction (leftward).}  While the convergence of these two solutions confirms that multi-spiral solutions respect locally discrete translational symmetries, the \rgedit{Goldstone} modes associated with continuous translational symmetry remain absent in the spectrum for both solutions. Solutions with smaller shifts (for example, 5 grid points instead of 30) also converge and have similar spectral properties.

\begin{figure}[t]
  \begin{tabular}{cc}
    	\subfloat[]{\hspace{1.5mm}\includegraphics[width=0.49\columnwidth]{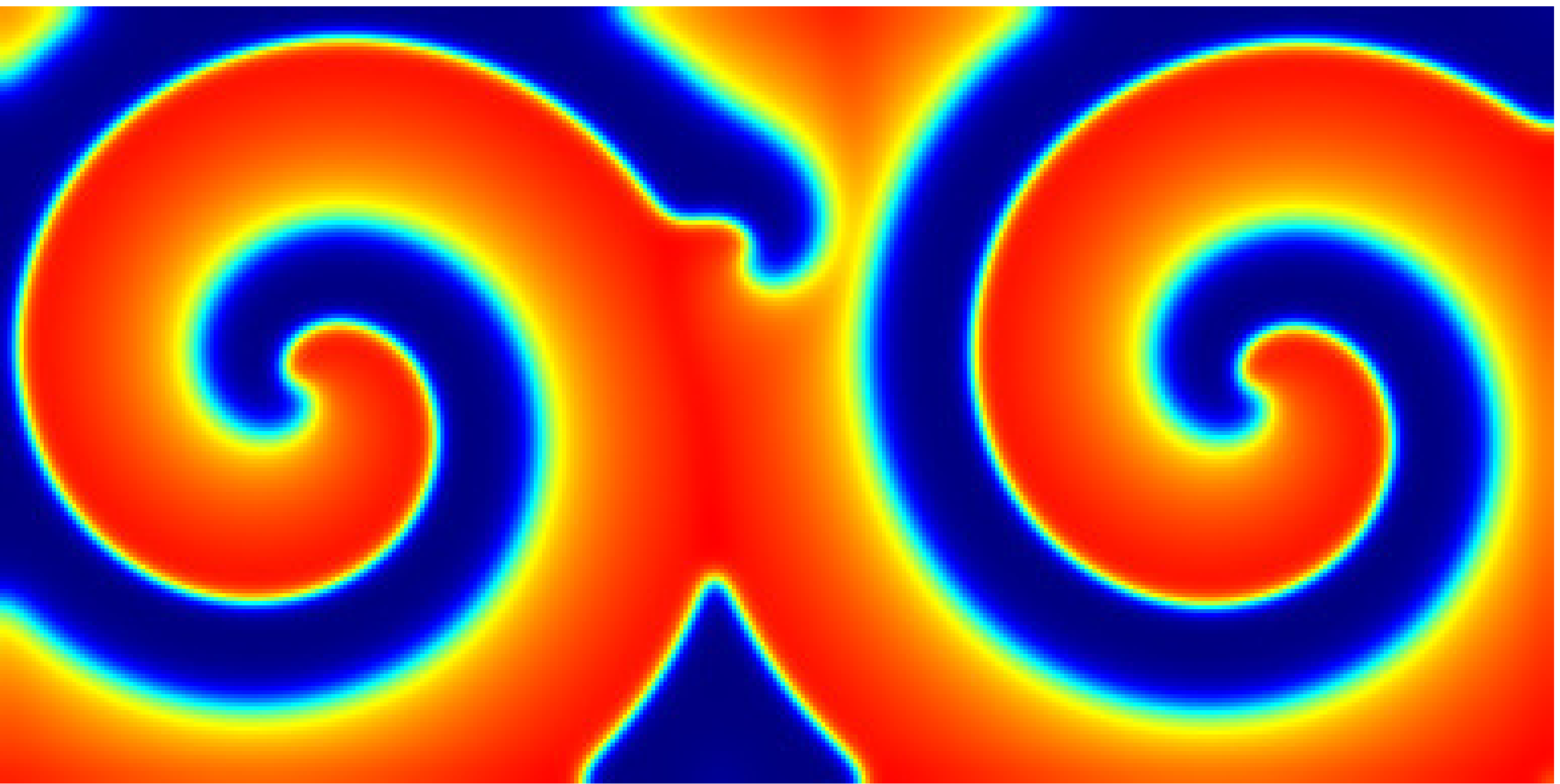}} 
	&
	\subfloat[]{\hspace{1mm}\includegraphics[width=0.49\columnwidth]{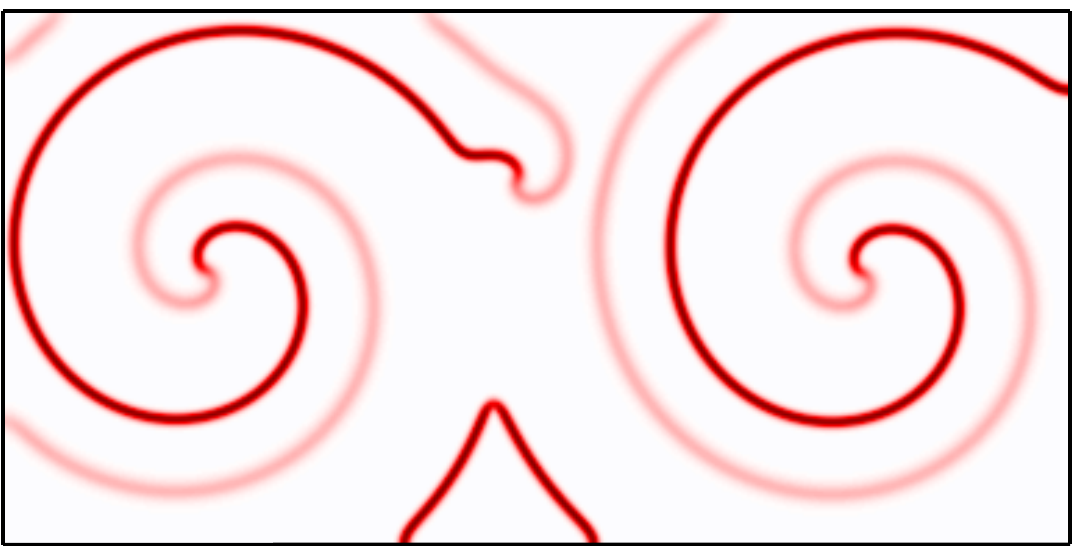}}
        \\
	\subfloat[]{
	\begin{minipage}[t][][s]{0.49\columnwidth}
	\centering
	\includegraphics[width=\columnwidth]{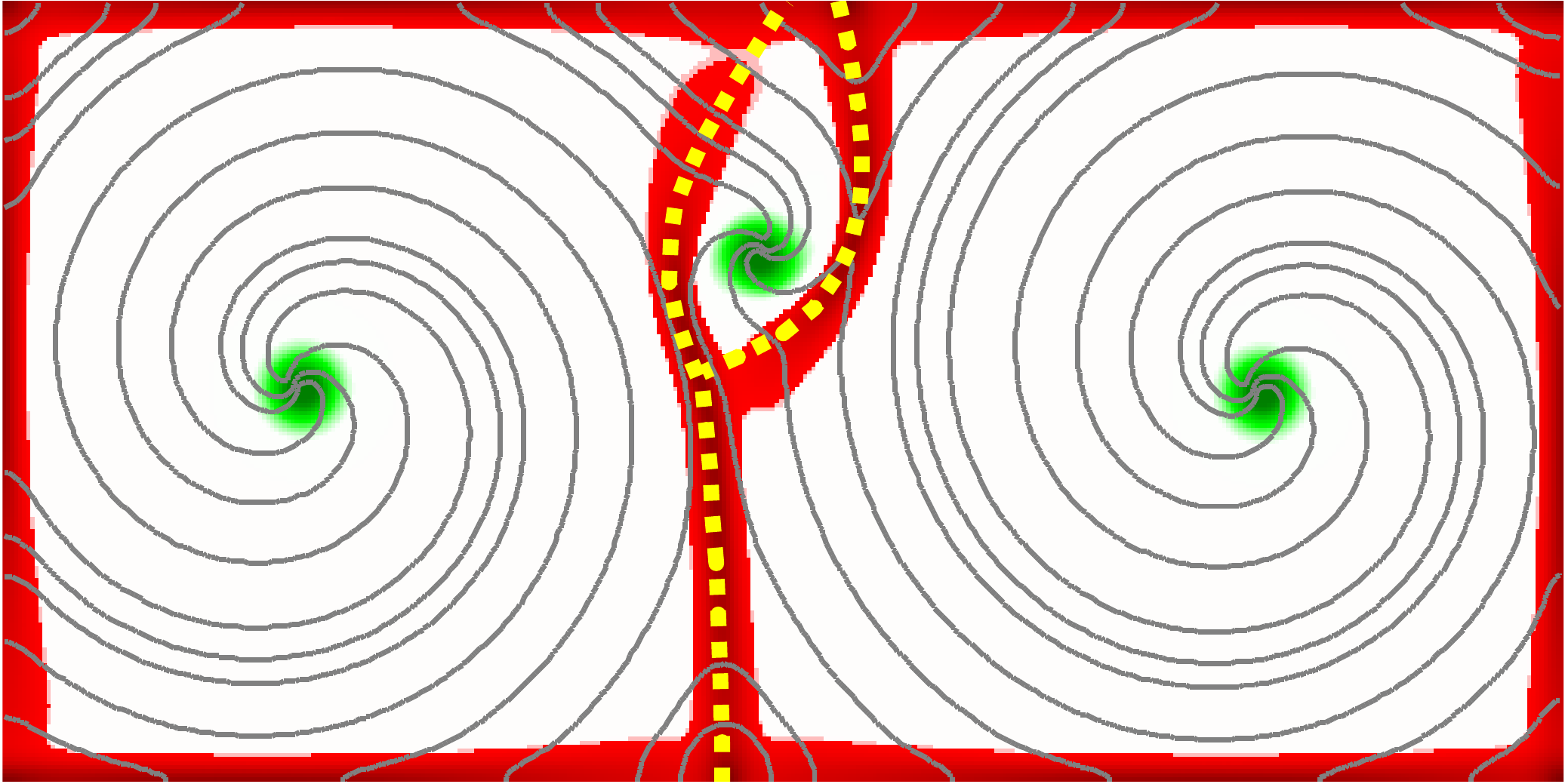}
	\\
	\includegraphics[width=\columnwidth]{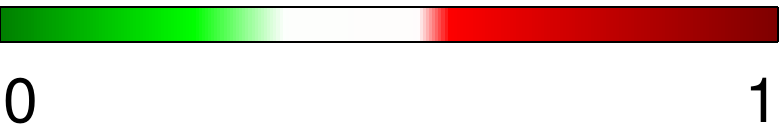}
	\end{minipage}}
	&
	\subfloat[]{
	\begin{minipage}[t][][s]{0.49\columnwidth}
	\centering
	\includegraphics[width=\columnwidth]{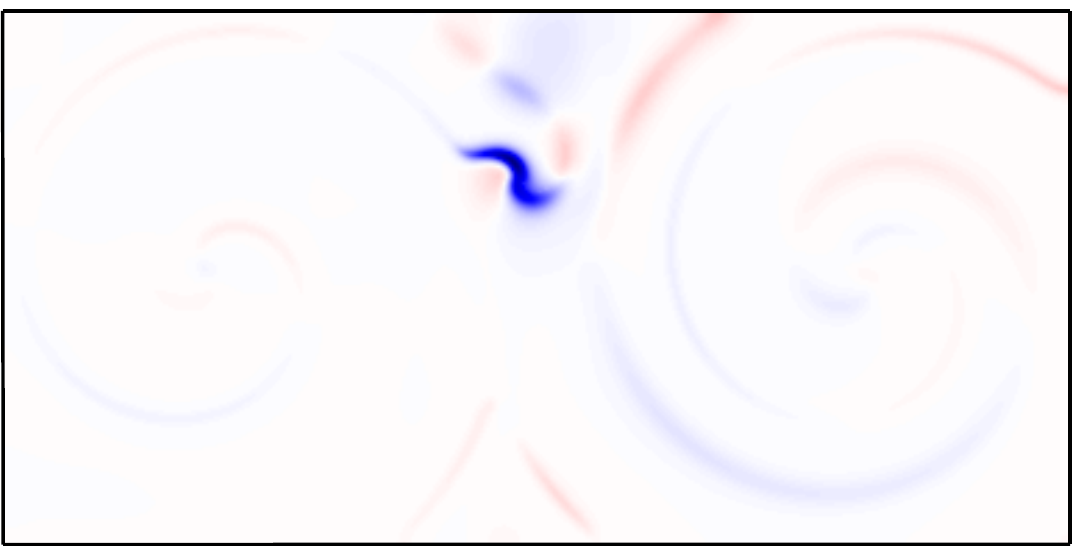}
	\\ \vspace{0.15mm}
	\includegraphics[width=0.9\columnwidth]{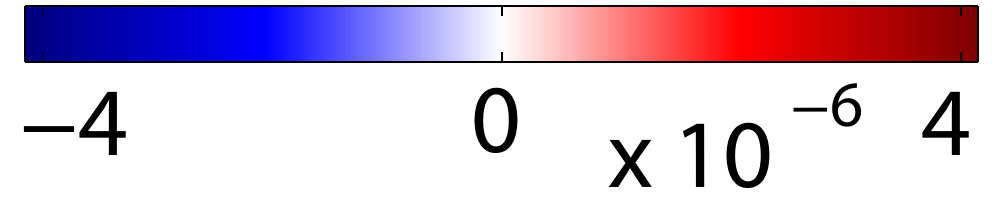}
	\end{minipage}}
  \end{tabular}
  \caption{(a) Snapshot of the voltage $u$ for a three-spiral \rgedit{transient} solution on a rectangular domain of size $50.3$ mm $\times 100.6$ mm. (b) The normalized magnitude of the voltage gradient $|\nabla u|$. (c) The cycle area $I_1$. The analytical solutions \rgedit{(see Sect.~\ref{sec:sixSS1})} for the tile boundaries are shown as dashed yellow lines and the level sets of $v$ as solid gray lines. (d) The relative residual $[u(t)-u(t-\tau)]/\|u(t)\|_\infty$.}
\label{fig9}
\end{figure}

The second type of solutions used to explore local translational symmetry is shown in Fig.~\ref{fig9}(a).  It involves a three-spiral configuration in which a small spiral wave is formed between two larger spiral waves.  Under time evolution, the small spiral undergoes a drift in the mostly vertical direction.  This drift is caused by interaction with the neighboring spirals, as we explain below. Drifting spirals in bounded geometries are associated with relative periodic solutions which respect (local) continuous translational symmetries \cite{Marcotte2014}.  The two big spirals do not drift, so the Newton-Krylov solver fails to converge onto a time-periodic \rgedit{(or relative periodic)} solution, with the smallest relative residual magnitude equal to $3\times10^{-3}$.  Figure~\ref{fig9}(d) shows that the voltage component of the residual (with $\tau=125.98$ ms) is concentrated in the vicinity of the drifting spiral and \rgedit{(just like for the multi-spiral states shown in Fig.~\ref{fig4})} is localized near the front and back of the excitation wave (cf.  Figs.~\ref{fig9}(a)), which is consistent both with the slow drift and the slow rotation of the small spiral.

This simple three-spiral solution is important since it is qualitatively representative of spiral interactions that take place during more complex chaotic behavior.  It provides valuable insight into the dynamics and local symmetries of multi-spiral solutions such as those shown in Fig.~\ref{fig4}(a-b).  While each of the three individual spirals can be represented by either periodic or relative periodic solutions in the vicinity of its core, the solution on the entire domain is neither a periodic nor a relative periodic orbit. Similarly, one would expect \rgedit{multi-spiral ECS} to be described by periodic and/or relative periodic solutions locally, but not globally.

\section{Strategy for Local Symmetry Reduction}
\label{sec:six}

A new class (or classes) of nonchaotic solutions are required to define a skeleton for the chaotic set associated with spiral turbulence.  These nonchaotic solutions are associated with near-recurrences such as those shown in Figs.~\ref{fig4}(a-b) that are found frequently, and hence play an important dynamical role, in spiral turbulence. In this Section we discuss how such nonchaotic solutions can be constructed starting with nearly-recurrent solutions that can be identified directly by exploring the natural measure on the chaotic set.  As we have shown previously, spatial coherence in spiral turbulence is limited to areas (or tiles) associated with one spiral. Hence, we start by describing the process of \rgedit{decomposing} the computational domain into individual tiles, each of which contains a single-spiral solution.  We then discuss which boundary conditions these local single-spiral solutions should satisfy and how they can be assembled into a global \rgedit{nonchaotic} solution.

\subsection{Tiling the domain}
\label{sec:sixSS1}

\begin{figure}
\begin{tabular}{cc}
      \subfloat[]{\raisebox{6mm}{\includegraphics[width=0.4\columnwidth]{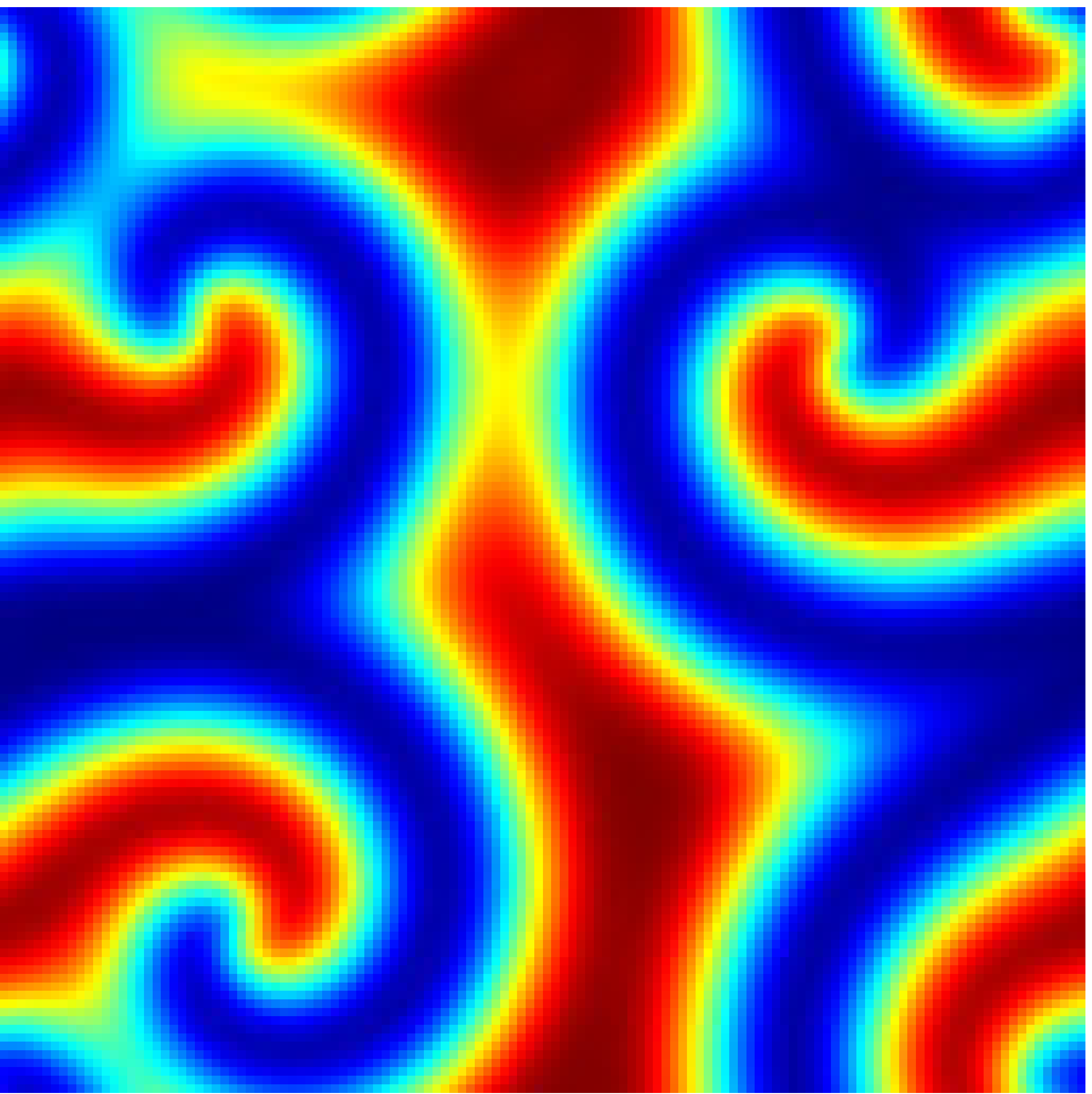}}}
      \hspace{1mm}{\raisebox{6mm}{\includegraphics[height=35mm]{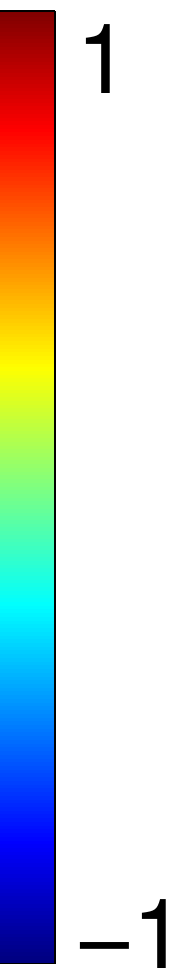}}}
&     \hspace{-2.5mm} \subfloat[]{\includegraphics[width=0.46\columnwidth]{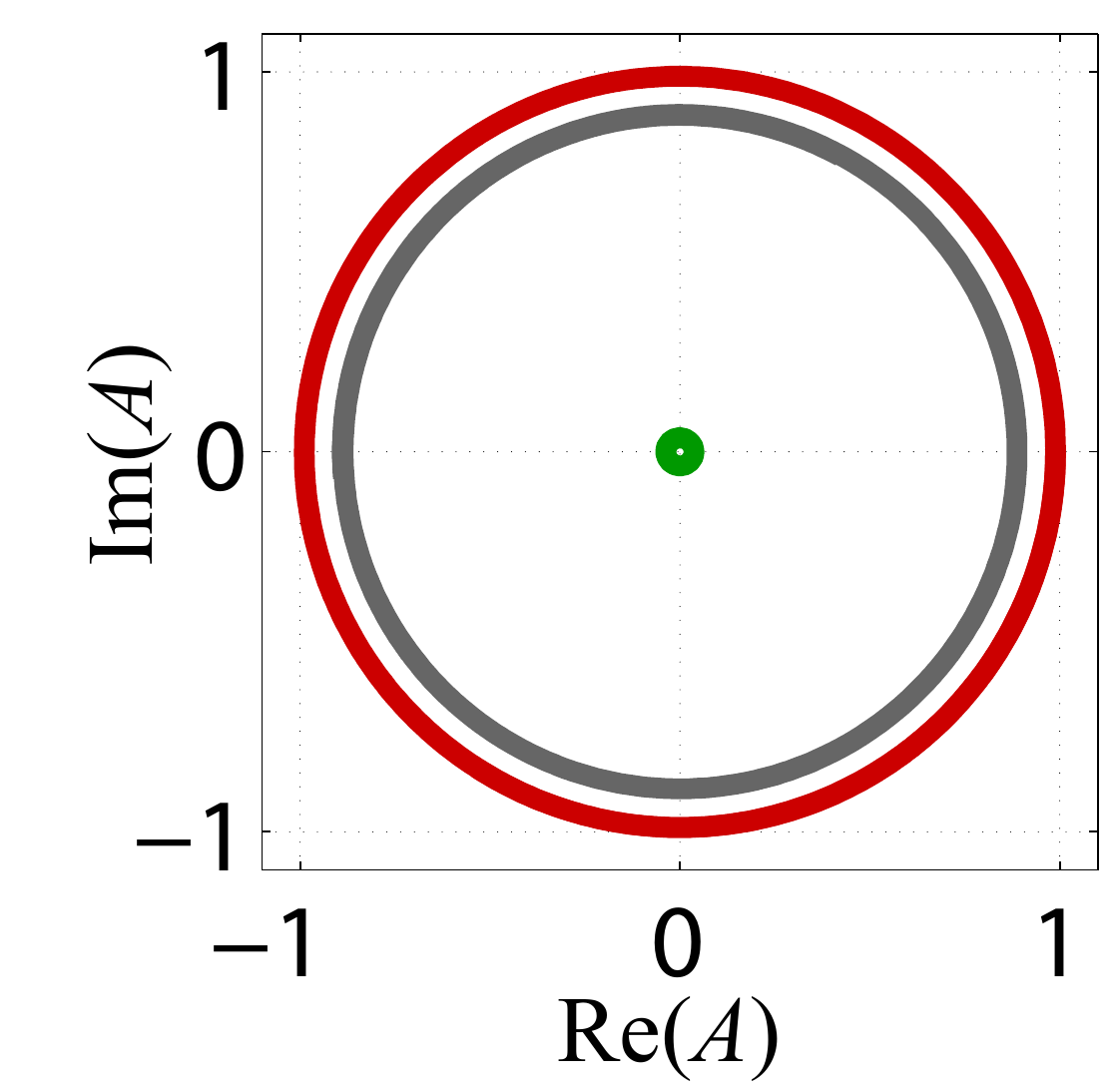}} \vspace{0.4cm}\\    
   \hspace{-2mm}\subfloat[]{\includegraphics[width=0.4\columnwidth]{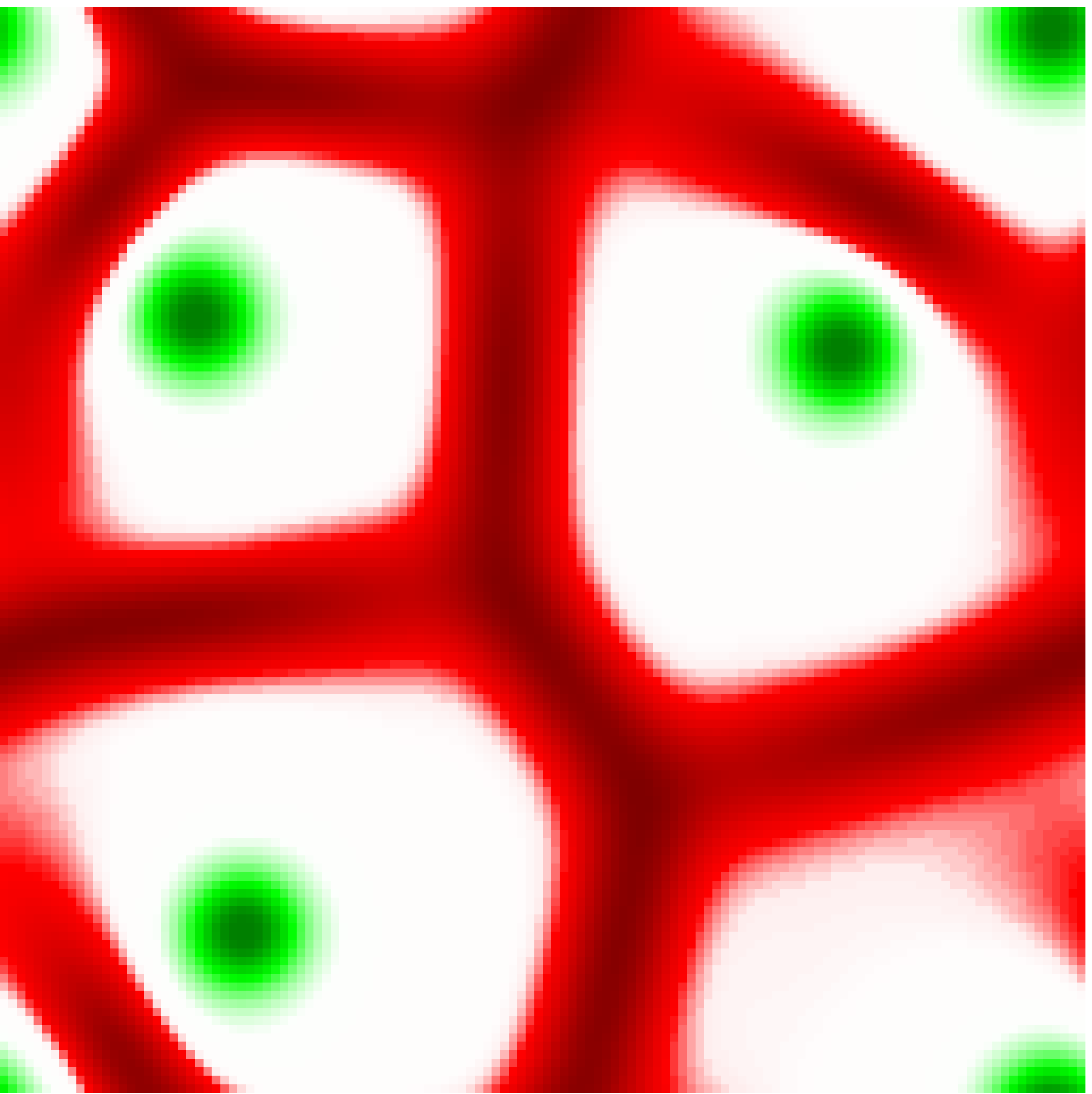}}
   \hspace{1mm}{{\includegraphics[height=35mm]{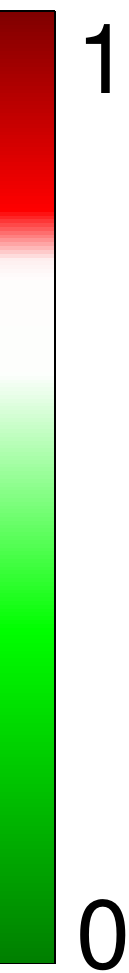}}}
& \hspace{2.5mm} 
   \subfloat[]{\includegraphics[width=0.4\columnwidth]{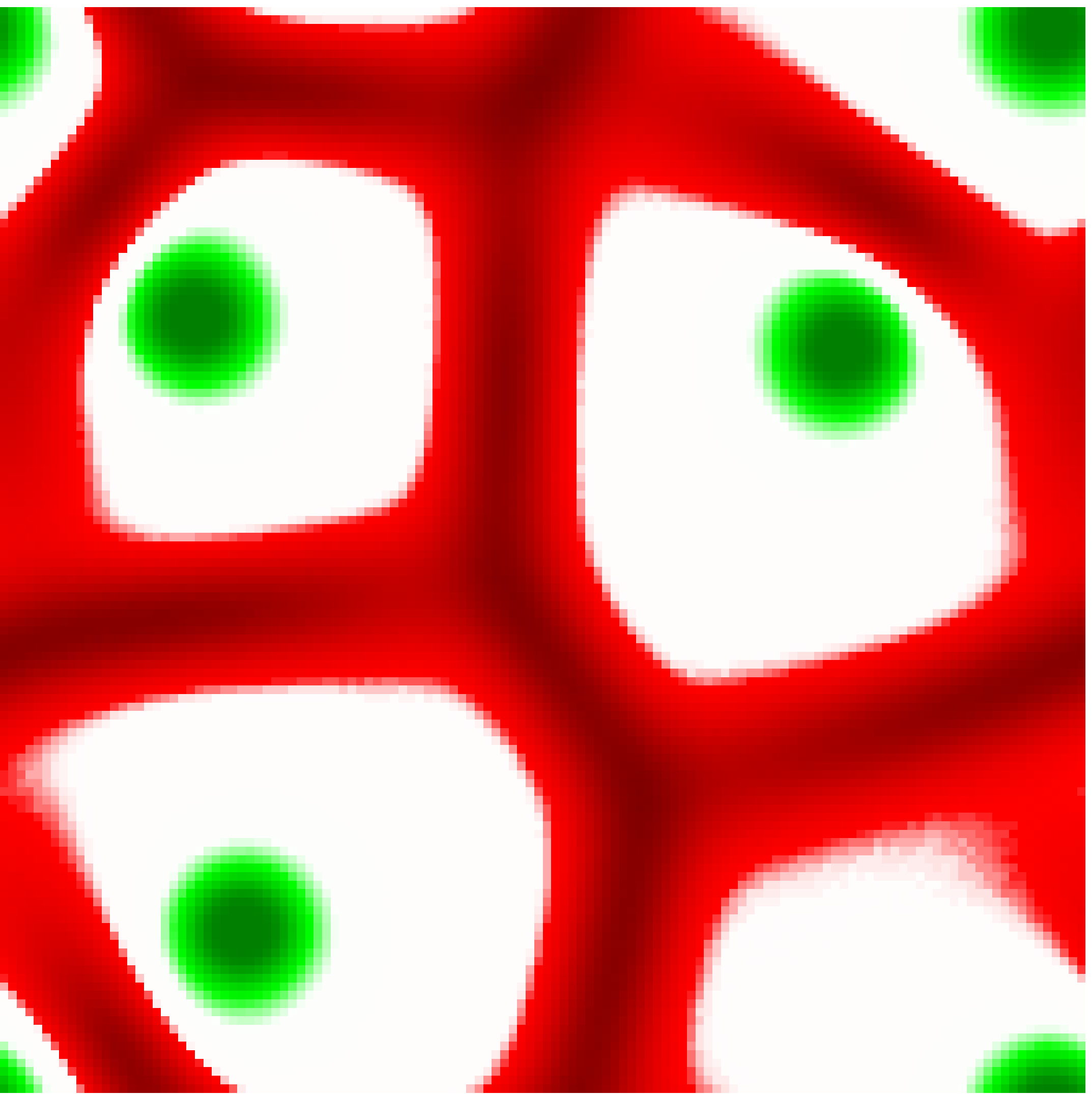}}
\hspace{1mm}{{\includegraphics[height=35mm]{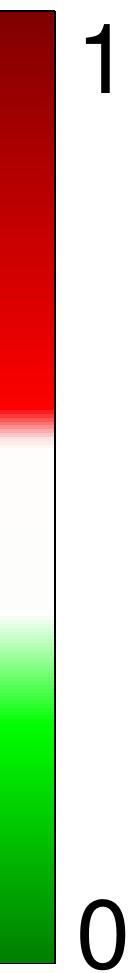}}}
\end{tabular}
  \caption{Complex Ginzburg-Landau equation.  (a) Snapshot of Re($A$) for a frozen spiral state with $\alpha=0$ and $\beta=1.2$. (b) The cycles in the complex plane for three representative spatial locations: spiral core (green), interior of a tile (gray), and a shock separating two tiles (red). \rgedit{The amplitude of the gray cycle is slightly less than unity because the tiles are small (the size is comparable to the wavelength $\lambda$).}
(c) The normalized amplitude $|A|$. (d) The normalized cycle area $I$.}
  \label{fig10}
\end{figure}

The concept of domain tiling was introduced by Bohr {\it et al.}~\cite{BohHubOtt96,BohHubOtt97} to describe frozen spiral waves, sometimes referred to as vortex glasses, in the complex Ginzburg-Landau equation (CGLE)
\begin{equation}
\partial_t A = A + (1+i\alpha)\nabla^2 A - (1+i\beta)|A|^2 A,
\end{equation}
where $A$ is a complex field and $\alpha$ and $\beta$ are control parameters.  Each tile contains exactly one spiral core and the dynamics on each tile is controlled almost entirely by that core. The boundaries of individual tiles were identified with the ridges (or shocks) of the field $|A|$ which describes the amplitude of local oscillation. \rgedit{Fig.~\ref{fig10}(a) shows} the real part of a representative solution \rgedit{$A=\rho e^{i\phi}$ which can be used to identify individual spirals. In most of the domain the phase $\phi$ of the oscillation varies slowly in space, so according to the amplitude equation~\cite{Aranson2002}
\begin{equation}
\partial_t\rho=[\nabla^2-(\nabla\phi)^2]\rho-\alpha(2\nabla\phi\cdot\nabla\rho+\rho\nabla^2\phi)+(1-\rho^2)\rho,
\end{equation}
the amplitude of oscillation is essentially constant, $\rho\approx 1$, which corresponds to the gray cycle in the complex-$A$ plane shown in Fig.~\ref{fig10}(b). The spiral cores are associated with phase singularities and are characterized by small values of the amplitude ($\rho\ll 1$, green cycle). Similarly, the phase changes quickly at the boundaries of the tiles, where the amplitude increases ($\rho\gtrsim 1$, red cycle). Hence, the maxima (red) and minima (green) of $|A|$ \rgedit{shown in Fig.~\ref{fig10}(c)} can be used to identify, respectively, the spatial locations of the tile boundaries and spiral cores.}

For excitable media characterized by strongly nonlinear oscillations, a different representation has to be used, since the phase and amplitude of the oscillations can be difficult to define, let alone compute. In this case the local amplitude of oscillation can be characterized instead by the area $I(x,y)$ of the cycle $C$ that is traced out by the solution \rgedit{in an appropriate state space. For CGLE the area in the complex-$A$ plane is given by
\begin{equation}
I(x,y)=\oint_C \! \frac{\rho^2}{2}\,d\phi
=\int_0^T\frac{\rho^2}{2}\dot{\phi}\,dt,
\end{equation}
with the result shown in Fig.~\ref{fig10}(d). For frozen spirals $\dot{\phi}=2\pi/T$, so that $I(x,y)=\pi|A(x,y,t)|^2$ and the cycle area representation is equivalent to the amplitude representation.}

\begin{figure}
\centering
    \begin{tabular}{cc}
	\subfloat[]{\includegraphics[height=3cm]{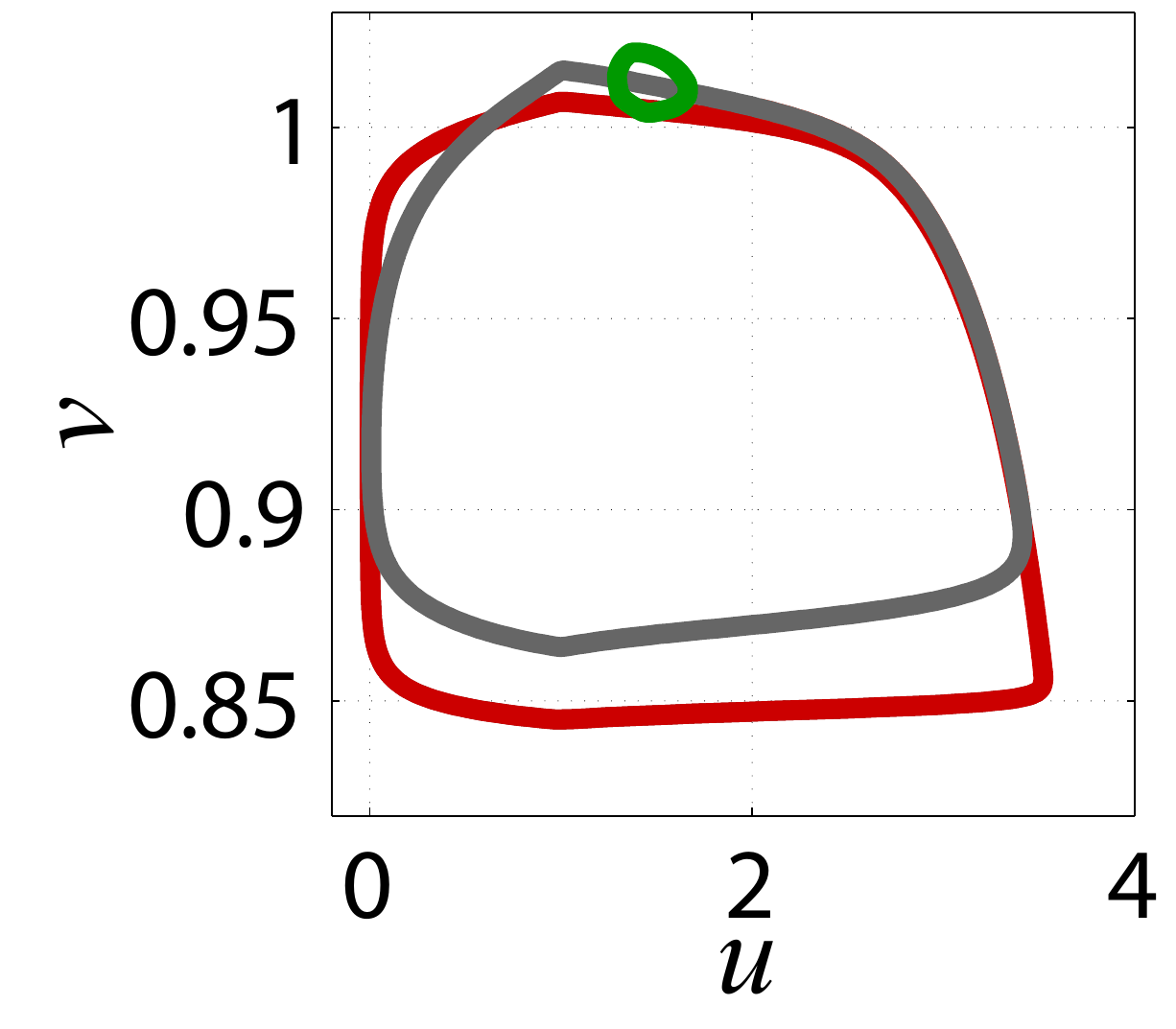}}
	&
	\subfloat[]{
	\raisebox{8mm}{
	\begin{minipage}[t][][s]{0.49\columnwidth}
	\centering
	\includegraphics[width=\columnwidth]{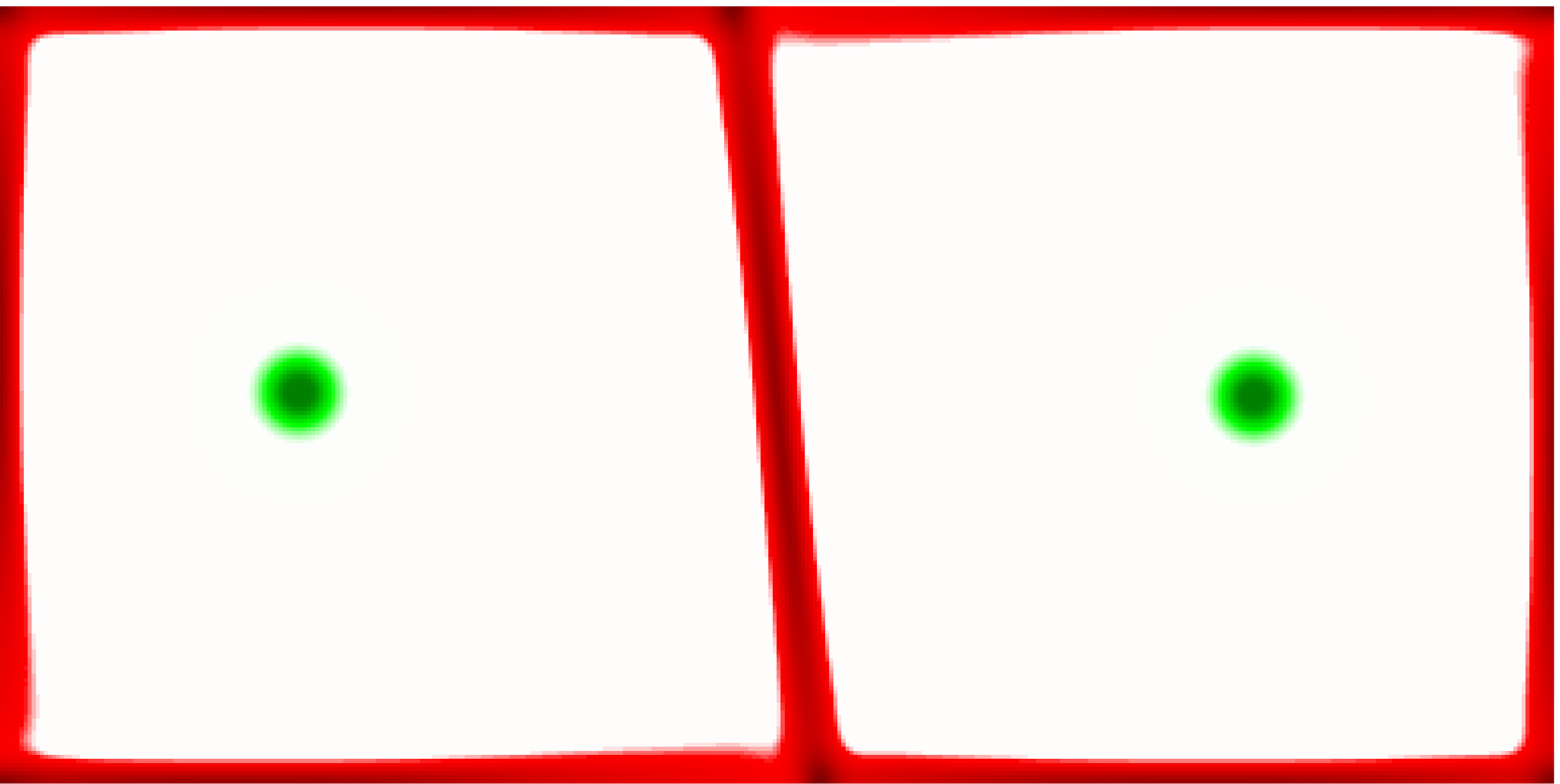}
	\\
	\includegraphics[width=\columnwidth]{figure11bcb}
	\end{minipage}}}
	\\
	\subfloat[]{\includegraphics[height=3cm]{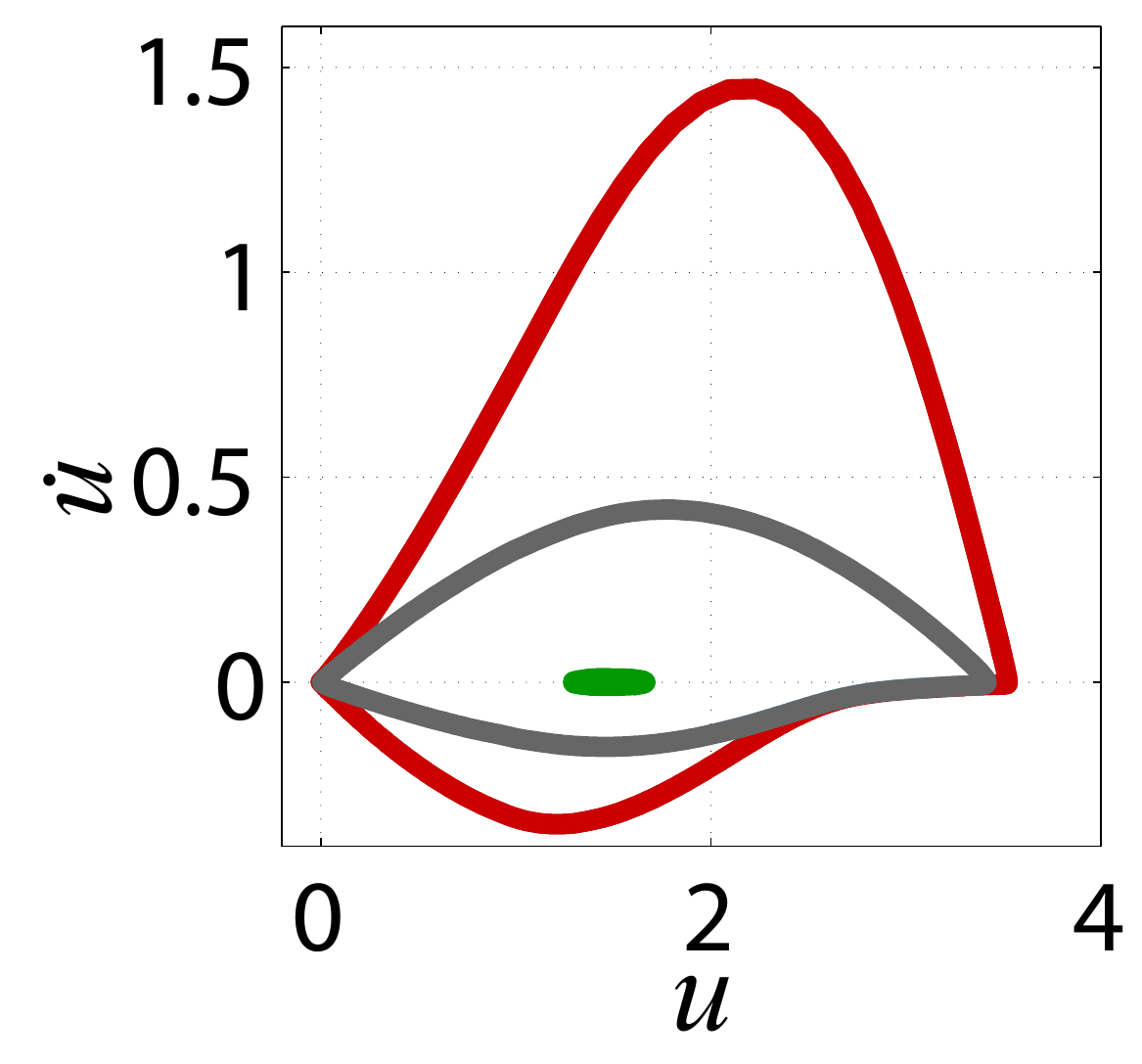}}
	&
	\subfloat[]{
	\raisebox{8mm}{
	\begin{minipage}[t][][s]{0.49\columnwidth}
	\centering
	\includegraphics[width=\columnwidth]{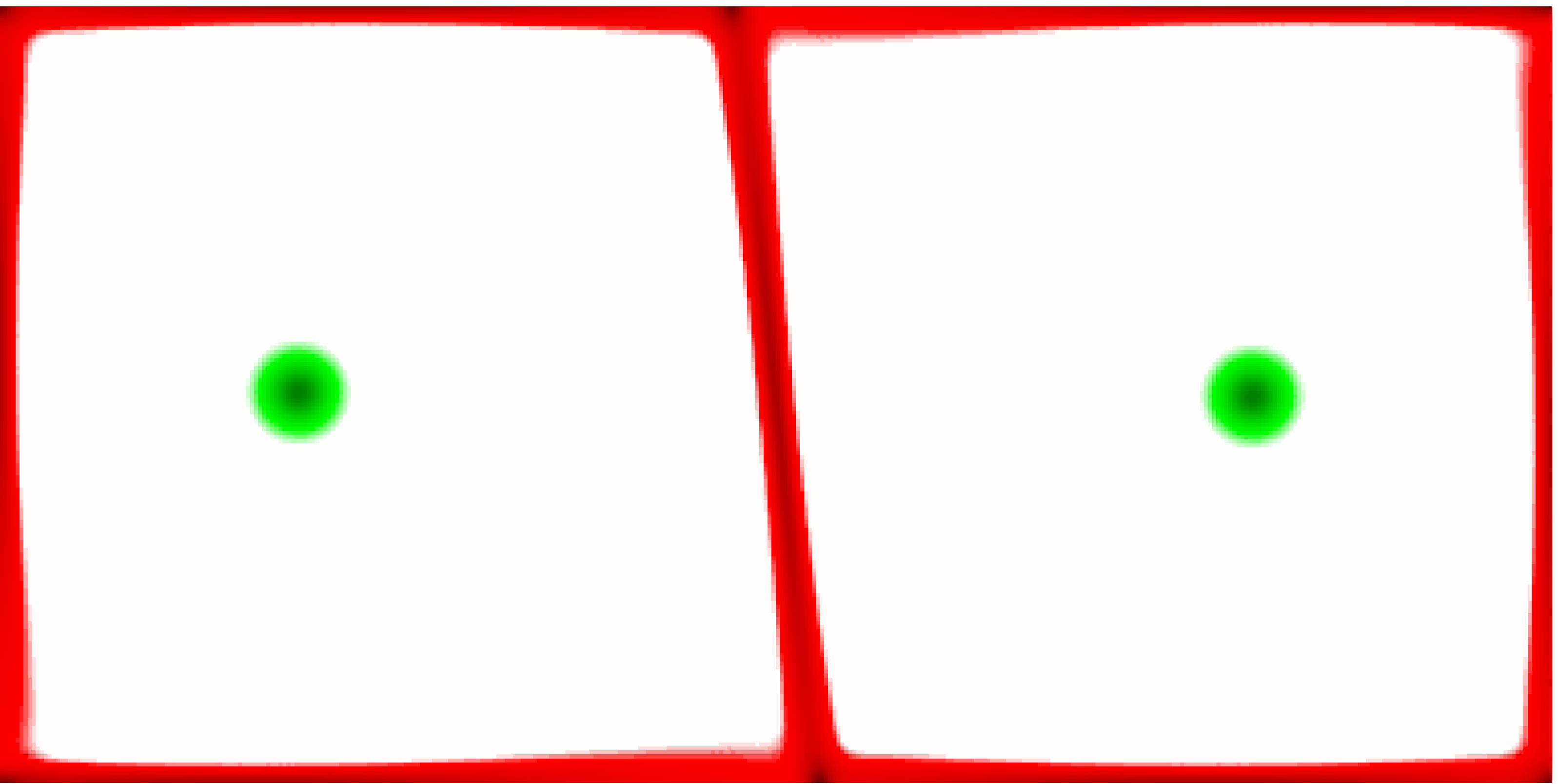}
	\\
	\includegraphics[width=\columnwidth]{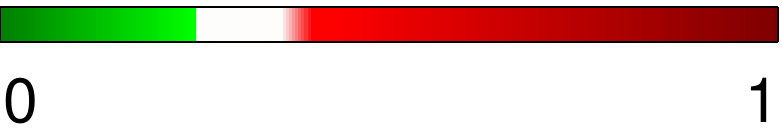}
	\end{minipage}}}
	\\
	\subfloat[]{\hspace{2mm}\includegraphics[height=3.1cm]{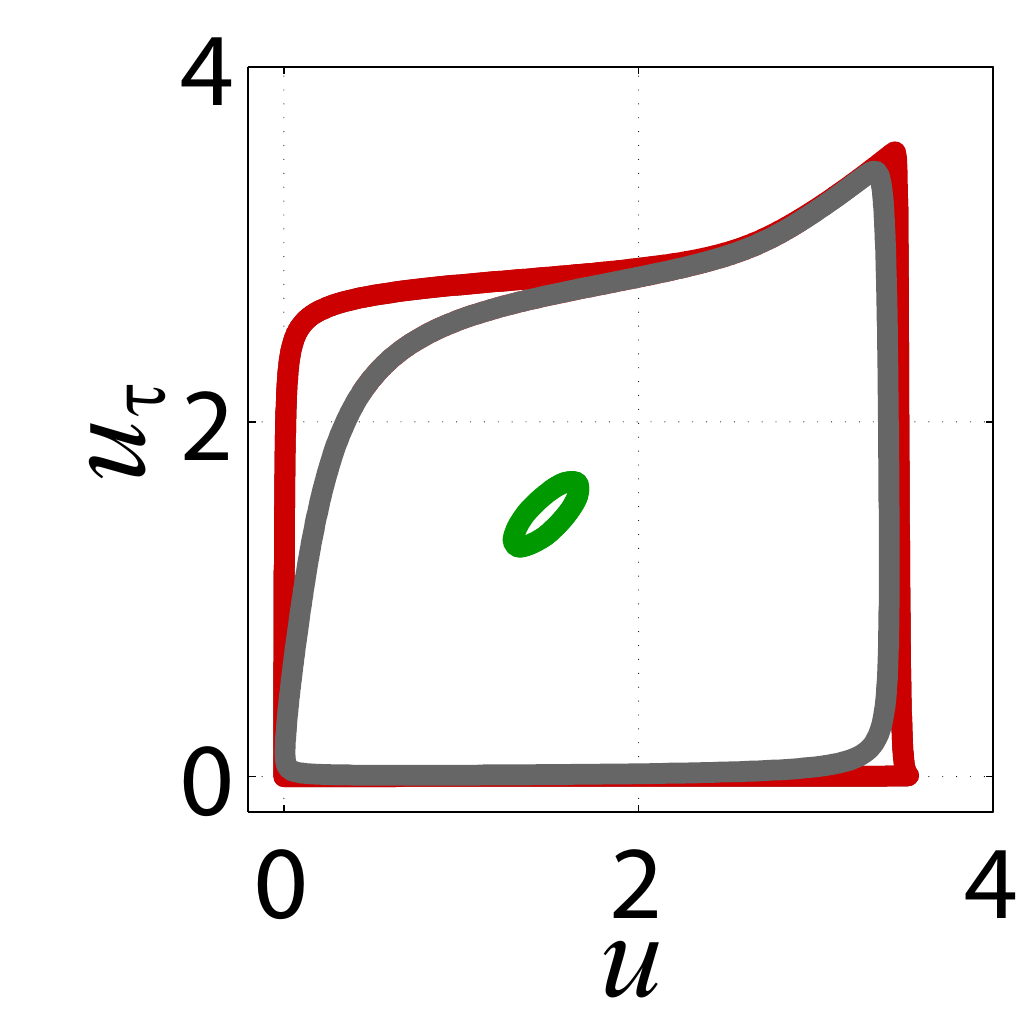}}
	&
	\subfloat[]{
	\raisebox{8mm}{
	\begin{minipage}[t][][s]{0.49\columnwidth}
	\centering
	\includegraphics[width=\columnwidth]{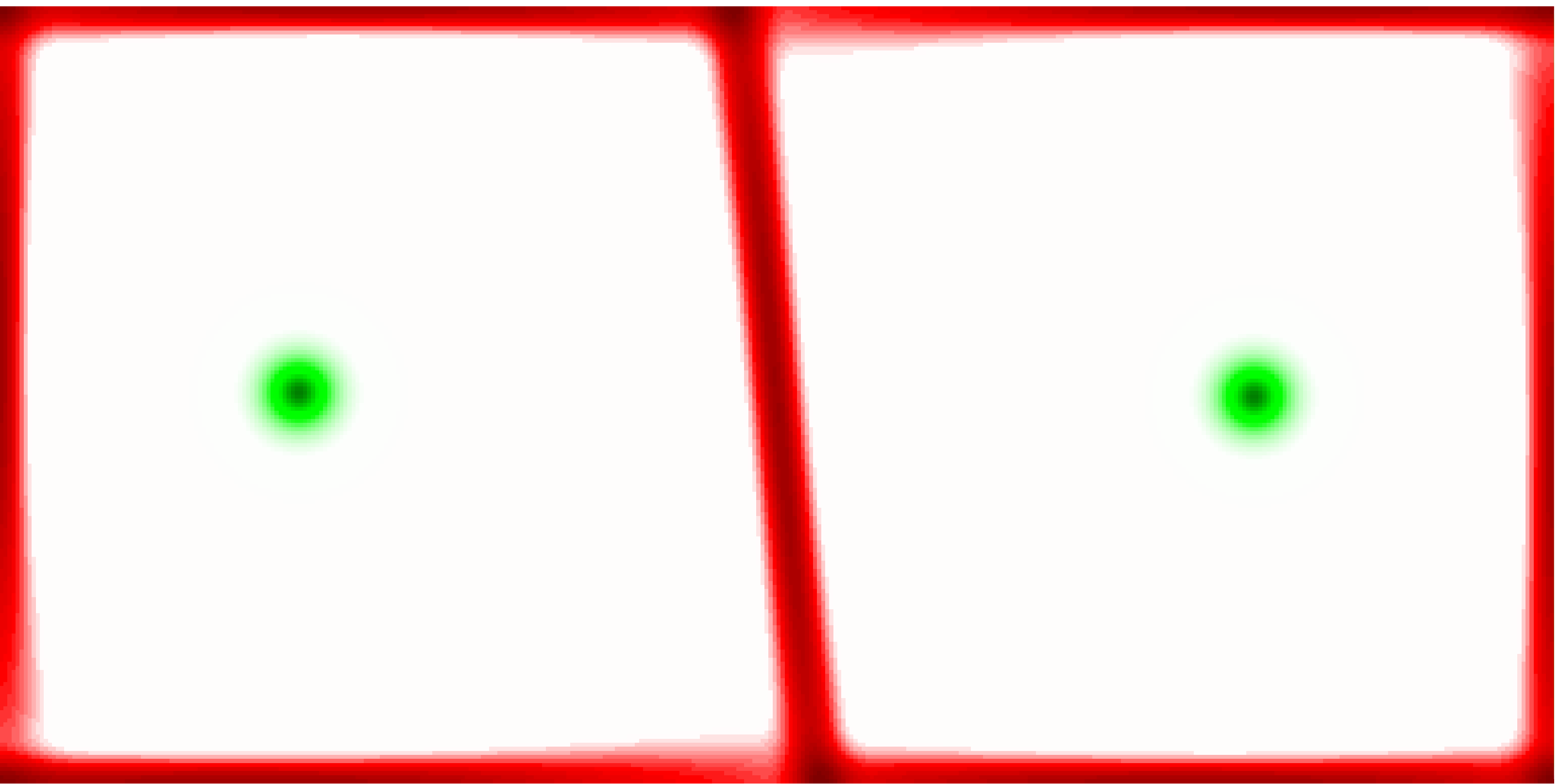}
	\\
	\includegraphics[width=\columnwidth]{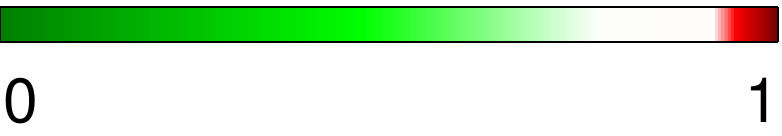}
	\end{minipage}}}
    \end{tabular}
    \caption{The cycles in the (a) $(u,v)$ plane, (c) $(u,\dot{u})$ plane, and (e) $(u,u_\tau)$ plane. The corresponding cycle areas $I_1$ (b), $I_2$ (d), and $I_3$ (f) for the two-spiral solution shown in Fig.~\ref{fig6}(c).}
  \label{fig11}
\end{figure}

A similar approach can be used to identify the tile boundaries for the modified Karma model. The easiest way to characterize the amplitude of the strongly nonlinear oscillations is by computing the area $I_1$ of the cycle $C$ in the $(u,v)$ plane,
\begin{equation}
I_1(x,y)=\left|\oint_C v\,du\right|=\left|\int_0^T v\dot{u}\,dt\right|.
\end{equation}
Several representative cycles $C$ computed for the co-rotating two-spiral solution \rgedit{depicted in Fig.~\ref{fig6}(c)} are shown in Figure~\ref{fig11}(a) and the corresponding spatial distribution $I_1(x,y)$ is shown in Figure~\ref{fig11}(b). The two spirals are separated by a shock which corresponds to the local maximum of $I_1(x,y)$. In addition, we also find shocks that form along the outer boundary, where no-flux boundary condition is imposed. Taken together, the shocks form a closed boundary for each of the spiral domains, defining the tiles on which the dynamics is controlled by one or the other core.

While computing cycle areas in the $(u,v)$ plane is convenient in the models where all variables are accessible, in experiment this is rarely the case. Most typically only one variable is easily accessible experimentally (e.g., voltage or calcium, if voltage- or calcium-sensitive \rgedit{dye} is used). In this case cycle areas can also be computed using alternative planar representations of the cellular dynamics based solely on one variable, e.g. the voltage $u$.  One possibility is to use a  $(u,\dot{u})$ plane, with the cycle area defined as
\begin{equation}
I_2(x,y)=\oint_C \dot{u}\,du=\int_0^T \dot{u}^2\,dt.
\end{equation}
Some representative cycles and the cycle area distribution are shown in Figs.~\ref{fig11}(c) and (d), respectively.

The expression for $I_2$, however, involves a derivative of $u$. Since experimental measurements are typically noisy, derivatives obtained using finite-differencing of a time series can be very inaccurate. To reduce the influence of noise, a time-delay embedding can be used instead, with the cycles defined in the plane spanned by $u$ evaluated at times $t$ and $t-\tau$ (the latter will be denoted $u_\tau$ for brevity).  The corresponding cycle area
\begin{equation}
I_3(x,y)=\left|\oint_C u_{\tau}\,du\right|
\end{equation}
can be computed without evaluating derivatives of any field. The choice of the time delay $\tau$ is not unique. We chose the \rgedit{value} $\tau=13.5$ ms \rgedit{which corresponds to} the first minimum of the mutual information function~\cite{kantz04}. The corresponding representative cycles in the $(u,u_{\tau})$ plane and the cycle area distribution are shown in Figs.~\ref{fig11}(e) and (f), respectively. Comparison of Figs.~\ref{fig11}(b), (d), and (f) shows that \rgedit{all} three cycle area representations are consistent and accurately capture the shock line separating the two spirals, as well as the shock lines along the domain boundaries.

In the rest of this study we use the cycle areas $I_1$ computed in the $(u,v)$ plane to identify tiles in the computational domain.  In particular, Figs.~\ref{fig4}(g) and (h) show the shocks that separate the tiles that form for the nearly-recurrent multi-spiral solutions, while Fig.~\ref{fig9}(b) shows the shocks that separate the tiles in the three-spiral solution. For time-periodic solutions (e.g., single- or two-spiral solutions of the Karma model) the period $T$ is well-defined and the cycles close perfectly. For non-periodic solutions such as those shown in Figs.~\ref{fig4} and \ref{fig9} the integration is instead performed between crossings of a convenient Poincar\'e section (we used $u=1.5$).

If the phase of the spiral solution is well-described by the Archimedian approximation, the tile boundaries can also be constructed analytically. Generally, the Archimedian approximation, and hence the analytic solutions for the tile boundaries, is only valid when the distance from each spiral core to the tile boundary is \rgedit{sufficiently large.}  Bohr {\it et al.}~\cite{BohHubOtt96,BohHubOtt97} showed that for CGLE \rgedit{the tile boundaries} are segments of hyperbolas with the two nearest spiral cores serving as foci and that the approximation is valid even when the separation between the cores is \rgedit{as small as one} wavelength $\lambda$. The hyperbolic solution, however, only applies to spirals of opposite chirality (i.e., counter-rotating \rgedit{spiral waves}).

A more general equation for the tile boundary between spirals of any chirality was derived by Zhan {\it et al.}~\cite{zhan07}. \rgedit{Let the origins of the two spirals be ${\bf x}$ and ${\bf x}'$, their chiralities $\sigma$, $\sigma'=\pm 1$, ${\bf R}={\bf x'}-{\bf x}$, and let ${\bf x}+{\bf r}$ define a point on the boundary. Then the boundary is given by the solution to the differential equation
\begin{equation}\label{eq:zhanEq}
	\frac{dr}{d\varphi} = \frac{-\sigma\rho^2 - 
	\sigma^\prime r(R\cos\varphi - r)
       + 2\pi m \rho r\sin\varphi}
        {\sigma' R^2 \sin\varphi 
       + m\left(\rho^2 - \rho(r-R\cos\varphi)\right)},
\end{equation}
where $m = R/\lambda$, $\rho = \sqrt{R^2 + r^2 - 2rR\cos\varphi}$, and $\varphi$ is the angle between ${\bf r}$ and ${\bf R}$. This equation can be solved numerically given an initial condition that lies on the line connecting the origins of the two spirals. Equation (\ref{eq:zhanEq})} was shown to accurately capture the tile boundaries not only for weakly nonlinear oscillations found in CGLE, but also for the Barkley model~\cite{BaKnTu90} which, like the Karma model, describes an excitable medium supporting strongly nonlinear oscillations~\cite{luo09}.

\begin{figure}
    \begin{tabular}{cc}
      \subfloat[]{\includegraphics[width=0.49\columnwidth]{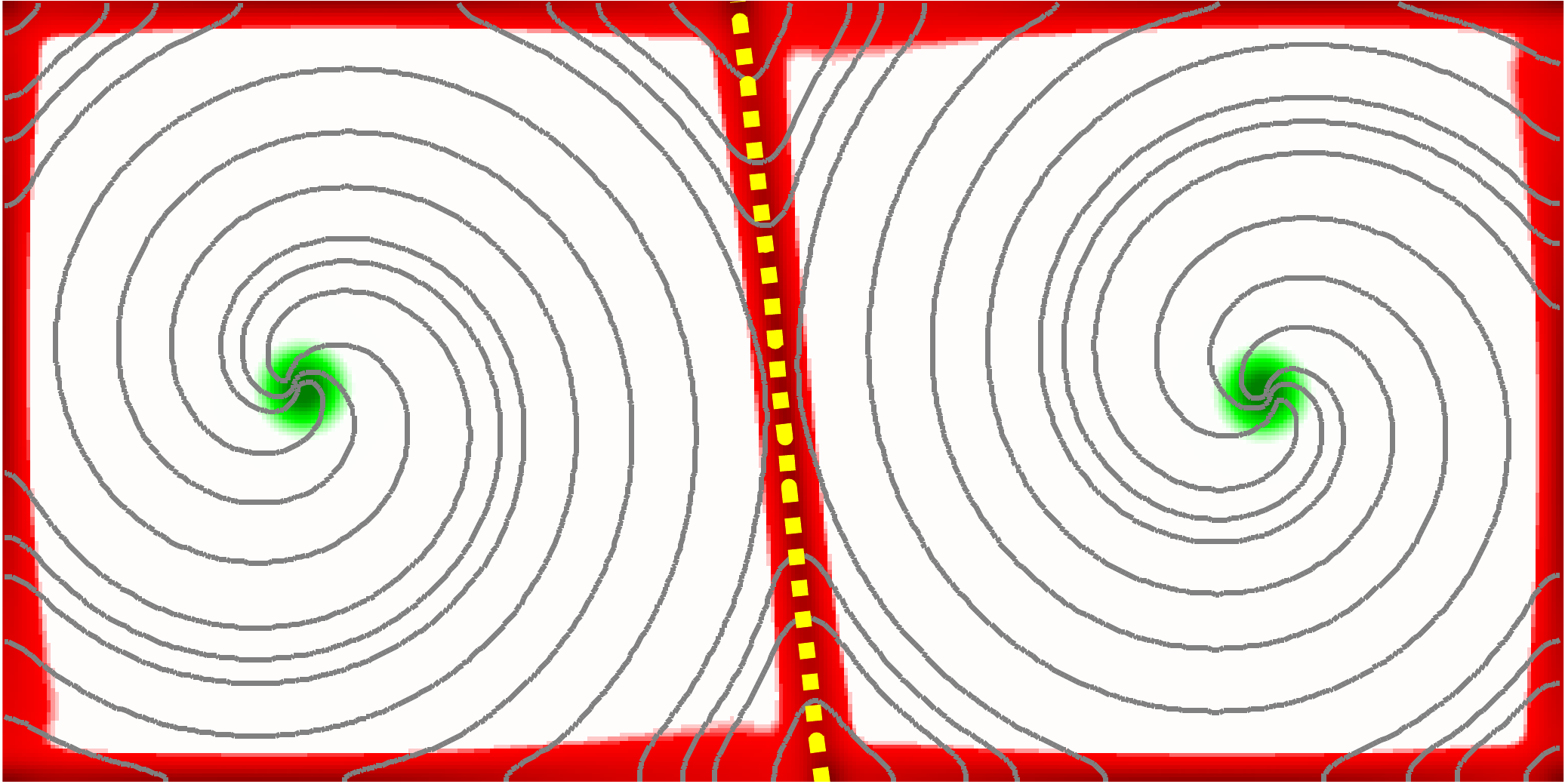}} &
      \subfloat[]{\includegraphics[width=0.49\columnwidth]{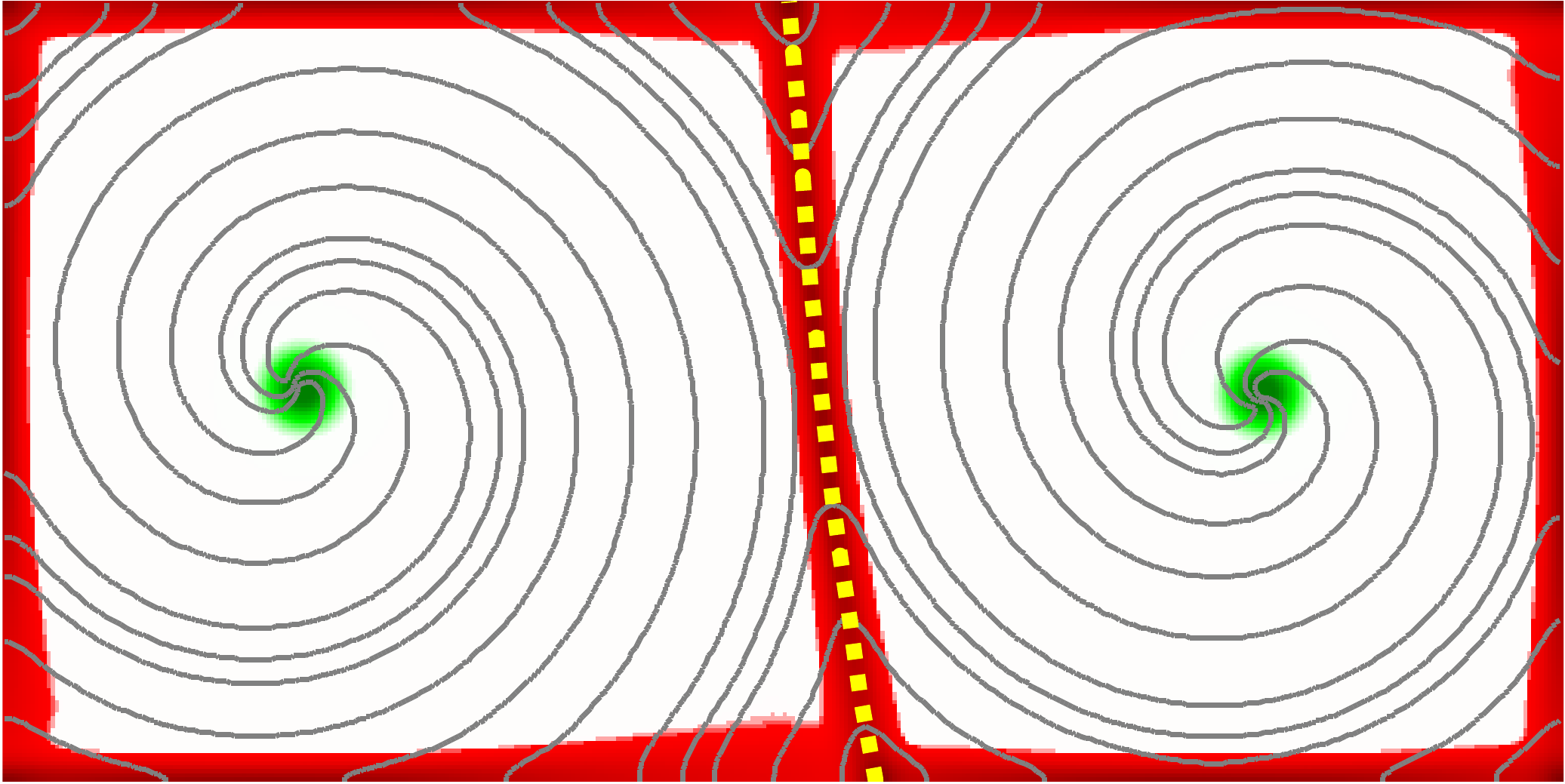}}\\
      \subfloat[]{\includegraphics[width=0.49\columnwidth]{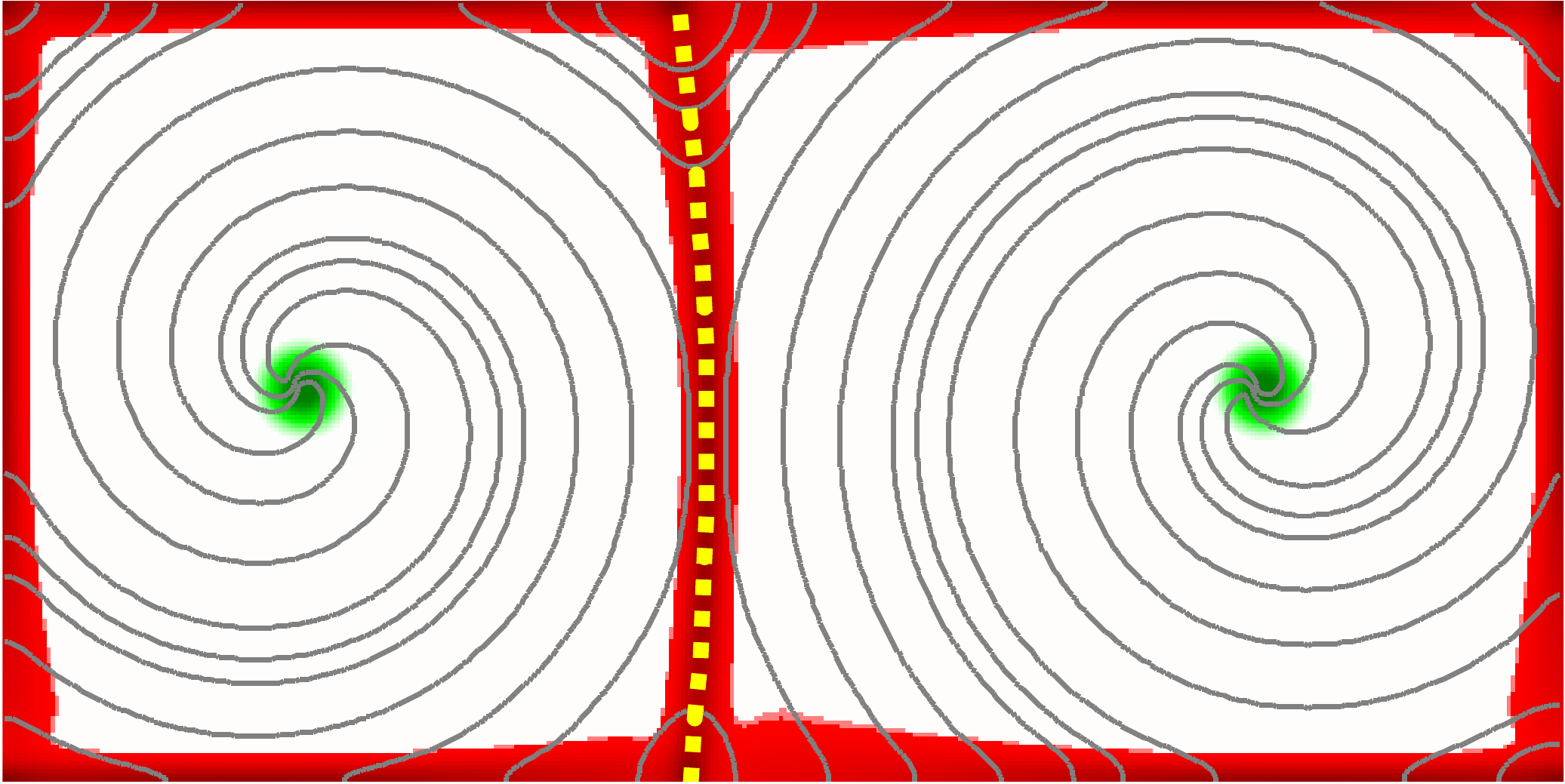}} &
      \subfloat[]{\includegraphics[width=0.49\columnwidth]{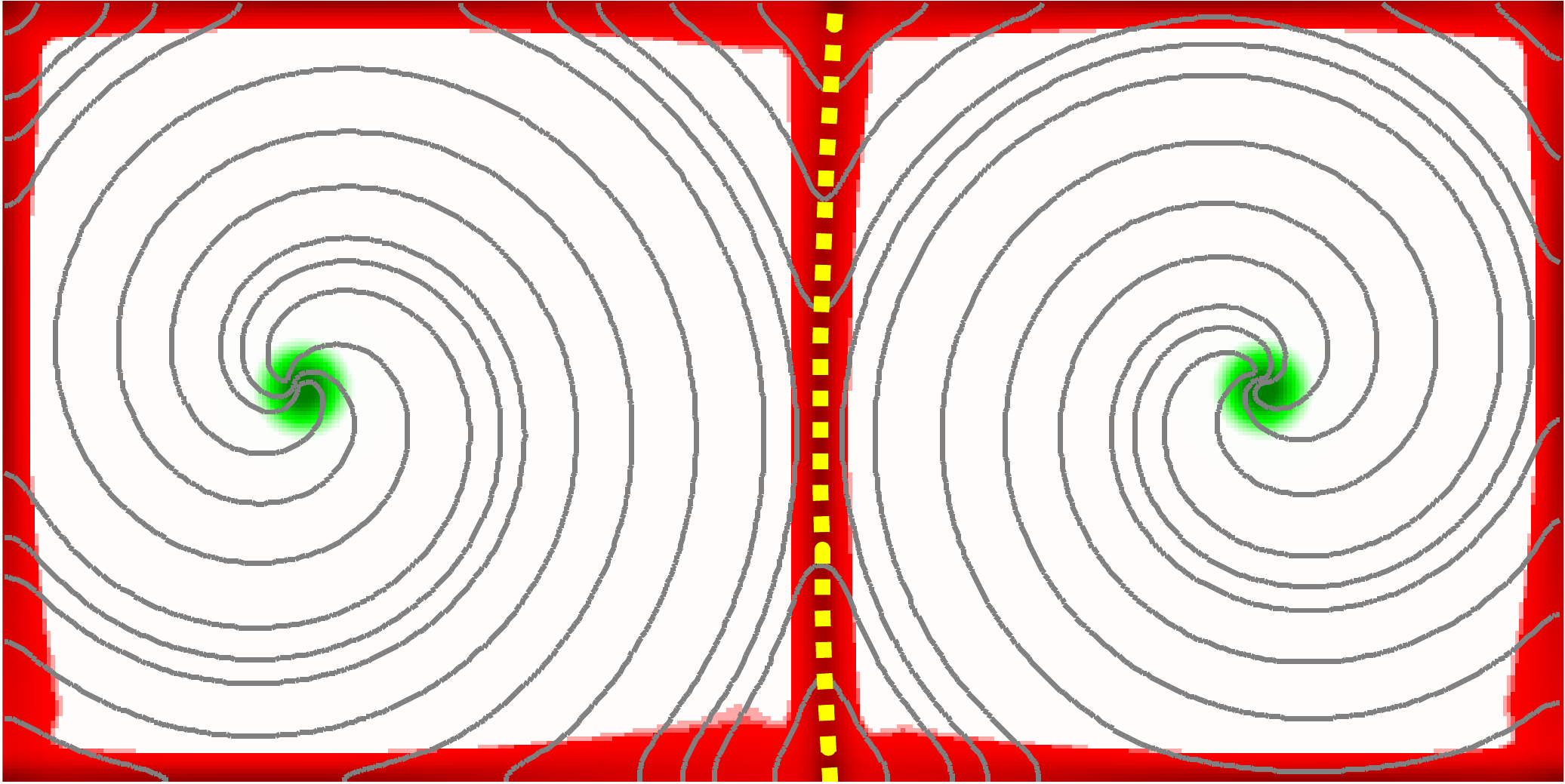}}
    \end{tabular}
    \caption{ The cycle areas $I_1$ for select \rgedit{unstable} two-spiral states. Also shown are the analytic solutions for the (internal) tile boundaries (dashed yellow line) and level sets of $v$ (gray lines).  The states correspond to: (a) the co-rotating spirals shown in Fig.~\ref{fig6}(c), (b) the co-rotating spirals shown in Fig.~\ref{fig6}(d), (c) the counter-rotating spirals shown in Fig.~\ref{fig7}(c), and (d) the counter-rotating spirals shown in Fig.~\ref{fig7}(d). The color bar from Fig.~\ref{fig11}(b) is used in all panels.
}
  \label{fig12}
\end{figure}

We checked the validity of this equation for the modified Karma model by comparing the analytic solutions to those computed using the cycle area method for the co- and counter-rotating phase-shifted solutions from Figs.~\ref{fig6} and~\ref{fig7}, for which $m=3.18$. \rgedit{Excellent} agreement was found in all cases. Four examples with the analytic solutions superimposed on the cycle area plots are shown in Fig.~\ref{fig12}. Numerical and analytical solutions also agree well for the three-spiral solution, as Fig.~\ref{fig9}(c) illustrates. 

It should be noted that both methods of computing the tile boundaries have advantages and drawbacks. Numerical solutions based on cycle areas do not require any assumptions (e.g., the Archimedean approximation for the phase) and can be computed in real time. On the other hand, they are only updated once per period (i.e., upon crossing of the Poincar\'e section), which may not be adequate for quickly drifting spirals. Furthermore, the shocks often do not entirely enclose each spiral \rgedit{(cf. Fig.~\ref{fig4}(g-h))}; an additional construction is needed to form a closed boundary or determine the precise position of the boundary based on the transverse profile of the shock. Analytical solutions form closed boundaries, but require identification of the position of the cores and the initial condition (e.g., the point where the boundary crosses the straight line connecting the two cores). Analytical construction also requires an algorithm for determining the endpoints of each smooth segment of the boundary where three (or more) different tiles meet.

\subsection{Assembling a Global Solution}
\label{sec:sixSS2}

For computational domains with no-flux boundary conditions, shocks form naturally along the boundaries. \rgedit{Additionally, empirical observations for states with tile boundaries evolving slowly compared with the spiral rotation demonstrate that the level sets of both $v$ (cf. Figs.~\ref{fig4}(g-h), \ref{fig9}(c), and \ref{fig12}) and $u$ (not shown) are orthogonal to the shocks.}
Hence, the single-spiral solution on each of the tiles satisfies the no-flux boundary conditions on its entire boundary, whether it is external or internal with respect to the computational domain. We can therefore reduce the interaction between different spirals to the effect of boundary conditions, which only affects the solution in the interior of the tile through its shape (and the dynamics of the tile boundaries, if any). The effect of the boundaries on the enclosed spiral \rgedit{wave} should be qualitatively the same regardless of the tile geometry.  In particular, the temporal period of each spiral \rgedit{wave} should depend on the size of the tile. \rgedit{For example, the period of unstable single-spiral wave solutions decreases from the asymptotic value $T_0$ as the domain size $L$ decreases for the Karma model on square domains~\cite{Marcotte2014}.} For domains of size comparable to the smallest tiles in Figs.~\ref{fig4}(g-h) the period is estimated to decrease by $O(10^{-4}T_0)$. Such differences in the periods of different spirals would lead to a relative residual of $O(10^{-4})$, which is consistent with the residuals reported in Figs.~\ref{fig4}(c-d).

It is well known \cite{Howard1977,Krinsky1983} that if the frequencies $\omega_1$ and $\omega_2$ of two neighboring spirals differ, the boundary between them moves with velocity
\begin{equation}
{\bf c}=(\omega_1-\omega_2)\frac{{\bf k}_1-{\bf k}_2}{|{\bf k}_1-{\bf k}_2|^2},
\end{equation}
where ${\bf k}_1$ and ${\bf k}_2$ are the wave vectors the two spirals would have on an unbounded domain at the location of the tile boundary. This is a consequence of the phase continuity at the tile boundaries: the frequencies of the two spirals become equal in a reference frame moving with velocity ${\bf c}$. In particular, small differences in the frequencies (equivalently, periods) of two neighboring spirals lead to a slow motion of the boundary. 

\rgedit{As we have shown elsewhere~\cite{Marcotte2014}, in the Karma model the spiral cores are repelled from no-flux boundaries at distances of $O(2\ell_c$), where $\ell_c$ is the size of the spiral core. This appears to be a general result, as repulsive interaction was also found in other excitable systems~\cite{LanBar13,LanBar14}.
Hence, when the distance between a spiral core and the nearest tile boundary decreases to about $O(2\ell_c)$, the spiral core starts to drift away from the boundary.}
This is the mechanism that forces the core inside the small tile in Fig.~\ref{fig9} to drift relative to the cores in big tiles. Such core drift is also unavoidable for multi-spiral solutions with tiles of different sizes and \rgedit{will lead to the growth of big spirals at the expense of small ones, which can be considered as one of the mechanisms contributing to maintenance of spiral turbulence.}

The tile-based decomposition suggests a natural approach to constructing global multi-spiral solutions from single-spiral segments satisfying local Euclidean symmetries.
The results obtained for single-spiral solutions \cite{Marcotte2014} suggest that the solution on each tile would be defined either by a periodic solution or a relative periodic solution. If the tile boundaries do not move significantly during a typical period $T_0$, we can compute the solutions locally on each tile using weighted Newton-Krylov method~\cite{Marcotte2014}. The algorithm, however, will need to be generalized in such a way that updates of the initial condition at each step of Newton iteration preserve the continuity of both $u$ and $v$ fields (or equivalently the phase and amplitude) on the tile boundaries. This constraint, however, is not expected to present any conceptual difficulties.

\section{Conclusions}
\label{sec:eight}

To summarize, we have applied numerical methods originally developed for fluid turbulence to search for the exact coherent structures that may form a skeleton for the spatiotemporally chaotic dynamics produced by a prototypical monodomain model of cardiac tissue. We showed that these methods, designed to identify recurrent solutions in the presence of global symmetries, fail rather spectacularly for an excitable reaction-diffusion system whose dynamics is characterized by local, rather than global Euclidean symmetries. The origin of the failure was traced to the weak correlation between the dynamics of individual spirals which underpin spiral turbulence. As a result of this weak correlation, typical multi-spiral states display recurrent dynamics locally, but not globally. Locally the dynamics can be represented, to numerical accuracy, by periodic or relative periodic solutions, but globally neither periodic nor relative periodic solutions play a dynamically important role. Nonchaotic unstable solutions \rgedit{embedded in the chaotic set would} have a more complicated nature and require development of novel computational approaches.

We propose one such approach based on the decomposition of the computational domain into sub-domains, or tiles, that each support one spiral \rgedit{wave}. Over short time scales (before individual spiral \rgedit{waves} are destroyed by local instabilities) the dynamics for each near-recurrent multi-spiral state can be decomposed into the dynamics {\it of the tiles} (relative motion of tiles and changes in their shape associated with the differences in the spiral frequencies) and the dynamics of individual spiral waves {\it on the tiles} subject to no-flux boundary conditions at the tile boundaries. \rgedit{In particular, the dynamics of spiral waves on the tiles} would be described by periodic or relative periodic solutions which correspond, respectively, to pinned and drifting cores. 

\rgedit{It should be emphasized that the formalism based on decomposition into tiles is only expected to describe spiral turbulence during relatively quiescent intervals when no breakups or mergers of different spiral waves occur. Breakups and mergers are driven, respectively, by the alternans and meandering instabilities of individuals spirals~\cite{Karma1994,Marcotte2014} and involve birth or annihilation of pairs of spiral cores with opposite chirality and the associated changes in the number of tiles. During such events the cores move on the time scale comparable to the rotation period (or even faster) and it may not be possible to even define the tiles on the entire domain. These relatively active intervals should not be described using the proposed formalism and likely correspond to heteroclinic connections between different exact coherent structures. This can be rephrased in terms of the of skew-decomposition of the dynamics \cite{FiSaScWu96,FiTu98,SaScWu97,SaScWu99} in the vicinity of relative solutions induced by local symmetries: the quiescent intervals correspond to the evolution mainly along the group manifold, while active intervals correspond to motion mainly transverse to the group manifold.}

Future work will focus on a numerical implementation of the procedure we have described for constructing global solutions composed of locally symmetry-reduced solutions.  Specifically, the method for computing single-spiral \rgedit{wave} solutions on square domains should be validated for tiles with an arbitrary shape.  Furthermore, it remains to be determined how individual solutions should be recombined such that the phases and amplitudes of neighboring spirals match. If successful, this framework may provide valuable new insight not only into spatiotemporally chaotic dynamics of cardiac tissue, but also other systems that exhibit local symmetries.

The definition of locality is, of course, relative. In the model considered here, local symmetries survive on domains, or tiles, whose dimensions \rgedit{significantly} exceed the characteristic correlation length \rgedit{$\ell_c$} defined by the spatial extent of the adjoint eigenmodes for spiral \rgedit{wave} solutions \cite{BBBBF09,Biktasheva:2010co}. For \rgedit{excitable systems, such as the FitzHugh-Nagumo, Barkley, and Karma models,} these eigenmodes decay exponentially, reflecting the lack of any long-range correlations. Short-range correlations, however, \rgedit{may not describe} all cardiac tissue models. For instance, bidomain models \cite{Sepulveda1989} also include an additional Poisson equation for the extracellular potential, generating long-range correlations. Similarly, long-range correlations can arise as a result of stretch-activated feedback \cite{Alvarez:2009}. Investigation of the relation between symmetries and the structure of exact coherent structures in bidomain models is of particular interest both because they provide a more realistic description of cardiac tissue, compared with the monodomain models, and because of the analogy with fluid dynamics where long-range coupling is due to the pressure field, which also satisfies a Poisson equation.

\begin{acknowledgments}
The authors would like to thank Vadim Biktashev for helpful discussions and comments. This material is based upon work supported by the National Science Foundation under Grant No. CMMI-1028133.  The Tesla K20 GPUs used for this research were donated by the ``NVIDIA Corporation'' through the academic hardware donation program.
\end{acknowledgments}


\bibliography{cardiac}

\begin{thebibliography}{41}%
\makeatletter
\providecommand \@ifxundefined [1]{%
 \@ifx{#1\undefined}
}%
\providecommand \@ifnum [1]{%
 \ifnum #1\expandafter \@firstoftwo
 \else \expandafter \@secondoftwo
 \fi
}%
\providecommand \@ifx [1]{%
 \ifx #1\expandafter \@firstoftwo
 \else \expandafter \@secondoftwo
 \fi
}%
\providecommand \natexlab [1]{#1}%
\providecommand \enquote  [1]{``#1''}%
\providecommand \bibnamefont  [1]{#1}%
\providecommand \bibfnamefont [1]{#1}%
\providecommand \citenamefont [1]{#1}%
\providecommand \href@noop [0]{\@secondoftwo}%
\providecommand \href [0]{\begingroup \@sanitize@url \@href}%
\providecommand \@href[1]{\@@startlink{#1}\@@href}%
\providecommand \@@href[1]{\endgroup#1\@@endlink}%
\providecommand \@sanitize@url [0]{\catcode `\\12\catcode `\$12\catcode
  `\&12\catcode `\#12\catcode `\^12\catcode `\_12\catcode `\%12\relax}%
\providecommand \@@startlink[1]{}%
\providecommand \@@endlink[0]{}%
\providecommand \url  [0]{\begingroup\@sanitize@url \@url }%
\providecommand \@url [1]{\endgroup\@href {#1}{\urlprefix }}%
\providecommand \urlprefix  [0]{URL }%
\providecommand \Eprint [0]{\href }%
\providecommand \doibase [0]{http://dx.doi.org/}%
\providecommand \selectlanguage [0]{\@gobble}%
\providecommand \bibinfo  [0]{\@secondoftwo}%
\providecommand \bibfield  [0]{\@secondoftwo}%
\providecommand \translation [1]{[#1]}%
\providecommand \BibitemOpen [0]{}%
\providecommand \bibitemStop [0]{}%
\providecommand \bibitemNoStop [0]{.\EOS\space}%
\providecommand \EOS [0]{\spacefactor3000\relax}%
\providecommand \BibitemShut  [1]{\csname bibitem#1\endcsname}%
\let\auto@bib@innerbib\@empty
\bibitem [{\citenamefont {Poincar\'e}(1899)}]{poincare1899}%
  \BibitemOpen
  \bibfield  {author} {\bibinfo {author} {\bibfnamefont {H.}~\bibnamefont
  {Poincar\'e}},\ }\href@noop {} {\emph {\bibinfo {title} {Les M\'ethodes
  Nouvelles de la M\'echanique C\'eleste}}}\ (\bibinfo  {publisher}
  {Gauthier-Villars},\ \bibinfo {year} {1899})\BibitemShut {NoStop}%
\bibitem [{\citenamefont {Gutzwiller}(1971)}]{gutzwiller71}%
  \BibitemOpen
  \bibfield  {author} {\bibinfo {author} {\bibfnamefont {M.~C.}\ \bibnamefont
  {Gutzwiller}},\ }\bibfield  {title} {\enquote {\bibinfo {title} {Periodic
  orbits and classical quantization conditions},}\ }\href@noop {} {\bibfield
  {journal} {\bibinfo  {journal} {J. Math. Phys.}\ }\textbf {\bibinfo {volume}
  {12}},\ \bibinfo {pages} {343--358} (\bibinfo {year} {1971})}\BibitemShut
  {NoStop}%
\bibitem [{\citenamefont {Christiansen}, \citenamefont {Cvitanovi\'{c}},\ and\
  \citenamefont {Putkaradze}(1997)}]{Christiansen97}%
  \BibitemOpen
  \bibfield  {author} {\bibinfo {author} {\bibfnamefont {F.}~\bibnamefont
  {Christiansen}}, \bibinfo {author} {\bibfnamefont {P.}~\bibnamefont
  {Cvitanovi\'{c}}}, \ and\ \bibinfo {author} {\bibfnamefont {V.}~\bibnamefont
  {Putkaradze}},\ }\bibfield  {title} {\enquote {\bibinfo {title}
  {Spatiotemporal chaos in terms of unstable recurrent patterns},}\ }\href@noop
  {} {\bibfield  {journal} {\bibinfo  {journal} {Nonlinearity}\ }\textbf
  {\bibinfo {volume} {10}},\ \bibinfo {pages} {55--70} (\bibinfo {year}
  {1997})},\ \bibinfo {note} {\arXiv{chao-dyn/9606016}}\BibitemShut {NoStop}%
\bibitem [{\citenamefont {Lan}\ and\ \citenamefont
  {Cvitanovi{\'c}}(2008)}]{lanCvit07}%
  \BibitemOpen
  \bibfield  {author} {\bibinfo {author} {\bibfnamefont {Y.}~\bibnamefont
  {Lan}}\ and\ \bibinfo {author} {\bibfnamefont {P.}~\bibnamefont
  {Cvitanovi{\'c}}},\ }\bibfield  {title} {\enquote {\bibinfo {title} {Unstable
  recurrent patterns in {Kuramoto-Sivashinsky} dynamics},}\ }\href@noop {}
  {\bibfield  {journal} {\bibinfo  {journal} {Phys. Rev. E}\ }\textbf {\bibinfo
  {volume} {78}},\ \bibinfo {pages} {026208} (\bibinfo {year} {2008})},\
  \bibinfo {note} {\arXiv{0804.2474}}\BibitemShut {NoStop}%
\bibitem [{\citenamefont {L{\'o}pez}\ \emph {et~al.}(2006)\citenamefont
  {L{\'o}pez}, \citenamefont {Boyland}, \citenamefont {Heath},\ and\
  \citenamefont {Moser}}]{lop05rel}%
  \BibitemOpen
  \bibfield  {author} {\bibinfo {author} {\bibfnamefont {V.}~\bibnamefont
  {L{\'o}pez}}, \bibinfo {author} {\bibfnamefont {P.}~\bibnamefont {Boyland}},
  \bibinfo {author} {\bibfnamefont {M.~T.}\ \bibnamefont {Heath}}, \ and\
  \bibinfo {author} {\bibfnamefont {R.~D.}\ \bibnamefont {Moser}},\ }\bibfield
  {title} {\enquote {\bibinfo {title} {Relative periodic solutions of the
  complex {Ginzburg-Landau} equation},}\ }\href {\doibase 10.1137/040618977}
  {\bibfield  {journal} {\bibinfo  {journal} {SIAM J. Appl. Dyn. Syst.}\
  }\textbf {\bibinfo {volume} {4}},\ \bibinfo {pages} {1042--1075} (\bibinfo
  {year} {2006})},\ \bibinfo {note} {\arXiv{nlin/0408018}}\BibitemShut
  {NoStop}%
\bibitem [{\citenamefont {Viswanath}(2007)}]{Viswanath:2007re}%
  \BibitemOpen
  \bibfield  {author} {\bibinfo {author} {\bibfnamefont {D.}~\bibnamefont
  {Viswanath}},\ }\bibfield  {title} {\enquote {\bibinfo {title} {Recurrent
  motions within plane {C}ouette turbulence},}\ }\href@noop {} {\bibfield
  {journal} {\bibinfo  {journal} {J. Fluid Mech.}\ }\textbf {\bibinfo {volume}
  {580}},\ \bibinfo {pages} {339} (\bibinfo {year} {2007})}\BibitemShut
  {NoStop}%
\bibitem [{\citenamefont {Gibson}, \citenamefont {Halcrow},\ and\ \citenamefont
  {Cvitanovi{\'c}}(2008)}]{GHCW07}%
  \BibitemOpen
  \bibfield  {author} {\bibinfo {author} {\bibfnamefont {J.~F.}\ \bibnamefont
  {Gibson}}, \bibinfo {author} {\bibfnamefont {J.}~\bibnamefont {Halcrow}}, \
  and\ \bibinfo {author} {\bibfnamefont {P.}~\bibnamefont {Cvitanovi{\'c}}},\
  }\bibfield  {title} {\enquote {\bibinfo {title} {Visualizing the geometry of
  state-space in plane {Couette} flow},}\ }\href@noop {} {\bibfield  {journal}
  {\bibinfo  {journal} {J. Fluid Mech.}\ }\textbf {\bibinfo {volume} {611}},\
  \bibinfo {pages} {107--130} (\bibinfo {year} {2008})},\ \bibinfo {note}
  {\arXiv{0705.3957}}\BibitemShut {NoStop}%
\bibitem [{\citenamefont {Meseguer}\ \emph {et~al.}(2009)\citenamefont
  {Meseguer}, \citenamefont {Mellibovsky}, \citenamefont {Avila},\ and\
  \citenamefont {Marques}}]{Meseguer2009}%
  \BibitemOpen
  \bibfield  {author} {\bibinfo {author} {\bibfnamefont {A.}~\bibnamefont
  {Meseguer}}, \bibinfo {author} {\bibfnamefont {F.}~\bibnamefont
  {Mellibovsky}}, \bibinfo {author} {\bibfnamefont {M.}~\bibnamefont {Avila}},
  \ and\ \bibinfo {author} {\bibfnamefont {F.}~\bibnamefont {Marques}},\
  }\bibfield  {title} {\enquote {\bibinfo {title} {Families of subcritical
  spirals in highly counter-rotating {Taylor-Couette} flow},}\ }\href@noop {}
  {\bibfield  {journal} {\bibinfo  {journal} {Phys. Rev. E}\ }\textbf {\bibinfo
  {volume} {79}} (\bibinfo {year} {2009})}\BibitemShut {NoStop}%
\bibitem [{\citenamefont {de~Lozar}\ \emph {et~al.}(2012)\citenamefont
  {de~Lozar}, \citenamefont {Mellibovsky}, \citenamefont {Avila},\ and\
  \citenamefont {Hof}}]{deLozar2012}%
  \BibitemOpen
  \bibfield  {author} {\bibinfo {author} {\bibfnamefont {A.}~\bibnamefont
  {de~Lozar}}, \bibinfo {author} {\bibfnamefont {F.}~\bibnamefont
  {Mellibovsky}}, \bibinfo {author} {\bibfnamefont {M.}~\bibnamefont {Avila}},
  \ and\ \bibinfo {author} {\bibfnamefont {B.}~\bibnamefont {Hof}},\ }\bibfield
   {title} {\enquote {\bibinfo {title} {Edge state in pipe flow experiments},}\
  }\href@noop {} {\bibfield  {journal} {\bibinfo  {journal} {Phys. Rev. Lett.}\
  }\textbf {\bibinfo {volume} {108}},\ \bibinfo {pages} {214502} (\bibinfo
  {year} {2012})}\BibitemShut {NoStop}%
\bibitem [{\citenamefont {Chandler}\ and\ \citenamefont
  {Kerswell}(2013)}]{Chandler2013}%
  \BibitemOpen
  \bibfield  {author} {\bibinfo {author} {\bibfnamefont {G.~J.}\ \bibnamefont
  {Chandler}}\ and\ \bibinfo {author} {\bibfnamefont {R.~R.}\ \bibnamefont
  {Kerswell}},\ }\bibfield  {title} {\enquote {\bibinfo {title} {Invariant
  recurrent solutions embedded in a turbulent two-dimensional {Kolmogorov}
  flow},}\ }\href@noop {} {\bibfield  {journal} {\bibinfo  {journal} {J. Fluid
  Mech.}\ }\textbf {\bibinfo {volume} {722}},\ \bibinfo {pages} {554--595}
  (\bibinfo {year} {2013})}\BibitemShut {NoStop}%
\bibitem [{\citenamefont {Ideker}, \citenamefont {Zhou},\ and\ \citenamefont
  {Knisley}(1995)}]{IdZhKn95}%
  \BibitemOpen
  \bibfield  {author} {\bibinfo {author} {\bibfnamefont {R.~E.}\ \bibnamefont
  {Ideker}}, \bibinfo {author} {\bibfnamefont {X.}~\bibnamefont {Zhou}}, \ and\
  \bibinfo {author} {\bibfnamefont {S.~B.}\ \bibnamefont {Knisley}},\
  }\bibfield  {title} {\enquote {\bibinfo {title} {Correlation among
  fibrillation, defibrillation, and cardiac pacing},}\ }\href {\doibase
  10.1111/j.1540-8159.1995.tb02562.x} {\bibfield  {journal} {\bibinfo
  {journal} {Pacing and Clinical Electrophysiology}\ }\textbf {\bibinfo
  {volume} {18}},\ \bibinfo {pages} {512--525} (\bibinfo {year}
  {1995})}\BibitemShut {NoStop}%
\bibitem [{\citenamefont {Walcott}, \citenamefont {Killingsworth},\ and\
  \citenamefont {Ideker}(2003)}]{WaKilIde03}%
  \BibitemOpen
  \bibfield  {author} {\bibinfo {author} {\bibfnamefont {G.~P.}\ \bibnamefont
  {Walcott}}, \bibinfo {author} {\bibfnamefont {C.~R.}\ \bibnamefont
  {Killingsworth}}, \ and\ \bibinfo {author} {\bibfnamefont {R.~E.}\
  \bibnamefont {Ideker}},\ }\bibfield  {title} {\enquote {\bibinfo {title} {Do
  clinically relevant transthoracic defibrillation energies cause myocardial
  damage and dysfunction?}}\ }\href {\doibase 10.1016/S0300-9572(03)00161-8}
  {\bibfield  {journal} {\bibinfo  {journal} {Resuscitation}\ }\textbf
  {\bibinfo {volume} {59}},\ \bibinfo {pages} {59--70} (\bibinfo {year}
  {2003})}\BibitemShut {NoStop}%
\bibitem [{\citenamefont {Luther}\ \emph {et~al.}(2011)\citenamefont {Luther},
  \citenamefont {Fenton}, \citenamefont {Kornreich}, \citenamefont {Squires},
  \citenamefont {Bittihn}, \citenamefont {Hornung}, \citenamefont {Zabel},
  \citenamefont {Flanders}, \citenamefont {Gladuli}, \citenamefont {Campoy},
  \citenamefont {Cherry}, \citenamefont {Luther}, \citenamefont {Hasenfuss},
  \citenamefont {Krinsky}, \citenamefont {Pumir}, \citenamefont {Gilmour},\
  and\ \citenamefont {Bodenschatz}}]{Luther2011}%
  \BibitemOpen
  \bibfield  {author} {\bibinfo {author} {\bibfnamefont {S.}~\bibnamefont
  {Luther}}, \bibinfo {author} {\bibfnamefont {F.~H.}\ \bibnamefont {Fenton}},
  \bibinfo {author} {\bibfnamefont {B.~G.}\ \bibnamefont {Kornreich}}, \bibinfo
  {author} {\bibfnamefont {A.}~\bibnamefont {Squires}}, \bibinfo {author}
  {\bibfnamefont {P.}~\bibnamefont {Bittihn}}, \bibinfo {author} {\bibfnamefont
  {D.}~\bibnamefont {Hornung}}, \bibinfo {author} {\bibfnamefont
  {M.}~\bibnamefont {Zabel}}, \bibinfo {author} {\bibfnamefont
  {J.}~\bibnamefont {Flanders}}, \bibinfo {author} {\bibfnamefont
  {A.}~\bibnamefont {Gladuli}}, \bibinfo {author} {\bibfnamefont
  {L.}~\bibnamefont {Campoy}}, \bibinfo {author} {\bibfnamefont {E.~M.}\
  \bibnamefont {Cherry}}, \bibinfo {author} {\bibfnamefont {G.}~\bibnamefont
  {Luther}}, \bibinfo {author} {\bibfnamefont {G.}~\bibnamefont {Hasenfuss}},
  \bibinfo {author} {\bibfnamefont {V.~I.}\ \bibnamefont {Krinsky}}, \bibinfo
  {author} {\bibfnamefont {A.}~\bibnamefont {Pumir}}, \bibinfo {author}
  {\bibfnamefont {R.~F.}\ \bibnamefont {Gilmour}}, \ and\ \bibinfo {author}
  {\bibfnamefont {E.}~\bibnamefont {Bodenschatz}},\ }\bibfield  {title}
  {\enquote {\bibinfo {title} {Low-energy control of electrical turbulence in
  the heart},}\ }\href@noop {} {\bibfield  {journal} {\bibinfo  {journal}
  {Nature}\ }\textbf {\bibinfo {volume} {475}},\ \bibinfo {pages} {235--9}
  (\bibinfo {year} {2011})}\BibitemShut {NoStop}%
\bibitem [{\citenamefont {Barkley}(1992)}]{barkley1992}%
  \BibitemOpen
  \bibfield  {author} {\bibinfo {author} {\bibfnamefont {D.}~\bibnamefont
  {Barkley}},\ }\bibfield  {title} {\enquote {\bibinfo {title} {Linear
  stability analysis of rotating spiral waves in excitable media},}\ }\href
  {\doibase 10.1103/PhysRevLett.68.2090} {\bibfield  {journal} {\bibinfo
  {journal} {Phys. Rev. Lett.}\ }\textbf {\bibinfo {volume} {68}},\ \bibinfo
  {pages} {2090--2093} (\bibinfo {year} {1992})}\BibitemShut {NoStop}%
\bibitem [{\citenamefont {Henry}\ and\ \citenamefont
  {Hakim}(2002)}]{Henry2002}%
  \BibitemOpen
  \bibfield  {author} {\bibinfo {author} {\bibfnamefont {H.}~\bibnamefont
  {Henry}}\ and\ \bibinfo {author} {\bibfnamefont {V.}~\bibnamefont {Hakim}},\
  }\bibfield  {title} {\enquote {\bibinfo {title} {Scroll waves in isotropic
  excitable media: Linear instabilities, bifurcations, and restabilized
  states},}\ }\href@noop {} {\bibfield  {journal} {\bibinfo  {journal} {Phys.
  Rev. E}\ }\textbf {\bibinfo {volume} {65}},\ \bibinfo {pages} {046235}
  (\bibinfo {year} {2002})}\BibitemShut {NoStop}%
\bibitem [{\citenamefont {Beyn}\ and\ \citenamefont
  {Th{\"u}mmler}(2004)}]{BeTh04}%
  \BibitemOpen
  \bibfield  {author} {\bibinfo {author} {\bibfnamefont {W.-J.}\ \bibnamefont
  {Beyn}}\ and\ \bibinfo {author} {\bibfnamefont {V.}~\bibnamefont
  {Th{\"u}mmler}},\ }\bibfield  {title} {\enquote {\bibinfo {title} {Freezing
  solutions of equivariant evolution equations},}\ }\href@noop {} {\bibfield
  {journal} {\bibinfo  {journal} {SIAM J. Appl. Dyn. Syst.}\ }\textbf {\bibinfo
  {volume} {3}},\ \bibinfo {pages} {85--116} (\bibinfo {year}
  {2004})}\BibitemShut {NoStop}%
\bibitem [{\citenamefont {Marcotte}\ and\ \citenamefont
  {Grigoriev}(2014)}]{Marcotte2014}%
  \BibitemOpen
  \bibfield  {author} {\bibinfo {author} {\bibfnamefont {C.~D.}\ \bibnamefont
  {Marcotte}}\ and\ \bibinfo {author} {\bibfnamefont {R.~O.}\ \bibnamefont
  {Grigoriev}},\ }\href@noop {} {\enquote {\bibinfo {title} {Unstable spiral
  waves and local euclidean symmetry in a model of cardiac tissue},}\ }
  (\bibinfo {year} {2014}),\ \bibinfo {note} {\arXiv{1412.4731}, submitted to
  Chaos}\BibitemShut {NoStop}%
\bibitem [{\citenamefont {Bohr}, \citenamefont {Huber},\ and\ \citenamefont
  {Ott}(1996)}]{BohHubOtt96}%
  \BibitemOpen
  \bibfield  {author} {\bibinfo {author} {\bibfnamefont {T.}~\bibnamefont
  {Bohr}}, \bibinfo {author} {\bibfnamefont {G.}~\bibnamefont {Huber}}, \ and\
  \bibinfo {author} {\bibfnamefont {E.}~\bibnamefont {Ott}},\ }\bibfield
  {title} {\enquote {\bibinfo {title} {The structure of spiral domain
  patterns},}\ }\href@noop {} {\bibfield  {journal} {\bibinfo  {journal}
  {Europhys. Lett.}\ }\textbf {\bibinfo {volume} {33}},\ \bibinfo {pages} {589}
  (\bibinfo {year} {1996})}\BibitemShut {NoStop}%
\bibitem [{\citenamefont {Bohr}, \citenamefont {Huber},\ and\ \citenamefont
  {Ott}(1997)}]{BohHubOtt97}%
  \BibitemOpen
  \bibfield  {author} {\bibinfo {author} {\bibfnamefont {T.}~\bibnamefont
  {Bohr}}, \bibinfo {author} {\bibfnamefont {G.}~\bibnamefont {Huber}}, \ and\
  \bibinfo {author} {\bibfnamefont {E.}~\bibnamefont {Ott}},\ }\bibfield
  {title} {\enquote {\bibinfo {title} {The structure of spiral-domain patterns
  and shocks in the {2D} complex {Ginzburg-Landau} equation},}\ }\href@noop {}
  {\bibfield  {journal} {\bibinfo  {journal} {Physica D}\ }\textbf {\bibinfo
  {volume} {106}},\ \bibinfo {pages} {95--112} (\bibinfo {year}
  {1997})}\BibitemShut {NoStop}%
\bibitem [{\citenamefont {Karma}(1994)}]{Karma1994}%
  \BibitemOpen
  \bibfield  {author} {\bibinfo {author} {\bibfnamefont {A.}~\bibnamefont
  {Karma}},\ }\bibfield  {title} {\enquote {\bibinfo {title} {Electrical
  alternans and spiral wave breakup in cardiac tissue},}\ }\href {\doibase
  10.1063/1.166024} {\bibfield  {journal} {\bibinfo  {journal} {Chaos}\
  }\textbf {\bibinfo {volume} {4}},\ \bibinfo {pages} {461--472} (\bibinfo
  {year} {1994})}\BibitemShut {NoStop}%
\bibitem [{\citenamefont {Bevans}\ \emph {et~al.}(1998)\citenamefont {Bevans},
  \citenamefont {Kordel}, \citenamefont {Rhee},\ and\ \citenamefont
  {Harris}}]{Bevans98}%
  \BibitemOpen
  \bibfield  {author} {\bibinfo {author} {\bibfnamefont {C.~G.}\ \bibnamefont
  {Bevans}}, \bibinfo {author} {\bibfnamefont {M.}~\bibnamefont {Kordel}},
  \bibinfo {author} {\bibfnamefont {S.~K.}\ \bibnamefont {Rhee}}, \ and\
  \bibinfo {author} {\bibfnamefont {A.~L.}\ \bibnamefont {Harris}},\ }\bibfield
   {title} {\enquote {\bibinfo {title} {Isoform composition of connexin
  channels determines selectivity among second messengers and uncharged
  molecules},}\ }\href {\doibase {10.1074/jbc.273.5.2808}} {\bibfield
  {journal} {\bibinfo  {journal} {J. Biol. Chem.}\ }\textbf {\bibinfo {volume}
  {273}},\ \bibinfo {pages} {2808--2816} (\bibinfo {year} {1998})}\BibitemShut
  {NoStop}%
\bibitem [{\citenamefont {Garcia-Dorado}, \citenamefont {Rodriguez-Sinovas},\
  and\ \citenamefont {Ruiz-Meana}(2004)}]{Garcia04}%
  \BibitemOpen
  \bibfield  {author} {\bibinfo {author} {\bibfnamefont {D.}~\bibnamefont
  {Garcia-Dorado}}, \bibinfo {author} {\bibfnamefont {A.}~\bibnamefont
  {Rodriguez-Sinovas}}, \ and\ \bibinfo {author} {\bibfnamefont
  {M.}~\bibnamefont {Ruiz-Meana}},\ }\bibfield  {title} {\enquote {\bibinfo
  {title} {Gap junction-mediated spread of cell injury and death during
  myocardial ischemia-reperfusion},}\ }\href {\doibase
  {10.1016/j.cardiores.2003.11.039}} {\bibfield  {journal} {\bibinfo  {journal}
  {Cardiovasc. Res.}\ }\textbf {\bibinfo {volume} {61}},\ \bibinfo {pages}
  {386--401} (\bibinfo {year} {2004})}\BibitemShut {NoStop}%
\bibitem [{\citenamefont {Auerbach}\ \emph {et~al.}(1987)\citenamefont
  {Auerbach}, \citenamefont {Cvitanovi{\'c}}, \citenamefont {Eckmann},
  \citenamefont {Gunaratne},\ and\ \citenamefont {Procaccia}}]{pchaot}%
  \BibitemOpen
  \bibfield  {author} {\bibinfo {author} {\bibfnamefont {D.}~\bibnamefont
  {Auerbach}}, \bibinfo {author} {\bibfnamefont {P.}~\bibnamefont
  {Cvitanovi{\'c}}}, \bibinfo {author} {\bibfnamefont {J.-P.}\ \bibnamefont
  {Eckmann}}, \bibinfo {author} {\bibfnamefont {G.}~\bibnamefont {Gunaratne}},
  \ and\ \bibinfo {author} {\bibfnamefont {I.}~\bibnamefont {Procaccia}},\
  }\bibfield  {title} {\enquote {\bibinfo {title} {Exploring chaotic motion
  through periodic orbits},}\ }\href@noop {} {\bibfield  {journal} {\bibinfo
  {journal} {Phys. Rev. Lett.}\ }\textbf {\bibinfo {volume} {58}},\ \bibinfo
  {pages} {23} (\bibinfo {year} {1987})}\BibitemShut {NoStop}%
\bibitem [{\citenamefont {Cvitanovi{\'c}}\ and\ \citenamefont
  {Gibson}(2010)}]{CviGib10}%
  \BibitemOpen
  \bibfield  {author} {\bibinfo {author} {\bibfnamefont {P.}~\bibnamefont
  {Cvitanovi{\'c}}}\ and\ \bibinfo {author} {\bibfnamefont {J.~F.}\
  \bibnamefont {Gibson}},\ }\bibfield  {title} {\enquote {\bibinfo {title}
  {Geometry of turbulence in wall-bounded shear flows: {Periodic} orbits},}\
  }\href@noop {} {\bibfield  {journal} {\bibinfo  {journal} {Phys. Scr. T}\
  }\textbf {\bibinfo {volume} {142}},\ \bibinfo {pages} {014007} (\bibinfo
  {year} {2010})}\BibitemShut {NoStop}%
\bibitem [{\citenamefont {Aranson}\ and\ \citenamefont
  {Kramer}(2002)}]{Aranson2002}%
  \BibitemOpen
  \bibfield  {author} {\bibinfo {author} {\bibfnamefont {I.~S.}\ \bibnamefont
  {Aranson}}\ and\ \bibinfo {author} {\bibfnamefont {L.}~\bibnamefont
  {Kramer}},\ }\bibfield  {title} {\enquote {\bibinfo {title} {The world of the
  complex ginzburg-landau equation},}\ }\href@noop {} {\bibfield  {journal}
  {\bibinfo  {journal} {Rev. Mod. Phys.}\ }\textbf {\bibinfo {volume} {74}},\
  \bibinfo {pages} {99--143} (\bibinfo {year} {2002})}\BibitemShut {NoStop}%
\bibitem [{\citenamefont {Kantz}(2004)}]{kantz04}%
  \BibitemOpen
  \bibfield  {author} {\bibinfo {author} {\bibfnamefont {H.}~\bibnamefont
  {Kantz}},\ }\href@noop {} {\emph {\bibinfo {title} {Nonlinear Time Series
  Analysis}}},\ \bibinfo {edition} {2nd}\ ed.\ (\bibinfo  {publisher}
  {Cambridge Univ. Press},\ \bibinfo {address} {Cambridge, {UK} ; New York},\
  \bibinfo {year} {2004})\BibitemShut {NoStop}%
\bibitem [{\citenamefont {Zhan}, \citenamefont {Luo},\ and\ \citenamefont
  {J.}(2007)}]{zhan07}%
  \BibitemOpen
  \bibfield  {author} {\bibinfo {author} {\bibfnamefont {M.}~\bibnamefont
  {Zhan}}, \bibinfo {author} {\bibfnamefont {J.}~\bibnamefont {Luo}}, \ and\
  \bibinfo {author} {\bibfnamefont {G.}~\bibnamefont {J.}},\ }\bibfield
  {title} {\enquote {\bibinfo {title} {Chirality effect on the global structure
  of spiral-domain patterns in the two-dimensional complex {Ginzburg-Landau}
  equation},}\ }\href {\doibase 10.1103/PhysRevE.75.016214} {\bibfield
  {journal} {\bibinfo  {journal} {Phys. Rev. E}\ }\textbf {\bibinfo {volume}
  {75}},\ \bibinfo {pages} {016214} (\bibinfo {year} {2007})}\BibitemShut
  {NoStop}%
\bibitem [{\citenamefont {Barkley}, \citenamefont {Kness},\ and\ \citenamefont
  {Tuckerman}(1990)}]{BaKnTu90}%
  \BibitemOpen
  \bibfield  {author} {\bibinfo {author} {\bibfnamefont {D.}~\bibnamefont
  {Barkley}}, \bibinfo {author} {\bibfnamefont {M.}~\bibnamefont {Kness}}, \
  and\ \bibinfo {author} {\bibfnamefont {L.~S.}\ \bibnamefont {Tuckerman}},\
  }\bibfield  {title} {\enquote {\bibinfo {title} {Spiral wave dynamics in a
  simple model of excitable media: {Transition} from simple to compound
  rotation},}\ }\href@noop {} {\bibfield  {journal} {\bibinfo  {journal} {Phys.
  Rev. A}\ }\textbf {\bibinfo {volume} {42}},\ \bibinfo {pages} {2489--2492}
  (\bibinfo {year} {1990})}\BibitemShut {NoStop}%
\bibitem [{\citenamefont {Luo}, \citenamefont {Zhang},\ and\ \citenamefont
  {Zhan}(2009)}]{luo09}%
  \BibitemOpen
  \bibfield  {author} {\bibinfo {author} {\bibfnamefont {J.}~\bibnamefont
  {Luo}}, \bibinfo {author} {\bibfnamefont {B.}~\bibnamefont {Zhang}}, \ and\
  \bibinfo {author} {\bibfnamefont {M.}~\bibnamefont {Zhan}},\ }\bibfield
  {title} {\enquote {\bibinfo {title} {Frozen state of spiral waves in
  excitable media},}\ }\href {\doibase 10.1063/1.3224034} {\bibfield  {journal}
  {\bibinfo  {journal} {Chaos}\ }\textbf {\bibinfo {volume} {19}},\ \bibinfo
  {pages} {033133} (\bibinfo {year} {2009})}\BibitemShut {NoStop}%
\bibitem [{\citenamefont {Howard}\ and\ \citenamefont
  {Kopell}(1977)}]{Howard1977}%
  \BibitemOpen
  \bibfield  {author} {\bibinfo {author} {\bibfnamefont {L.~N.}\ \bibnamefont
  {Howard}}\ and\ \bibinfo {author} {\bibfnamefont {N.}~\bibnamefont
  {Kopell}},\ }\bibfield  {title} {\enquote {\bibinfo {title} {Slowly varying
  waves and shock structures in reaction-diffusion equations},}\ }\href@noop {}
  {\bibfield  {journal} {\bibinfo  {journal} {Studies Appl. Math.}\ }\textbf
  {\bibinfo {volume} {56}},\ \bibinfo {pages} {95--145} (\bibinfo {year}
  {1977})}\BibitemShut {NoStop}%
\bibitem [{\citenamefont {Krinsky}\ and\ \citenamefont
  {Agladze}(1983)}]{Krinsky1983}%
  \BibitemOpen
  \bibfield  {author} {\bibinfo {author} {\bibfnamefont {V.~I.}\ \bibnamefont
  {Krinsky}}\ and\ \bibinfo {author} {\bibfnamefont {K.~I.}\ \bibnamefont
  {Agladze}},\ }\bibfield  {title} {\enquote {\bibinfo {title} {Interaction of
  rotating waves in an active chemical medium},}\ }\href@noop {} {\bibfield
  {journal} {\bibinfo  {journal} {Physica D}\ }\textbf {\bibinfo {volume}
  {8}},\ \bibinfo {pages} {50--56} (\bibinfo {year} {1983})}\BibitemShut
  {NoStop}%
\bibitem [{\citenamefont {Langham}\ and\ \citenamefont
  {Barkley}(2013)}]{LanBar13}%
  \BibitemOpen
  \bibfield  {author} {\bibinfo {author} {\bibfnamefont {J.}~\bibnamefont
  {Langham}}\ and\ \bibinfo {author} {\bibfnamefont {D.}~\bibnamefont
  {Barkley}},\ }\bibfield  {title} {\enquote {\bibinfo {title} {Non-specular
  reflections in a macroscopic system with wave-particle duality: {Spiral}
  waves in bounded media},}\ }\href {\doibase 10.1063/1.4793783} {\bibfield
  {journal} {\bibinfo  {journal} {Chaos}\ }\textbf {\bibinfo {volume} {23}},\
  \bibinfo {pages} {013134} (\bibinfo {year} {2013})},\ \bibinfo {note}
  {\arXiv{1304.0591}}\BibitemShut {NoStop}%
\bibitem [{\citenamefont {Langham}, \citenamefont {Biktasheva},\ and\
  \citenamefont {Barkley}(2014)}]{LanBar14}%
  \BibitemOpen
  \bibfield  {author} {\bibinfo {author} {\bibfnamefont {J.}~\bibnamefont
  {Langham}}, \bibinfo {author} {\bibfnamefont {I.}~\bibnamefont {Biktasheva}},
  \ and\ \bibinfo {author} {\bibfnamefont {D.}~\bibnamefont {Barkley}},\
  }\href@noop {} {\enquote {\bibinfo {title} {Asymptotic theory for spiral wave
  reflections},}\ } (\bibinfo {year} {2014}),\ \bibinfo {note}
  {\arXiv{11401.7626}}\BibitemShut {NoStop}%
\bibitem [{\citenamefont {Fiedler}\ \emph {et~al.}(1996)\citenamefont
  {Fiedler}, \citenamefont {Sandstede}, \citenamefont {Scheel},\ and\
  \citenamefont {Wulff}}]{FiSaScWu96}%
  \BibitemOpen
  \bibfield  {author} {\bibinfo {author} {\bibfnamefont {B.}~\bibnamefont
  {Fiedler}}, \bibinfo {author} {\bibfnamefont {B.}~\bibnamefont {Sandstede}},
  \bibinfo {author} {\bibfnamefont {A.}~\bibnamefont {Scheel}}, \ and\ \bibinfo
  {author} {\bibfnamefont {C.}~\bibnamefont {Wulff}},\ }\bibfield  {title}
  {\enquote {\bibinfo {title} {Bifurcation from relative equilibria of
  noncompact group actions: {Skew} products, meanders, and drifts},}\
  }\href@noop {} {\bibfield  {journal} {\bibinfo  {journal} {Doc. Math.}\
  }\textbf {\bibinfo {volume} {141}},\ \bibinfo {pages} {479--505} (\bibinfo
  {year} {1996})}\BibitemShut {NoStop}%
\bibitem [{\citenamefont {Fiedler}\ and\ \citenamefont
  {Turaev}(1998)}]{FiTu98}%
  \BibitemOpen
  \bibfield  {author} {\bibinfo {author} {\bibfnamefont {B.}~\bibnamefont
  {Fiedler}}\ and\ \bibinfo {author} {\bibfnamefont {D.}~\bibnamefont
  {Turaev}},\ }\bibfield  {title} {\enquote {\bibinfo {title} {Normal forms,
  resonances, and meandering tip motions near relative equilibria of
  {Euclidean} group actions},}\ }\href@noop {} {\bibfield  {journal} {\bibinfo
  {journal} {Arch. Rational Mech. Anal.}\ }\textbf {\bibinfo {volume} {145}},\
  \bibinfo {pages} {129--159} (\bibinfo {year} {1998})}\BibitemShut {NoStop}%
\bibitem [{\citenamefont {Sandstede}, \citenamefont {Scheel},\ and\
  \citenamefont {Wulff}(1997)}]{SaScWu97}%
  \BibitemOpen
  \bibfield  {author} {\bibinfo {author} {\bibfnamefont {B.}~\bibnamefont
  {Sandstede}}, \bibinfo {author} {\bibfnamefont {A.}~\bibnamefont {Scheel}}, \
  and\ \bibinfo {author} {\bibfnamefont {C.}~\bibnamefont {Wulff}},\ }\bibfield
   {title} {\enquote {\bibinfo {title} {Dynamics of spiral waves on unbounded
  domains using center-manifold reductions},}\ }\href@noop {} {\bibfield
  {journal} {\bibinfo  {journal} {J. Diff. Eqn.}\ }\textbf {\bibinfo {volume}
  {141}},\ \bibinfo {pages} {122--149} (\bibinfo {year} {1997})}\BibitemShut
  {NoStop}%
\bibitem [{\citenamefont {Sandstede}, \citenamefont {Scheel},\ and\
  \citenamefont {Wulff}(1999)}]{SaScWu99}%
  \BibitemOpen
  \bibfield  {author} {\bibinfo {author} {\bibfnamefont {B.}~\bibnamefont
  {Sandstede}}, \bibinfo {author} {\bibfnamefont {A.}~\bibnamefont {Scheel}}, \
  and\ \bibinfo {author} {\bibfnamefont {C.}~\bibnamefont {Wulff}},\ }\bibfield
   {title} {\enquote {\bibinfo {title} {Dynamical behavior of patterns with
  {Euclidean} symmetry},}\ }in\ \href@noop {} {\emph {\bibinfo {booktitle}
  {Pattern Formation in Continuous and Coupled Systems}}}\ (\bibinfo
  {publisher} {Springer},\ \bibinfo {address} {New York},\ \bibinfo {year}
  {1999})\ pp.\ \bibinfo {pages} {249--264}\BibitemShut {NoStop}%
\bibitem [{\citenamefont {Biktasheva}\ \emph {et~al.}(2009)\citenamefont
  {Biktasheva}, \citenamefont {Barkley}, \citenamefont {Biktashev},
  \citenamefont {Bordyugov},\ and\ \citenamefont {Foulkes}}]{BBBBF09}%
  \BibitemOpen
  \bibfield  {author} {\bibinfo {author} {\bibfnamefont {I.~V.}\ \bibnamefont
  {Biktasheva}}, \bibinfo {author} {\bibfnamefont {D.}~\bibnamefont {Barkley}},
  \bibinfo {author} {\bibfnamefont {V.~N.}\ \bibnamefont {Biktashev}}, \bibinfo
  {author} {\bibfnamefont {G.~V.}\ \bibnamefont {Bordyugov}}, \ and\ \bibinfo
  {author} {\bibfnamefont {A.~J.}\ \bibnamefont {Foulkes}},\ }\bibfield
  {title} {\enquote {\bibinfo {title} {Computation of the response functions of
  spiral waves in active media},}\ }\href@noop {} {\bibfield  {journal}
  {\bibinfo  {journal} {Phys. Rev. E}\ }\textbf {\bibinfo {volume} {79}},\
  \bibinfo {pages} {056702} (\bibinfo {year} {2009})}\BibitemShut {NoStop}%
\bibitem [{\citenamefont {Biktasheva}, \citenamefont {Biktashev},\ and\
  \citenamefont {Foulkes}(2010)}]{Biktasheva:2010co}%
  \BibitemOpen
  \bibfield  {author} {\bibinfo {author} {\bibfnamefont {I.~V.}\ \bibnamefont
  {Biktasheva}}, \bibinfo {author} {\bibfnamefont {V.~N.}\ \bibnamefont
  {Biktashev}}, \ and\ \bibinfo {author} {\bibfnamefont {A.~J.}\ \bibnamefont
  {Foulkes}},\ }\bibfield  {title} {\enquote {\bibinfo {title} {Computation of
  the drift velocity of spiral waves using response functions},}\ }\href@noop
  {} {\bibfield  {journal} {\bibinfo  {journal} {Phys. Rev. E}\ }\textbf
  {\bibinfo {volume} {81}},\ \bibinfo {pages} {066202} (\bibinfo {year}
  {2010})}\BibitemShut {NoStop}%
\bibitem [{\citenamefont {Sepulveda}, \citenamefont {Roth},\ and\ \citenamefont
  {Jr.}(1989)}]{Sepulveda1989}%
  \BibitemOpen
  \bibfield  {author} {\bibinfo {author} {\bibfnamefont {N.~G.}\ \bibnamefont
  {Sepulveda}}, \bibinfo {author} {\bibfnamefont {B.~J.}\ \bibnamefont {Roth}},
  \ and\ \bibinfo {author} {\bibfnamefont {J.~P.~W.}\ \bibnamefont {Jr.}},\
  }\bibfield  {title} {\enquote {\bibinfo {title} {Current injection into a
  two-dimensional anisotropic bidomain},}\ }\href@noop {} {\bibfield  {journal}
  {\bibinfo  {journal} {Biophys. J.}\ }\textbf {\bibinfo {volume} {55}},\
  \bibinfo {pages} {987--999} (\bibinfo {year} {1989})}\BibitemShut {NoStop}%
\bibitem [{\citenamefont {Alvarez-Lacalle}\ and\ \citenamefont
  {Echebarria}(2009)}]{Alvarez:2009}%
  \BibitemOpen
  \bibfield  {author} {\bibinfo {author} {\bibfnamefont {E.}~\bibnamefont
  {Alvarez-Lacalle}}\ and\ \bibinfo {author} {\bibfnamefont {B.}~\bibnamefont
  {Echebarria}},\ }\bibfield  {title} {\enquote {\bibinfo {title} {{Global
  coupling in excitable media provides a simplified description of
  mechanoelectrical feedback in cardiac tissue}},}\ }\href {\doibase
  {10.1103/PhysRevE.79.031921}} {\bibfield  {journal} {\bibinfo  {journal}
  {{Phys. Rev. E}}\ }\textbf {\bibinfo {volume} {{79}}},\ \bibinfo {pages}
  {031921} (\bibinfo {year} {{2009}})}\BibitemShut {NoStop}%
\end{thebibliography}%

\end{document}